\documentclass[fleqn,usenatbib]{mnras}

\usepackage{newtxtext,newtxmath}
\usepackage{multirow}
\usepackage{breqn}

\usepackage[T1]{fontenc}
\usepackage{ae,aecompl}

\usepackage{graphicx}	
\usepackage{amsmath}	
\usepackage{amssymb}	
\usepackage{epstopdf}	
\usepackage[usenames, dvipsnames]{color} 
\usepackage{pdflscape}	

\newcommand{\angstrom}{\textup{\AA}}
\newcommand{\kms}{\,km\,s$^{-1}$} 

\newcommand{\hii}{\mbox{H\,{\scshape ii}}\,}
\newcommand{\oii}{\mbox{O\,{\scshape ii}}\,}
\newcommand{\oiii}{\mbox{O\,{\scshape iii}}\,}
\newcommand{\neiii}{\mbox{Ne\,{\scshape iii}}\,}
\newcommand{\nii}{\mbox{N\,{\scshape ii}}\,}
\newcommand{\feii}{\mbox{Fe\,{\scshape ii}}\,}
\newcommand{\mgii}{\mbox{Mg\,{\scshape ii}}\,}

\title[MUSE Gravitational Arcs]{Kinematics, Turbulence and Star Formation of $z\sim$1 Strongly Lensed Galaxies seen with MUSE}

\author[Patr\'icio et al.]{
V. Patr\'icio$^{1,2}$\thanks{E-mail: vera.patricio@dark-cosmology.dk}, J. Richard$^{1}$, D. Carton$^{1}$, T. Contini$^{3}$, B. Epinat$^{3,4}$, J. Brinchmann$^{5,6}$, 
\newauthor
K. B. Schmidt$^{7}$, D. Krajnovi\'c$^{7}$, N. Bouch\'e$^{3}$, P. M. Weilbacher$^{7}$, R. Pell\'o$^{3}$, J. Caruana$^{8}$,
\newauthor
M. Maseda$^{5}$, H. Finley$^{3}$, F. E. Bauer$^{9,10,11}$, J. Martinez$^{1}$, G. Mahler$^{1}$, D. Lagattuta$^{1}$,
\newauthor
B. Cl\'ement$^{1}$, G. Soucail$^{3}$, and L. Wisotzki$^{7}$\\
$^{1}$ Univ Lyon, Univ Lyon1, Ens de Lyon, CNRS, Centre de Recherche Astrophysique de Lyon UMR5574, F-69230, Saint-Genis-Laval, France\\
$^{2}$ Dark Cosmology Centre, Niels Bohr Institute, University of Copenhagen, Juliane Maries Vej 30, 2100 Copenhagen, Denmark\\
$^{3}$ Institut de Recherche en Astrophysique et Plan\'etologie (IRAP), Universit\'e de Toulouse, CNRS, UPS, F-31400 Toulouse, France\\
$^{4}$ Aix Marseille Univ, CNRS, LAM, Laboratoire d'Astrophysique de Marseille, Marseille, France \\
$^{5}$ Leiden Observatory, Leiden University, P.O. Box 9513, 2300 RA Leiden, The Netherlands \\
$^{6}$ Instituto de Astrof\'isica e Ci\^encias do Espa\c{c}o, Universidade do Porto, CAUP, Rua das Estrelas, PT4150-762 Porto, Portugal\\
$^{7}$ Leibniz-Institut f\"ur Astrophysik Potsdam, AIP, An der Sternwarte 16, D-14482 Potsdam, Germany \\
$^{8}$ University of Malta, Msida MSD 2080, Malta\\
$^{9}$ Instituto de Astrof\'isica and Centro de Astroingenier\'ia, Facultad de F\'isica, Pontificia Universidad Cat\'olica de Chile,
Casilla 306, Chile\\ 
$^{10}$ Millennium Institute of Astrophysics (MAS), Nuncio Monse\~nor S\'otero Sanz 100, Providencia, Santiago, Chile\\
$^{11}$ Space Science Institute, 4750 Walnut Street, Suite 205, Boulder, Colorado 80301 
}

\date{Accepted XXX. Received YYY; in original form ZZZ}

\pubyear{2018}

\begin{document}
\label{firstpage}
\pagerange{\pageref{firstpage}--\pageref{lastpage}}
\maketitle
\begin{abstract} 
We analyse a sample of 8 highly magnified galaxies at redshift $0.6<z<1.5$ observed with MUSE, exploring the resolved properties of these galaxies at sub-kiloparsec scales. Combining multi-band \emph{HST} photometry and MUSE spectra, we derive the stellar mass, global star formation rates, extinction and metallicity from multiple nebular lines, concluding that our sample is representative of $z\sim$1 star-forming galaxies. We derive the 2D kinematics of these galaxies from the [\oii] emission and model it with a new method that accounts for lensing effects and fits multiple images simultaneously. 
We use these models to calculate the 2D beam-smearing correction and derive intrinsic velocity dispersion maps. We find them to be fairly homogeneous, with relatively constant velocity dispersions between 15 - 80 \kms and Gini coefficent of $\la0.3$. We do not find any evidence for higher (or lower) velocity dispersions at the positions of bright star-forming clumps. We derive resolved maps of dust attenuation and attenuation-corrected star formation rates from emission lines for two objects in the sample. We use this information to study the relation between resolved star formation rate and velocity dispersion. We find that these quantities are not correlated, and the high velocity dispersions found for relatively low star-forming densities seems to indicate that,  at sub-kiloparsec scales, turbulence in high-$z$ discs is mainly dominated by gravitational instability rather than
stellar feedback.
\end{abstract}



\section{Introduction}

High redshift disc galaxies display some striking differences when compared with their local counterparts: not only do they harbour giant \hii\ star-forming regions, but they also have higher gas velocity dispersions and higher gas fractions than local discs (see \citealt{Glazebrook+13} for a review).  The star-forming regions seen in these high-$z$ discs, referred to as clumps or knots (usually identified in rest-frame UV/optical images; e.g. \citealt{Elmegreen2006}), are also more extreme than the local star-forming regions. They can have sizes of up to one kiloparsec, star formation rates between 0.5 and 100 M$_\odot\,$yr$^{-1}$ and masses up to $10^8-10^{10}$ M$_\odot$ (e.g. \citealt{Swinbank2009,Jones2010,ForsterSchreiber2011,Dessauges-Zavadsky2011,Guo2012}), which makes them significantly more massive than giant clouds in local galaxies, with masses of $10^5-10^6$ M$_\odot$. While it was initially speculated that these features had their origin in merging episodes, the advent of Integral Field Spectrographs showed that most clumpy galaxies have smooth velocity fields and are rotationally supported (e.g. \citealt{Genzel2006,Bouche2007,Cresci2009,ForsterSchreiber2009,Wisnioski2015,Contini2016}), suggesting that these clumps may be part of the secular evolution of disc galaxies. 

The physical picture that explains why rotating discs are more clumpy and turbulent at $z\geq$1 is still being investigated, both observationally and theoretically. One possible scenario is that the highly turbulent interstellar medium is stirred by radiation pressure and winds of strong star formation taking place in these young galaxies (e.g. \citealt{Lehnert2009,Green2010,Lehnert2013}). Another possibility is that the inflow of gas into the galaxy from the cosmic web provides enough energy to sustain these high velocity dispersions, although some works seem to point that this energy is insufficient to maintain the high-velocity dispersions over long timescales (e.g. \citealt{Elmegreen2010}). Finally, a third scenario is gravitational instability: the high gas fractions of these galaxies make them gravitationally unstable, causing gas to spiral down to the centre of the galaxies, converting gravitational energy into turbulent motions and driving galaxies to stability (e.g. \citealt{Bournaud2007,Ceverino2010,Krumholz2010}). Possibly, a combination of factors is at play. Recent simulations by \citealt{Krumholz2017}, that include both stellar-feedback and gravitational instability (transport driven) turbulence, show that while both processes contribute to the gas turbulence, transport-driven mechanisms dominate, especially for $z>0.5$ and more massive galaxies.  

From the observational side, numerous surveys have provided some insights into these hypothesis. Recent work by \citealt{Johnson2017} studied the velocity dispersion of $\sim$450 star forming galaxies at $z\sim0.9$ from the KROSS survey, finding that the data are equally well explained by a scenario where turbulence is driven by stellar feedback or increased gas fractions. Previous studies have also not provided a definitive answer. In a analysis of $z\sim3$ galaxy analogues (the DYNAMO sample), \citealt{Green2014} found a correlation between the star formation rates and velocity dispersions that supports the idea that turbulence is driven by stellar feedback. On the other hand, at high redshift ($z\sim2$), the analysis of KMOS$^{3D}$ survey data by \citealt{Wisnioski2015} found only a weak correlation between gas velocity dispersion and SFR or gas fractions at each redshift.

Another possible path to understanding the properties of these young discs is to study their resolved properties at sub-kiloparsec scales. This is particularly challenging when trying to derive intrinsic velocity dispersions. Owing to the limited resolution of observations, the velocity gradient present in rotating discs artificially increases the observed velocity dispersion of galaxies. The effects of beam smearing are particularly problematic in the centres of galaxies, where the velocity field rapidly changes. A few works used adaptive optics to measured the resolved intrinsic velocity dispersion maps of high-$z$ disc galaxies, overcoming some of the issues caused by beam-smearing \cite[e.g][]{Genzel2011,Swinbank2012_kin,Newman2013}. \citealt{Genzel2011} find that the velocity dispersion maps are broadly compatible with a constant distribution, as seen in local disc galaxies, despite its higher value ($\sigma\sim$60 \kms compared with 10-20\kms in local galaxies). However, other studies find a relation between local star formation rate surface densities ($\Sigma_{\rm SFR}$) and velocity dispersions, which seems to point to some degree of structure (e.g. \citealt{Swinbank2012_kin,Lehnert2009}). 

Studying turbulence in the high-$z$ star-forming clumps could also provide some insights into how they differ from their local counterparts. High spatial resolution is necessary to study both the morphology of the velocity dispersion and the clumps' turbulence. However, many studies of high-redshift galaxies are still severely hampered by the relatively low spatial resolution achieved with the current facilities. A possible strategy to improve this is to target lensed galaxies, where even distant $z\sim1$ galaxies can be resolved down to a few hundred parsecs (e.g. \citealt{Jones2010,Livermore2015,Leethochawalit2016,Yuan2017}). Here, we target a sample of typical $z\sim$1 clumpy discs that, due to strong gravitational lensing, appear as extremely extended objects in the sky. Their high magnification factors, as well as the high quality MUSE\footnote{Multi Unit Spectroscopic Explorer} data acquired with excellent seeing, allows us to alleviate resolution issues, providing a sub-galactic view of the properties of these typical galaxies.

This paper is organized as follows. In Section \ref{sec:obs}, we present the sample and the MUSE observations used in this work as well as other ancillary data and the lensing models used to recover the intrinsic properties of these galaxies. In Section \ref{sec:sample} we derive the integrated physical properties of this sample from the MUSE spectra and \emph{HST} photometry, and in Section~\ref{sec:resolved} we study the resolved kinematic properties of the galaxies. We conclude with a discussion of the results and comparison with other samples and works in Section~\ref{sec:discussion}. Throughout this paper, we adopt a $\Lambda$-CDM cosmology with $\Omega$=0.7, $\Omega_m$=0.3 and H$_0$ = 70 km\,s$^{-1}$\,Mpc$^{-1}$. Magnitudes are provided in the AB photometric system \citep{Oke1974}. We adopt a solar metallicity of $12+\log$(O/H) = 8.69 \citep{AllendePrieto2001} and the \citealt{Chabrier2003} initial mass function. 


\section{Observations and Data Reduction}
\label{sec:obs} 

In this work, we analyse a sample of eight strongly lensed galaxies that lie behind eight different galaxy clusters (see Table~\ref{tab:muse_obs}). All of these clusters were observed with MUSE: Abell 370 (A370), Abell 2390 (A2390), MACSJ0416.12403 (M0416), MACSJ1206.2-0847 (M1206), Abell 2667 (A2667) and Abell 521 (A521) within the MUSE Guaranteed Time Observations (GTO) Lensing Clusters Programme (PI: Richard); Abell S1063/RXJ2248-4431 (AS1063) during MUSE science verification (PI: Caputi \& Cl\'ement); and MACSJ1149.5+2223 (M1149) during Director's Discretionary Time (PI: Grillo), targeting the newly discovered supernova in a lensed galaxy \citep{Kelly2015}. 

\begin{table*}
\caption{List of gravitational arcs. Coordinates $\alpha$ and $\delta$ correspond to the complete image position (see Section~\ref{subsec:multiple_images}). The Point Spread Function (PSF) FWHM was obtained by fitting a MUSE pseudo F814W image with a seeing convolved \emph{HST} F814W image. The magnification factor $\mu_{\rm MUSE}$ is the mean amplification factor within the MUSE aperture, predicted by the lensing model in Ref $\mu$. This  is merely indicative, since spectra were corrected pixel by pixel. }
\label{tab:muse_obs}
\centering
\tabcolsep=0.10cm
\begin{tabular}{|llccccccccc|} 
\hline
Object      & MUSE & $\alpha$ & $\delta$    & Exp.  & PSF    & $z$ & Ref $z$ & Size & $\mu_{\rm MUSE}$ & Ref $\mu$ \\
         	&Program & (J2000.0) &		  & (hrs) & (\arcsec)  &   &     & (\arcsec $^2$)  &        &		\\
\hline\hline
AS1063-arc   & 060.A-9345$^{(a)}$ & 22:48:42  & -44:31:57 & 3.25 & 1.03 & 0.611  & \citealt{Gomez2012}   & 33 &  4$\pm$1 & Cl\'ement et al. in prep. \\
A370-sys1    & 094.A-0115,        & 02:39:53  & -01:35:05 & 6.0 & 0.70 & 0.725  & \citealt{Soucail1988} & 30 &  17$\pm$1 & \citealt{Lagattuta2017}\\
         	 & 096.A-0710         &           &           &        &  &    &      &        \\                
A2390-arc    & 094.A-0115         & 21:53:34  & +17:41:59 & 2.0  & 0.56 &  0.913 &\citealt{Pello1991}    & 31 & 10$\pm$1 & Pello et al. in prep.\\
M0416-sys28  & 094.A-0115         & 04:16:10  & -24:04:16 & 2.0  & 0.45 &  0.940 & \citealt{Caminha2017}  & 15 & 29$\pm$2 &  Richard et al. in prep.     	\\ 
M1206-sys1   & 095.A-0181,        & 12:06:11  & -08:48:05 & 3.0  & 0.51 & 1.033  & \citealt{Ebeling2009}  & 50 & 18$\pm$1&   
\citealt{Cava2017}\\
             & 097.A-0269         &           &           &      &        &        &                     &    &&	\\
A2667-sys1   & 094.A-0115         & 23:51:39  & -26:04:50 & 2.0  & 0.62 & 1.033  & \citealt{Covone2006} & 89 & 30$\pm$2 &   Pello et al. in prep.\\
A521-sys1    & 100.A-0249 & 04:54:06  	  & -10:13:23 	  & 1.67  & 0.57$^{(c)}$ & 1.043  & \citealt{Richard2010a} & 33 & 40$\pm$3 &  \citealt{Richard2010a} \\
M1149-sys1   & 294.A-5032$^{(b)}$ & 11:49:35  & +22:23:45 & 4.8  & 0.57 & 1.491  & \citealt{Smith2009}  & 30 &  9$\pm$1 &      \citealt{Jauzac2016}\\
\hline
\end{tabular}\\
\begin{flushleft}
$^{(a)}$ see also \citealt{Karman2015} $^{(b)}$ see also 
\citealt{Grillo2016} $^{(c)}$ measured in F606W.
\end{flushleft}
\end{table*}

We base our selection criteria on the apparent (i.e. due to gravitational lensing magnification) size of the targets with the goal of resolving these galaxies at sub-kiloparsec scales. Within the sample of lensing clusters observed with MUSE, we select the highly magnified ($\mu>3$) and exceptionally extended ($>4$\,arcsec$^{2}$) objects, generally dubbed \textit{gravitational arcs}, that allow to probe properties at physical scales down to a few hundred parsec at $z=1$. For comparision, without lensing, at this redshift, a good seeing of 0.6" results in a resolution of $\sim$5 kpc. In order to investigate the physical properties of these gravitational arcs, such as kinematics and star formation rates, we require that strong, non-resonant emission lines are detected. Therefore, we target objects with visible [\oii]${\lambda3726,29}$ emission in the range 4650 and 9300 \AA. This limits our sample to eight galaxies between $0.6<z<1.5$. 

For clarity, we refer to these galaxies by the cluster name plus their respective number in the lensing models used in this work (see Table~\ref{tab:muse_obs}), for example "A370-sys1" or "A2667-sys1". Exceptions to this are the objects in cluster AS1063 and A2390, which do not have multiple images and are referred to as "AS1063-arc" and "A2390-arc". 

In the following subsections, we discuss the observations and data reduction and briefly summarise the ancillary data available for these targets, four of which are part of the Frontier Fields \emph{HST} programme \citep{Lotz2016}. We also present the lensing models and magnification factors used throughout the paper to derive the intrinsic properties of the gravitational arcs.

\subsection{MUSE observations and Data Reduction}
\label{subsec:muse_obs} 

The targets were observed with MUSE between 2014 and 2017. MUSE is an Integral Field Unit instrument with a field of view of 1'$\times$1', sampled at 0.2" per pixel, and covers the optical wavelength range between 4650 and 9300 \AA\, with a spectral sampling of 1.25\AA\, \citep{Bacon2014}. The GTO targets have a variety of integration depths (between 2 and 6h). Each observation is comprised of individual 1800\,s exposures. The position angle is rotated by 90 degrees between each exposure, minimising the stripe pattern that arises from the image slicer. In order to further minimise this pattern, a small dither ($<1\arcsec$) was added to each pointing. The observing strategy for the non-GTO targets (AS1063, MACS1149 and partially A370) is described in \citealt{Karman2015} and \citealt{Grillo2016} respectively. They follow very similar strategies to the one used within the GTO sample, including a small dithering and a 90 degrees rotation between exposures. A521 was observed in Oct. 2017 with the newly-commissioned Adaptive Optics Facility.

The data from all eight clusters was reduced using the ESO MUSE reduction pipeline version 1.2 \citep{Weilbacher2016}. The calibration files used to perform bias subtraction and flat fielding, including illumination and twilight exposures, were chosen to be the ones closest to the date of the exposure. To perform flux calibration and telluric correction, all 15 available standard stars were reduced and their flux and telluric response derived using the {\sc esorex} recipe {\sc muse\_standard}. The response curves were visually inspected and extreme responses (e.g. very high or low, or with unusual features compared with others) were removed before producing a median flux calibration and telluric correction with the 6 remaining response curves. These median calibration curves were applied to all individual cubes. 

The individual cubes were aligned with \emph{HST} F814W images using a combination of {\sc sextractor} \citep{Bertin1996}, to identify bright sources in individual the cubes and \emph{HST} image, and {\sc scamp} \citep{Bertin2006}, to calculate the offset between the two. 

Sky subtraction was performed using the ESO pipeline and the remaining sky subtraction residuals were removed from each individual exposure using the \textit{Zurich Atmosphere Purge} tool ({\sc ZAP} version 1; \citealt{Soto2016}), a principal component analysis that isolates and removes sky line residuals. Finally, the individual exposures were combined, rejecting voxels more than 3 $\sigma$ from the median, in order to eliminate cosmic rays. To normalize the exposures, we estimate the sky transparency using the median of the fluxes estimated with {\sc sextractor}. For each cluster, the exposure with the highest median flux was taken as the reference and the mean flux of the other exposures was rescaled to match this reference, before combining the data. A second sky residual subtraction using {\sc ZAP} was then performed on the combined cubes, and a median filter along the wavelength axis (box of 100 \AA) was applied in order to further smooth the background. During both operations, {\sc ZAP} and median subtraction, the brightest sources were masked when estimating the background. Finally, \citealt{Bacon2014} used the same data reduction and found that the variance propagated by the pipeline is underestimated by a factor of 1.6. We correct the variances in out cubes by this factor.

To determine the PSF size, the final cubes were compared to the available \emph{HST} data. By weighting the MUSE cubes with the F814W filter transmission curve, we produced MUSE F814W pseudo broad band images. We model the PSF following the \citealt{Bacon2017} method, which consists of assuming a Moffat profile, with a fixed power index of 2.8, and fitting the Full-Width Half Maximum (FWHM) by minimising the difference between the MUSE pseudo broad band and the \emph{HST} F814W image convolved with the Moffat kernel (F606W was used for A521). With the exception of AS1063, all targets were observed in excellent seeing conditions, always better than a FWHM of 0.70\arcsec (see Table~\ref{tab:muse_obs}).

\subsection{Ancillary Data and Lensing Models}
\label{subsec:lens_models}

The gravitational arcs used in this work were selected to be strongly magnified and are either multiply-imaged or very close to the multiple-image region near the cluster core. In order to recover their intrinsic (i.e. corrected for the lensing magnification) properties and morphology, accurate lensing mass models are required. 

The selected clusters have all been well studied in the past using numerous and deep \emph{HST} observations. All clusters except A370, A2390 and A521 are part of the Cluster Lensing And Supernova survey with Hubble (CLASH, \citealt{Postman2012}), for which photometry is available in 12 filters (see Table~\ref{tab:photom} for full list). Furthermore, four of them -- AS1063, A370, MACS0416 and MACS1149 -- are part of the Frontier Fields initiative \citep{Lotz2016}, that provided deep imaging in the F435W, F606W, F814W, F105W, F125W, F140W and F160W  bands. 

Only A2390 and A521 were not part of any of these programmes. For A2390, abundant \emph{HST} data is also available (see \citealt{Richard2008,Olmstead2014}, as well as Table \ref{tab:photom}). For A521 only WFPC2/F606W band is available plus a NIRC2/Ks band ground-based image retrieved from the Keck archive. This wealth of data has allowed a large number of multiple systems to be identified (between 5 and 60, creating up to 200 multiple images, e.g. \citealt{Caminha2017}) resulting in very well-constrained lensing models. The MUSE-GTO programme confirmed, as well as discovered, a significant number of these systems, improving the mass models in the process \citep{Richard2015,Bina2016,Lagattuta2017,Mahler2018}.

The mass models used in this work were all constructed using the {\sc Lenstool}\footnote{\url{https://projets.lam.fr/projects/lenstool/wiki}} software \citep{Jullo2007}, following the same methodology described in detail in earlier works (e.g. \citealt{Richard2009}). To summarise, we model the 2D-projected mass distribution of the cluster as a parametric combination of cluster-scale and galaxy-scale dPIE (double Pseudo Isothermal Elliptical, \citealt{Eliasdottir2007}) potentials. To limit the number of parameters in the model, the centres and shapes of the galaxy-scale components are tied to the centroid, ellipticity and position angle of cluster members as measured on the \emph{HST} images. The selection of cluster members is performed using the red sequence in a colour-magnitude diagram (see e.g. \citealt{Richard2014}) and confirmed with MUSE spectroscopy. Constraints on the lens model are robustly identified multiple images selected based on their morphology, colours and/or spectroscopic redshifts. Although previous {\sc Lenstool} models produced by our team were published for some of the clusters (e.g. \citealt{Richard2010a} for A2390 and A2667, \citealt{Richard2010b} for A370), we update them using the most recent spectroscopic information, in particular coming from MUSE \citep{Jauzac2016,Lagattuta2017}. We summarize the references of the lensing models used in Table \ref{tab:muse_obs}.


\section{Sample characterisation}
\label{sec:sample} 

We start by describing the lensed morphology of these galaxies and then proceed to derive their integrated properties, such as mass, gas-phase metallicity\footnote{Unless stated otherwise, we refer to \emph{gas-phase oxygen abundance} simply as \emph{metallicity}.} and star formation rate.

\subsection{Morphology}
\label{subsec:multiple_images}

Most galaxies studied in this work have multiple images, i.e., the same object can be seen in two (or more) different positions in the sky with different distortions due to lensing effects (see Fig.~\ref{fig:sample}). Often, the most extended objects do not contain the entire image of the galaxy, that is, only part of the galaxy is lensed into multiple images, which typically appear mirrored  with respect to the critical line. One advantage of these multiple images is that the integrated signal of these gravitational arcs is very high, since we observe the same galaxy twice or more, depending on the multiplicity of the image. However, a potential drawback is that properties measured in gravitational arcs, such as mass or morphology, may not be a correct representation of the full galaxy, since the galaxy images may be only partially lensed. To overcome this issue, less magnified but complete images, usually called \textit{counter-images}, can be used to recover the full morphology of the galaxy and to rescale the physical properties measured in the high signal-to-noise ratio data from gravitational arcs.

A good example of all these effects is A370-sys1, dubbed "the dragon" (see Fig.\ref{fig:sample}, upper right column). The "head" of the dragon is the complete (or counter) image of this galaxy, where the entire galaxy is lensed with an average magnification factor of 9 and the distortion is small. The "body" of the dragon (the elongated image) is composed of two to four multiple images \citep{Richard2010b}, each of them containing only a part of the galaxy. The magnification in these multiple images is much higher than in the complete image, reaching factors of almost 30. It is in these higher magnification images that the smallest physical scales can be resolved, down to 100 pc for A370-sys1. Similar structures can be seen in A2667-sys1, M1206-sys1, M0416-sys28 and A521-sys1, while AS1063-arc and A2390-arc are single images.

\begin{landscape}
\begin{figure}
\includegraphics[width=0.68\textwidth]{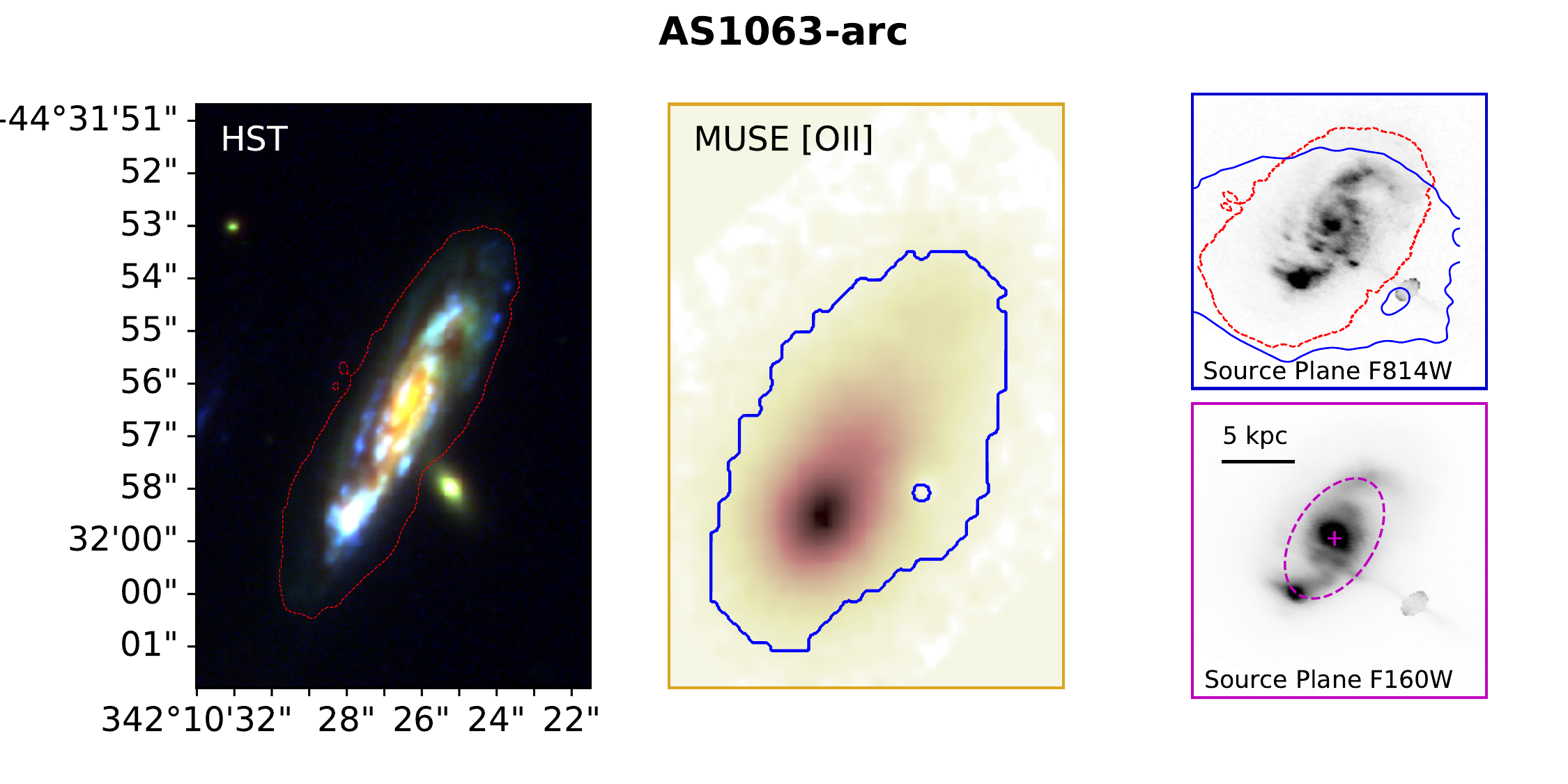}
\includegraphics[width=0.68\textwidth]{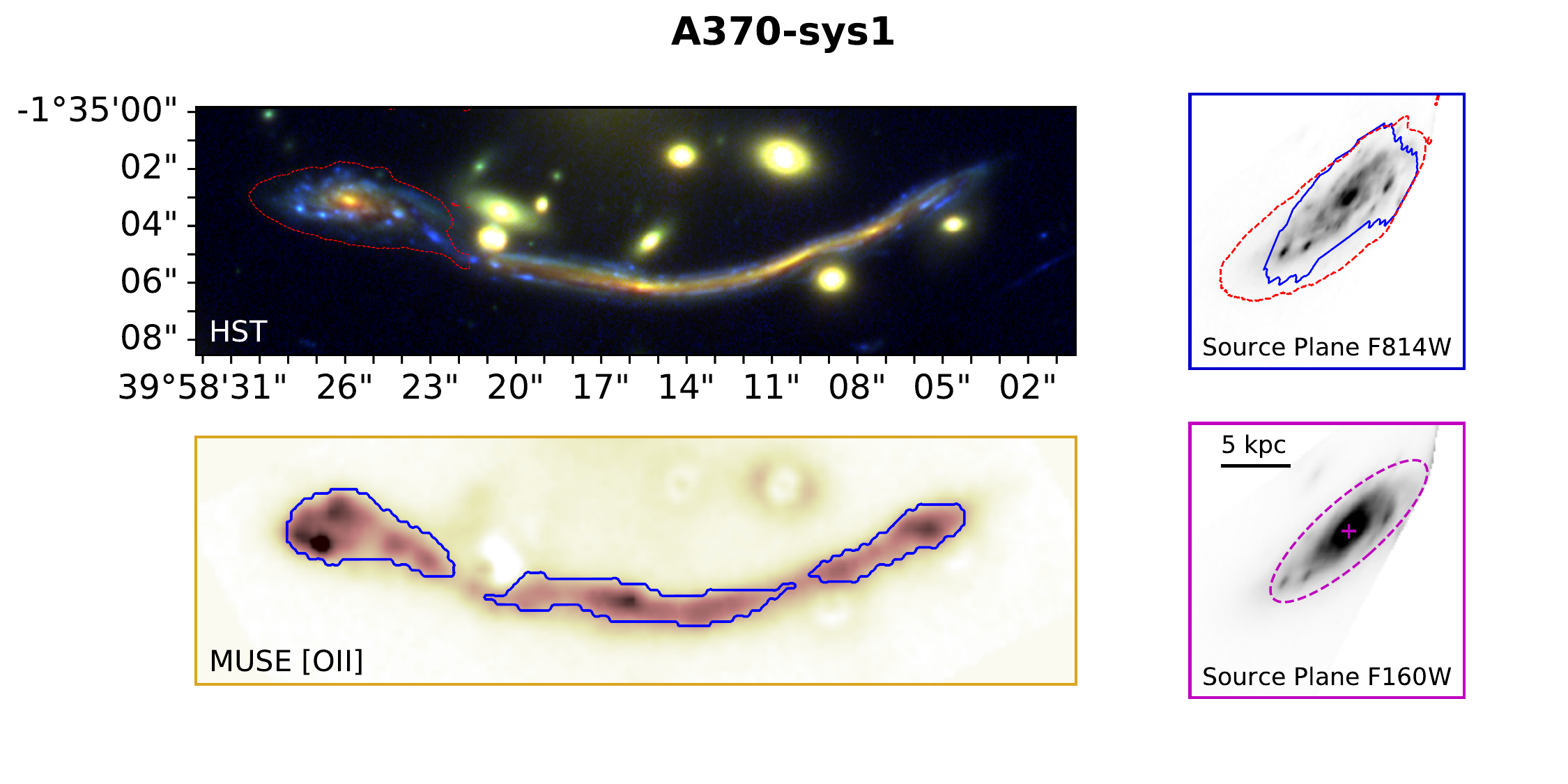}\\
\includegraphics[width=0.68\textwidth]{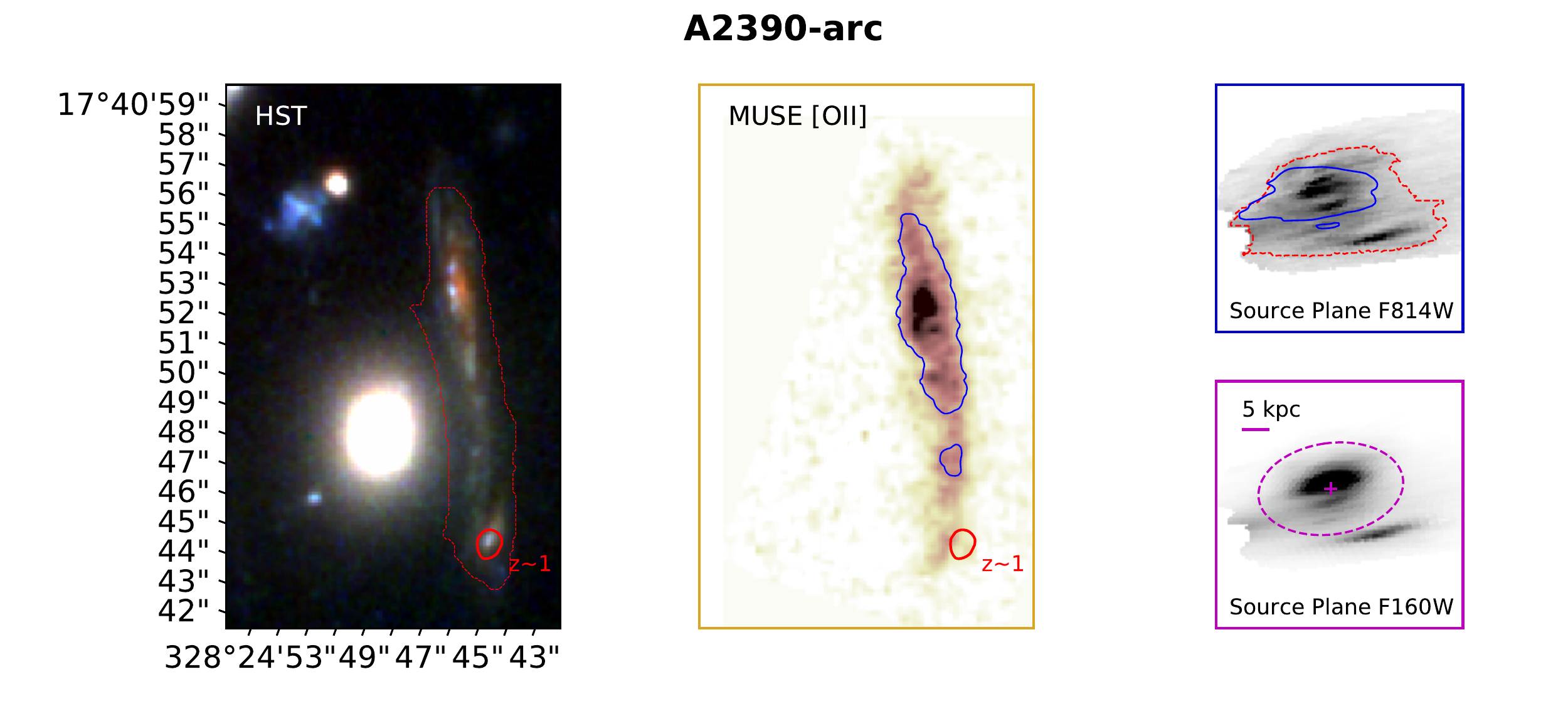} 
\includegraphics[width=0.68\textwidth]{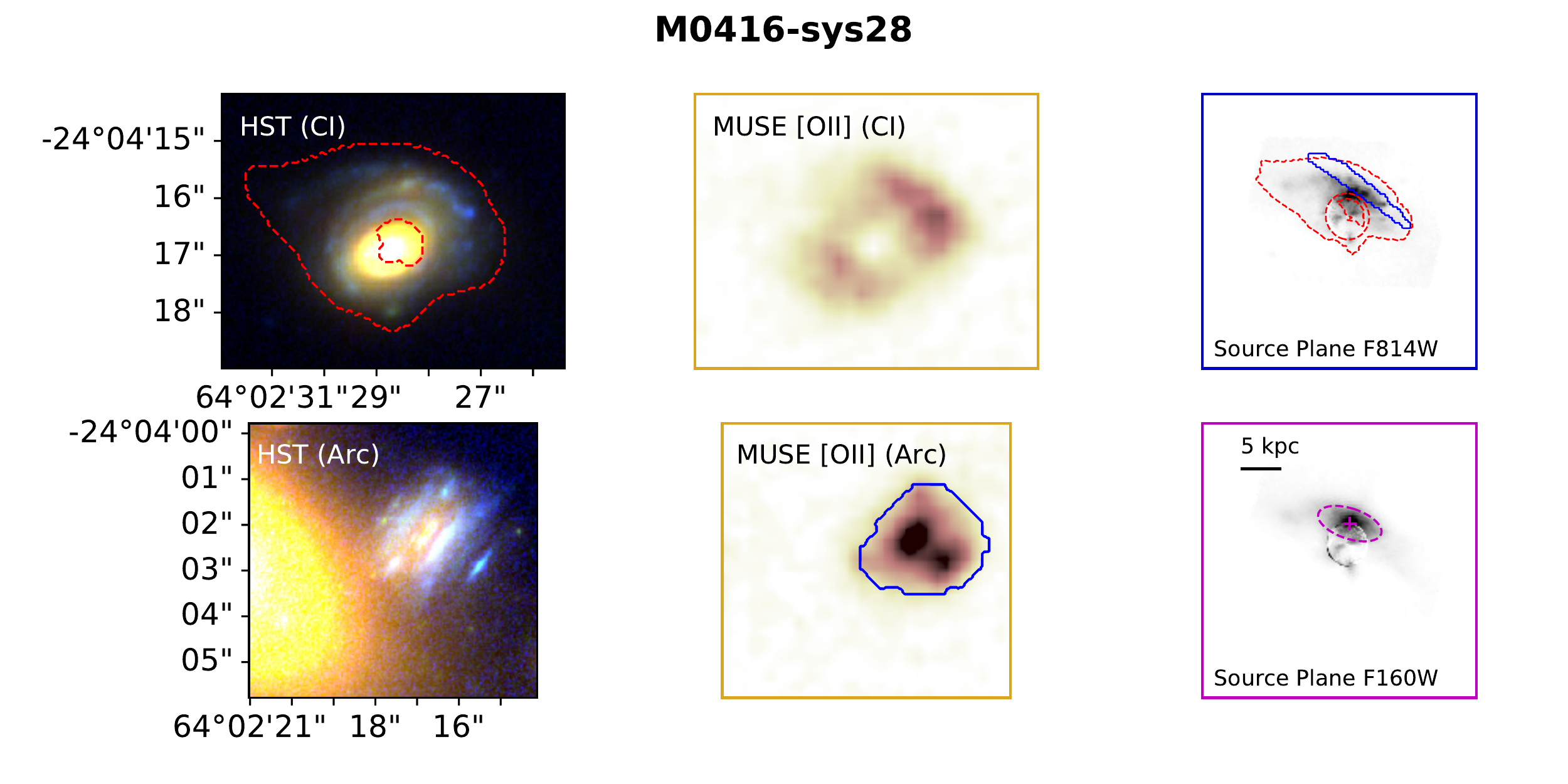}\\
\caption{Gravitational arcs sample. For each galaxy: \textit{No frame}: \emph{HST} composite image with filters F435W, F814W and F160W in blue, green and red, respectively (except A521-sys1 where F606W is green and blue, and NIRC2-K band in red).  The photometric aperture is plotted in dashed red. \textit{Yellow frame}: corresponding MUSE [\oii] pseudo-narrow band of the same sky region. The spectroscopic aperture is plotted in thick blue. \textit{Blue frame}:  F814W band source reconstruction. Both the photometric (\emph{HST}) and spectroscopic (MUSE) apertures are plotted in red and blue, respectively. \textit{Blue frame}: F160W band source reconstruction. The magenta cross marks the morphological centre of the galaxy, and the ellipse is placed at 1 effective radius with the inclination and position angle derived from the morphological fit. }
\label{fig:sample}
\end{figure}
\end{landscape}

\begin{landscape}
\begin{figure}
\includegraphics[width=0.68\textwidth]{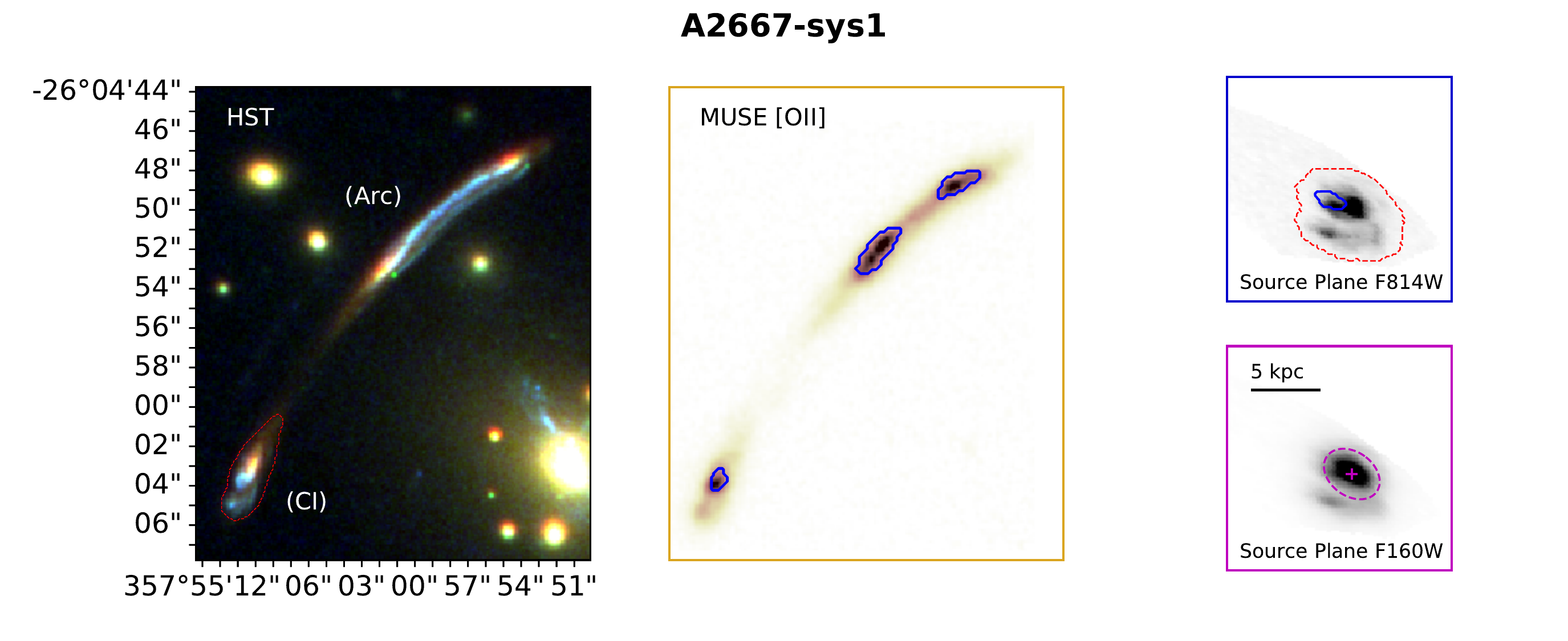}
\raisebox{-0.1\height}{\includegraphics[width=0.68\textwidth]{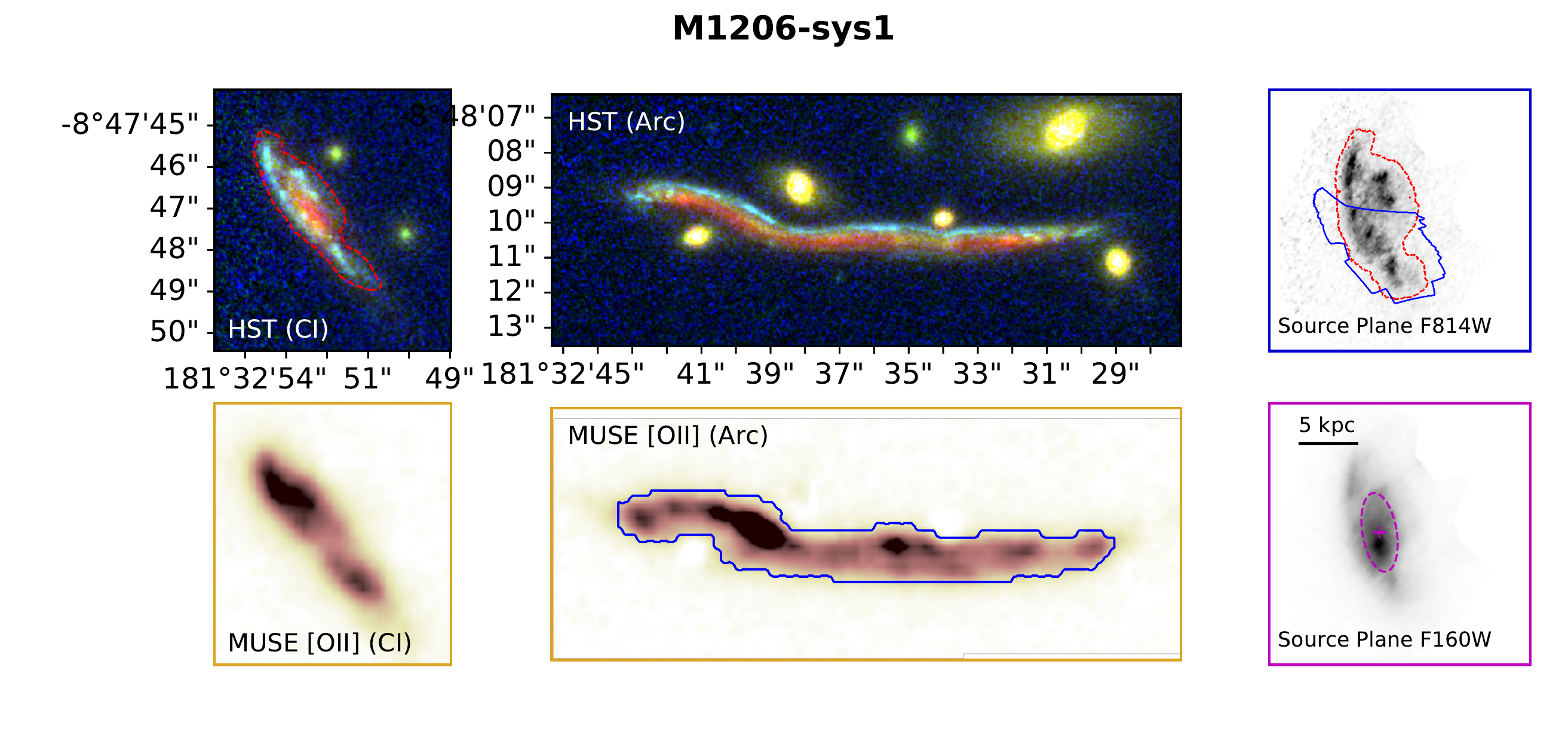}}\\
\raisebox{0.5\height}{\includegraphics[width=0.68\textwidth]{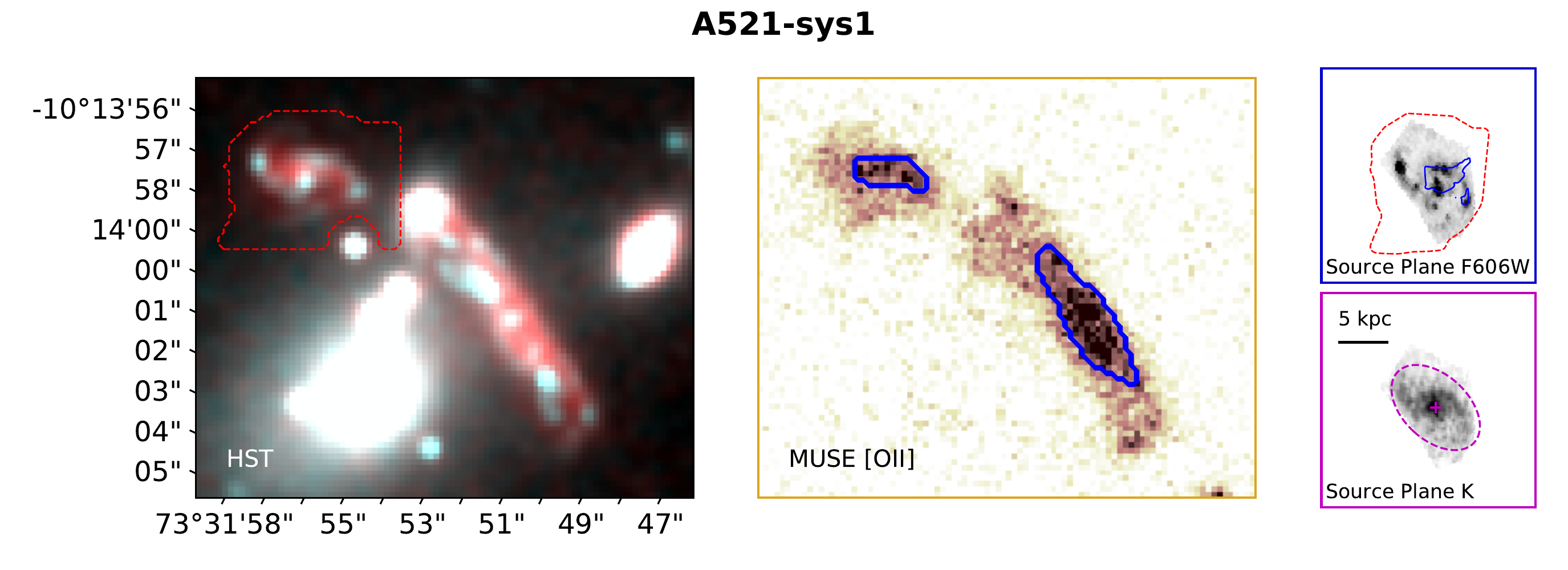}}
\includegraphics[width=0.68\textwidth]{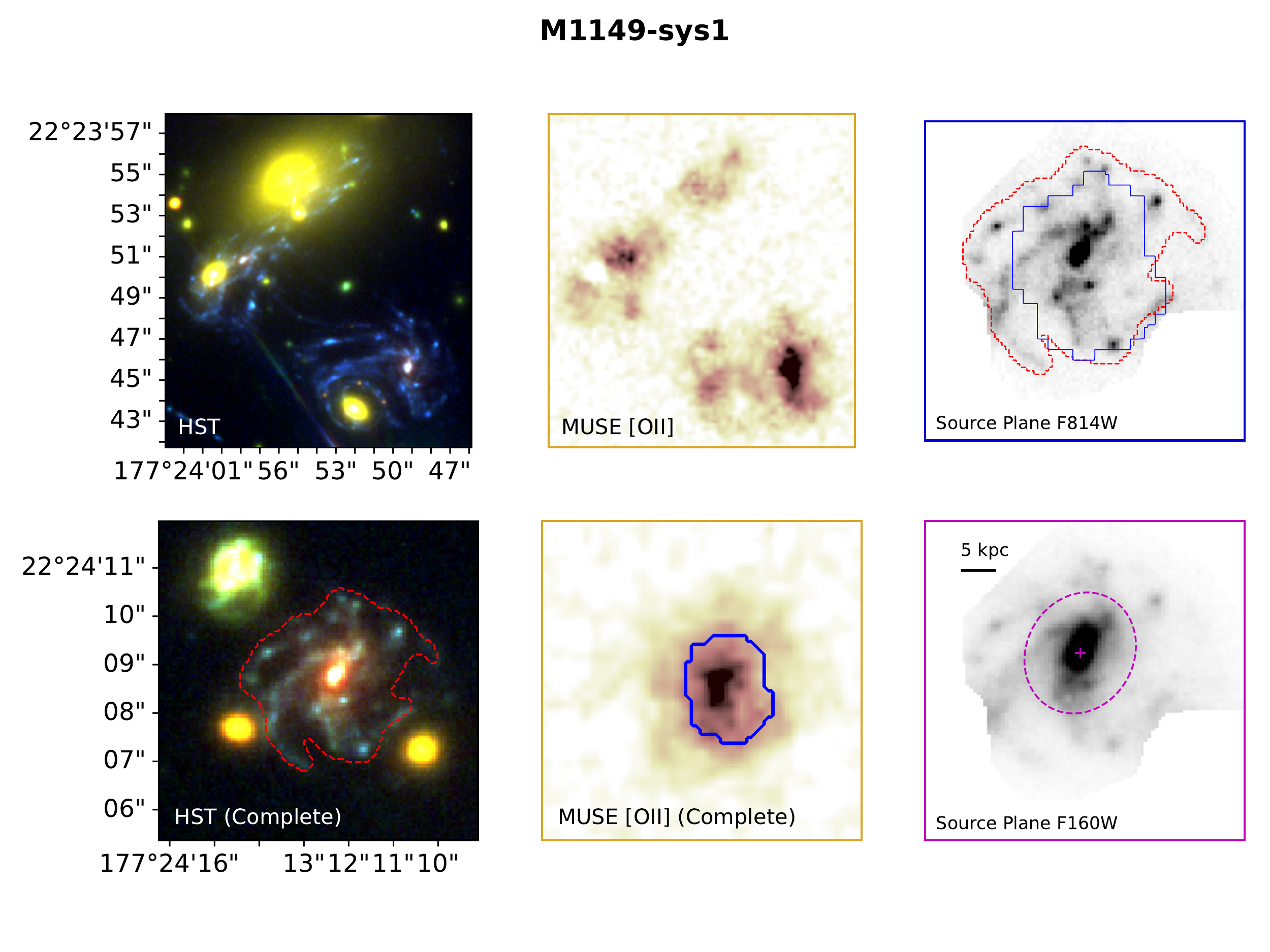}\\
\contcaption{}
\end{figure}
\end{landscape}

We apply the lensing model to each set of multiple images to obtain reconstructed images, or \textit{source plane images} images. From these source plane images, we then estimate the effective radius of these galaxies. We do this by fitting a 2D exponential profile to the reconstructed F160W \emph{HST} image of the complete images. We leave as free parameters the centre of the galaxy, the effective radius, the inclination and position angle. The fit was done using the {\sc AstroPy} 2D modelling package \citep{astropy}. We caution that these galaxies display complex morphologies, with visible spiral arms and bulges, that were not modelled in the fit. However, we masked the central bright region in A370-sys1, AS1063-arc, M1149-sys1 and A2390-arc, where a bulge is possibly present, to ensure it did not bias the fit. We list the fits parameters in Table~\ref{tab:morph} and plot ellipses with the best fit position angle, inclination and centre at one effective radius with magenta markings in the lower right hand panels of Fig.~\ref{fig:sample}.

\begin{table}
\centering
\tabcolsep=0.08cm
\caption{Best fit morphology parameters, from fitting a 2D exponential profile to the  lensed reconstructed F160W \emph{HST} images. Centre position ($\alpha$, $\delta$), effective radius (R$_e$), inclination (inc.) and position angle (PA)  for the reconstructed source. Position angle 0 corresponds to North and +90 to East.}
\label{tab:morph}
\begin{tabular}{|lrrrrrr|} 
\hline
Object  & $\alpha$ & $\delta$  				 & R$_e$ 	& inc. & PA \\ 
     &  (deg) & (deg)						 & (kpc)	& (deg) & (deg) \\   
\hline\hline
AS1063-arc  & 342.17851383 & -44.53256249 & 4.6$\pm$0.1  & 52$\pm$15  & $-33$ \\
A370-sys1   &  39.97151460 &  -1.57945147 & 7.2$\pm$0.4	 & 41$\pm$12 & $-48$\\
A2390-arc	& 328.39577581 &  17.69923321 & 10.4$\pm$0.9 & 51$\pm$27 & $-77$ \\
M0416-sys28 &  64.03737998 & -24.0710587  & 4.1$\pm$0.2	 & 57$\pm$15 & 72 \\
A2667-sys1  & 357.91536922 & -26.08347133 & 2.4$\pm$0.1  & 44$\pm$21 & 54 \\
M1206-sys1  & 181.55001582 &  -8.80075740 & 6.6$\pm$0.1	 & 64$\pm$9 & 10 \\ 
A521-sys1   &  73.52840603 & -10.22305071 & 10.2$\pm$0.8 & 51$\pm$24 & 47 \\  
M1149-sys1  & 177.40341077 &  22.40244261 & 15.4$\pm$0.9 & 51$\pm$43 & $-28$ \\
\hline
\end{tabular}
\end{table}

\subsection{Spectra Extraction and Photometry}
\label{subsec:spec_extraction}

The rest of the analysis is performed in the image plane, working directly in data cubes and \emph{HST} images. For the spectral characterisation of these galaxies, we are mainly interested in extracting high signal-to-noise ratio (S/N) spectra, where emission lines and continuum can be best analysed. Since the sources have an irregular shape, we do extract spectra from circular or elliptical apertures in the MUSE cube, but instead define extraction areas by imposing a flux threshold on MUSE pseudo narrow bands. To achieve this, we produce MUSE pseudo narrow band images centred on the [\oii] doublet, where we choose a spectral width that maximizes the S/N in the narrow band. We measure the background flux level and variance in these images, and select pixels that are above a 3$\sigma$ threshold, summing the spectra of the respective spaxels (see Fig.~\ref{fig:sample}). For M0416 and M1149 we exclude multiple images heavily contaminated by cluster members. For M1206, we did not include the counter-image, since it does not significantly increase the signal-to-noise ratio.

When coadding the spectra, magnification effects are corrected on a spaxel by spaxel basis, scaling the flux in each spaxel by 1/$\mu$. This is done in order to avoid differential magnification issues. Nonetheless, we note that applying a global (i.e. averaged) magnification factor leads to negligible differences on the overall sample. The average $\mu$ value for each spectrum is listed in Table~\ref{tab:muse_obs}. The variance propagated during data reduction is extracted in a similar way. 

Photometry was derived using the publicly available \emph{HST} data. Fluxes are measured over the multiple image which gives a complete coverage of the source in each system. Therefore, for  A370-sys1, A2667-sys1, M1206-sys1 and M0416-sys28, the photometry was measured in the counter image and not in the arc. To minimise contamination from cluster members that are too closet to the gravitational arcs, we performed a 2D fit to the light of close cluster galaxies. We do this separately for each photometric band, by modelling each cluster member with a S\'ersic profile and subtracting it from the image. The residuals of this subtraction were measured and included in the error budget of the photometry. Next, the F814W image, that matches the MUSE wavelength range, is used to define an extraction mask, by measuring the mean background and variance and placing a 3$\sigma$ threshold. For A521, the F606W image was used, since there no F814W observations available. We also mask strong residuals arising from the cluster member subtraction. The flux inside the aperture is summed and the background of the image subtracted. The luminosity-weighted magnification factors over the corresponding \emph{HST} apertures are calculated and the observed fluxes corrected. Uncertainties on the magnification factors are obtained from the Markov-Chain Monte-Carlo samples produced by {\sc Lenstool} as part of the optimisation of the mass model. The photometry and magnification factors of the \emph{HST} apertures are listed in Table \ref{tab:photom} and the spectra are shown in Fig.~\ref{fig:sample_spectra}, in Appendix.

The magnification-corrected photometry was used to renormalise the MUSE spectra, correcting both for aperture size and image multiplicity. We do this by measuring the flux in the extracted spectra (applying the F814W transmission curve), and normalising it to the value obtained from the \emph{HST} F814W image (discussed below). This process yields high signal to noise spectra, corrected for lensing and aperture effects, that we use to determine the integrated properties of the sample.

\subsection{Emission Line Measurements}
\label{subsec:emission_line_measurements}

Besides strong [\oii]$\lambda$3726,29 present in all galaxies due to the selection criteria, six out of the eight galaxies display [\neiii]$\lambda$3869 emission. Other strong lines, such as {[\oiii]$\lambda$4959,5007} and Balmer emission and absorption lines, as H$\beta$ and H$\gamma$, can also be found in the spectra depending on the object redshift. The narrow line profile of the emission lines, particularly at its centre (see in Section ~\ref{sec:resolved}), as well as the absence of [Mg II] emission, make the presence of broad-line AGN unlikely, although we cannot completely rule out this possibility.

In order to better constrain the properties of the emission lines (peak, flux and line width) we first fit and subtract the spectral continuum. To do this, we use the {\sc pPXF} code version 6.0.2 \citep{Cappellari2017} and a sample of stellar spectra from the Indo-US library \citep{Valdes2004}, which includes all stars with no wavelength gaps (448 stars). This library covers the entire MUSE observed wavelength range with a spectral resolution of 1.35 \AA\,(FWHM) at rest frame. This corresponds to a FWHM of 2.7 \AA\ in the observed frame at $z=1$. This is comparable to the MUSE LSF, about $\sim$2.5\AA, with a wavelength dependency, as determined in \citealt{Bacon2017}. The continuum fit is performed masking emission lines. To improve the fit, we add a low-order polynomial to the templates and multiply by a first order polynomial. We manually tune the degree of the additive polynomial by evaluating the quality of the fit as well as the distortion it causes in the shape of the original best fit (i.e., the combination of the stellar templates), since the polynomial can change the original continuum shape. Results of the fit are shown in Fig.~\ref{fig:sample_spectra}.

The continuum-subtracted spectra were used to measure emission lines fluxes. We use a Gaussian model implemented in the {\sc mpdaf} package \citep{mpdaf}. To estimate the error of the fit, we fit 500 realisations of the spectra, drawn randomly from a Gaussian distribution the with mean and variance corresponding to the observed spectrum flux and variance. We then take the error of the measurements (flux, FWHM and $z$) as the half-distance between the 16th and 84th percentile of the 500 fits. The flux values are presented in Table \ref{tab:line_fluxes}, and the redshift and line width of the spectra are listed in Table \ref{tab:emission_lines}.

\begin{table}
\centering
\tabcolsep=0.12cm
\caption{Kinematic properties from integrated spectra. Redshift ($z$) was measured from the emission lines. Uncertainties were calculated using the Monte-Carlo technique described in subsection~\ref{subsec:emission_line_measurements} and the error on the last digit is indicated. The uncertainties do not include systemic errors on the wavelength calibration ($\sim$0.030\AA). The velocity dispersion ($\sigma_{\rm EL}$) was measured from the FWHM of the emission lines, corrected for the instrument Line Spread Function (LSF), but not for beam smearing (which is done in Section~\ref{sec:resolved}). The velocity dispersion of the stellar population ($\sigma_{\rm \star}$), measured with {\sc pPXF} was also corrected for instrumental broadening, with a constant value of 2.5 \AA. Values for M1149-sys1 are not reported since there are no strong stellar features in the spectrum (see Appendix~\ref{app:prospector}). Errors of $\sigma_{\rm \star}$ are formal errors of the $\chi^2$ minimisation. Both velocity dispersions are presented in rest-frame.}
\label{tab:emission_lines}
\begin{tabular}{|lccc|} 
\hline
Object  & $z$  			& $\sigma_{\rm EL}$ &  $\sigma_{\rm \star}$ \\
     &           		& (km\,s$^{-1}$)    & (km\,s$^{-1}$)    \\       
\hline\hline
AS1063-arc   & 0.611532[7] & 69$\pm$5	& 108$\pm$6\\
A370-sys1    & 0.72505[1] & 61$\pm$2	& 83$\pm$6\\
A2390-arc    & 0.91280[4] & 88$\pm$10& 78$\pm$1\\
M0416-sys28  & 0.939667[8] & 48$\pm$2 & 374$\pm$39\\
A2667-sys1   & 1.034096[2] & 51$\pm$1 & 27$\pm$10 \\
M1206-sys1   & 1.036623[3] & 98$\pm$3 & 69$\pm$2\\
A521-sys1    & 1.04356[2] & 64$\pm$2 & 58$\pm$2\\
M1149-sys1   & 1.488971[1] & 55$\pm$2 & -- \\
\hline
\end{tabular}
\end{table}

\subsection{Metallicity and Star Formation Rates}
\label{subsec:integrated_properties}

From the emission line fluxes, we simultaneously derive the global metallicity and attenuation factor of the galaxies. We do this by fitting multiple metallicity and attenuation sensitive line ratios. This method is a modification to that presented by \citealt{Maiolino2008} (herein M08), whereby we extend their line ratio list and place it on a more formal Bayesian footing. We use the following line ratios:

\begin{enumerate}
\item [1] $[\oii]\lambda\,3727$/H$\beta$,  M08
\item [2] $[\oiii]\lambda\,5007$/{\rm H}$\beta$, M08
\item [3] $[\oiii]\lambda\,5007$/[\oii]$_{\lambda\,3727}$,  M08
\item [4] [\neiii]/[\oii]$\lambda\,3727$,  M08
\item [5] R23\footnote{([\oiii]$\lambda\,5007+[\oiii]\lambda\,4959+[\oii]\lambda\,3727)/{\rm H}\beta$},  M08
\item [6] [\oiii]$\lambda\,5007$/[\oiii]$\lambda\,4959$, 2.98 \citep{Storey+00}
\item [7] H$\gamma$/H$\beta$,  0.466 (\citealt{Osterbrock2006}, case B)
\item [8] H$\delta$/H$\beta$, 0.256 (\citealt{Osterbrock2006}, case B)
\item [9] H$\delta$/H$\gamma$, 0.549 (\citealt{Osterbrock2006}, case B)
\item [10] H7/H$\gamma$, 0.339 (\citealt{Osterbrock2006}, case B)
\end{enumerate}

The line ratios [1-5] are the same as those adopted by M08. These line ratios provide constraints on metallicity. But, because line ratios [1,3,5] compare lines at disparate wavelengths, they are also sensitive to the attenuation. So, to help alleviate the metallicity-dust degeneracy, we include ratios that are predominantly independent of metallicity [6-10].

We use all line ratios for which the lines are observed. However, since $H\beta$ is not available for galaxies at $z>$0.9, we replace the [\oii]/H$\beta$ in ratio [1] with  [\oii]/$H\gamma$ (assuming case B H$\gamma$/H$\beta$ ratio of 0.466, for T$_e=10000$ K and low electron density). In Table~\ref{tab:met_emission_lines}, we list the specific line ratios used for each galaxy.

Another modification to the M08 method is that we adopt the \citealt{Charlot2000} dust model to correct the observed line fluxes ($F_{\rm obs}$) for dust attenuation, assuming an index of $-1.3$ for the exponential attenuation law, proposed by the authors and found to be appropriate for star-forming regions (e.g. \citealt{Brinchmann2013}). The intrinsic fluxes are corrected in the following manner:

\begin{equation}
F_{\rm int} = F_{\rm obs}\,e^{\tau_v\,\left(\frac{\lambda }{5500 \scalebox{0.6}{\angstrom{}}}\right)^{-1.3}}
\end{equation} 


We derive metallicity ($Z$) and attenuation ($\tau_v$) from emission line ratios, fitting several ratios simultaneously in a Bayesian framework, using the {\sc emcee} \citep{Foreman-Mackey2013} Markov chain Monte Carlo Sampler to maximise the following a Gaussian ($\log$-)likelihood function:

\begin{equation}
\label{eq:lnprob}
ln\,p = - \frac{1}{2} \sum_r \left[ \frac{({{\rm M_r}(Z)} - {{\rm O_r}(\tau)})^2}{\sigma_r^2} + \ln(2\pi\sigma_r^2) \right]
\end{equation}

where ${\rm O_r}(\tau)$ are the observed line ratios corrected for attenuation factor $\tau$ and ${\rm M_r}(Z)$ are the predicted ratios using the above calibrations for the metallicity $Z$. $\sigma_r^2$ is the quadratic sum of the observed error and an additional model uncertainty. We adopt a model uncertainty of 10\% for the M08 calibrations, a 1\% uncertainty for the Balmer line ratios and 2\% for the [\oiii]$\lambda\,5007$/[\oiii]$\lambda\,4959$ ratio \citep{Storey+00}. We assume a solar abundance of 8.69, as in M08. We use wide flat priors for all galaxies with metallicity $7<12 + \log(Z)+12<10$  and attenuation $0<\tau_v<4$. The marginalised metallicity and attenuation distributions for the AS1063-arc can be found in Fig.~\ref{fig:as1063_met} and the remaining distributions in Appendix \ref{app:met_and_ext}. The median of these distributions are listed in Table~\ref{tab:met_emission_lines}, with the half-distance between the 16th and 84th percentiles adopted as the error. 

\begin{figure}
\includegraphics[width=0.5\textwidth]{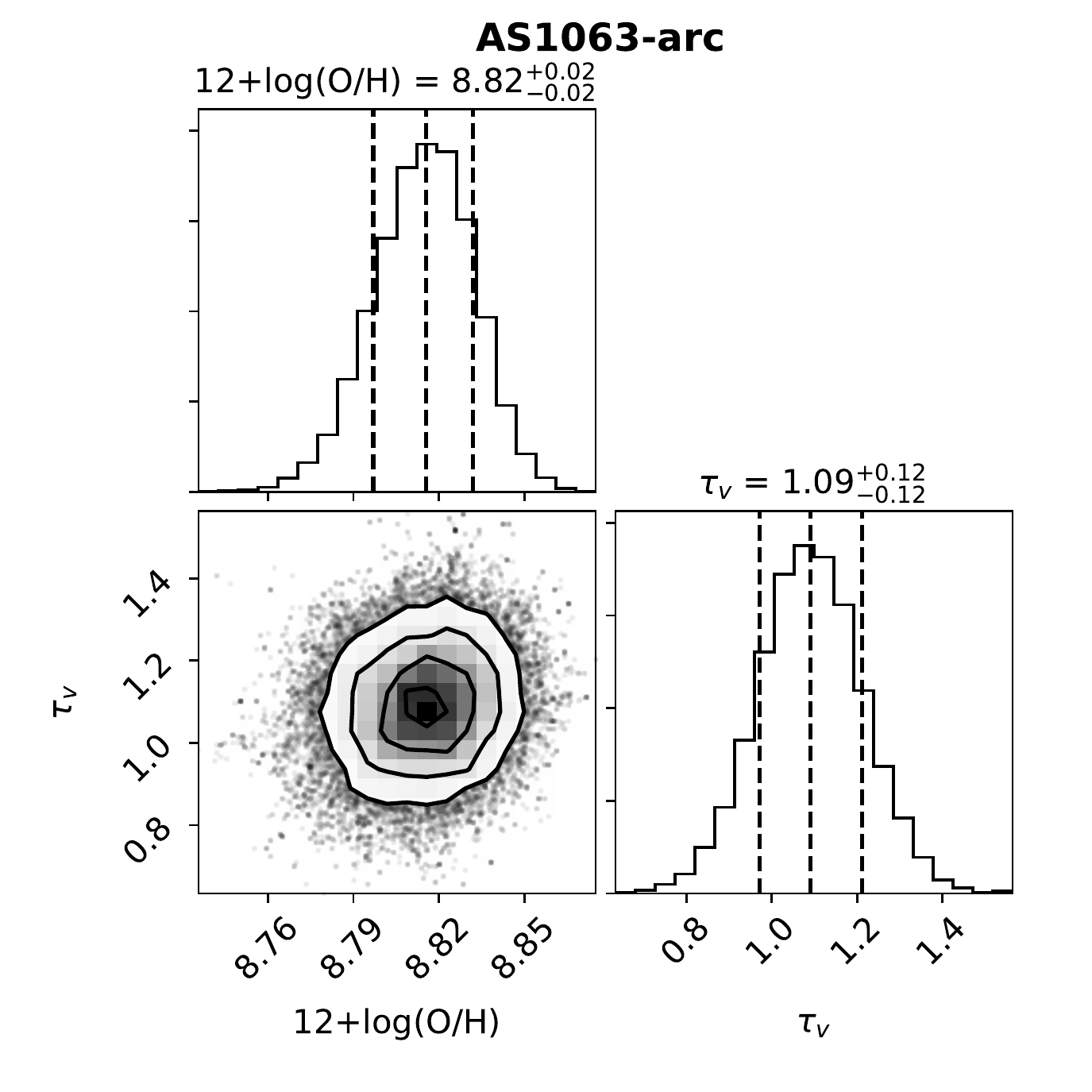}
\caption{Example of simultaneous determination of metallicity and extinction for galaxy AS1063-arc. The marginalised distributions for metallicity and extinction are displayed in the top left and bottom right panel, respectively. The bottom left panel displays the joint distribution. Similar plots for the remaining galaxies can be found in Appendix~\ref{app:met_and_ext}.}
\label{fig:as1063_met}
\end{figure}

M1149-sys1 was left out of this analysis, since only [\oii] is present in the MUSE spectrum. All the galaxies analysed have well determined metallicities and attenuations, ranging from 8.71 to 9.05 and 0.09 to 2.86 respectively, with symmetrical marginalised distributions (except for attenuation in M0416-sys28, A2390-arc and A521-sys1). 

For A2390, the only metallicity diagnostic that could be used was $[\oii]\lambda\,3727$/H$\gamma$, since the signal to noise ratio does not allow a robust measurement of the [\neiii] flux. As the $[\oii]\lambda\,3727$/H$\gamma$ diagnostic is degenerate, with both low and high metallicity solution, we investigate each branch by repeating the fit allowing only low or only high metallicity values ([7.0,8.6]  and [8.6,10.0] priors, respectively). For solutions found in each branch, the [\neiii] flux would be $\sim$6.2 $\times 10^{-18}$ erg\,cm$^{-2}$ s$^{-1}$ and $\sim$1.2 $\times 10^{-18}$ erg\,cm$^{-2}$\,s$^{-1}$ for the low and high metallicity solutions respectively. The low metallicity estimate approximately corresponds, for example, to the flux of H$\delta$ seen in this galaxy ($\sim$ 5.7 $\times 10^{-18}$ erg cm$^{-2}$ s$^{-1}$), so the low metallicity branch seems less likely than the high metallicity solution. The high metallicity solution would yield lower [Ne III] fluxes that would likely be undetected given the $\sim$0.3 $\times 10^{-18}$ erg cm$^{-2}$ s$^{-1}$ 1$\sigma$ spectral noise for this galaxy. Furthermore, the star formation rates derived via the Balmer lines and the [\oii] line (corrected for metallicity) are closer to the high metallicity case than the low one. These indications do not rule out the low metallicity hypothesis, but we adopt the high metallicity values as the most probable solution for A2390-arc throughout the rest of this work. 

The star formation rates (SFR) are calculated both from Balmer lines and the [\oii] doublet. For the Balmer lines, the \citealt{Kennicutt1998} calibration is adopted, adapting this calibration to be used with H$\beta$ or H$\gamma$ and assuming their ratios relative to H$\alpha$ are given by Case B theory at electronic temperature T$_e$=10000K and low electron density. For the [\oii], the metallicity dependent \citealt{Kewley2004} calibration is used. The SFR from Balmer lines is obtained by drawing 500 random values of H$\beta$ or $H\gamma$ fluxes (assuming a Gaussian distribution) and correcting them for dust attenuation using a sample of 500 $\tau_v$ values drawn from the marginalised distribution obtained from the metallicity and attenuation fit. The dust corrected line fluxes are converted to SFR using the \citealt{Kennicutt1998} calibration and the SFR taken as the median of the sample. For the SFR estimated from [\oii], an equivalent process is used. We produce a sample of [\oii] fluxes, $\tau_v$ and metallicity values and apply the \citealt{Kewley2004} calibrations. Finally, both SFR distributions are convolved with the magnification error, since they depend on the absolute flux of the emission lines. We list the calculated SFR in Table~\ref{tab:met_emission_lines}.

\begin{table}
\centering
\tabcolsep=0.03cm
\caption{Metallicity, attenuation and intrinsic SFR. Gas-phase metallicity and  attenuation ($\tau_v$) were calculated as described in subsection~\ref{subsec:integrated_properties} and using the diagnostics listed in the 'Line Ratios' column. Both star formation rates derived from the \citealt{Kennicutt1998} (SFR$_{\rm Balmer}$) and \citealt{Kewley2004} (SFR$_{[\oii]}$) relations are listed. M1149-sys1 was not included in this analysis, since only [\oii] is present in the MUSE spectra.}
\label{tab:met_emission_lines}
\begin{tabular}{|lccccccc|} 
\hline
Object  & Line 	& Gas Metallicity	  &	$\tau_v$ 	  & SFR$_{\rm Balmer}$ & SFR$_{[\oii]}$ \\
     &	Ratios		& ($12+\log$(O/H)) &          & (M$_\odot\,$yr$^{-1}$) & (M$_\odot\,$yr$^{-1}$) \\       
\hline\hline
AS1063-arc  & 1...10  &8.82$\pm$0.02 & 1.09$\pm$0.12 & 41.5$\pm$4.0 & 50.3$\pm$10.1\\
A370-sys1   & 1...10  &8.88$\pm$0.02 & 0.44$\pm$0.11 &  3.1$\pm$0.3 &  3.1$\pm$0.6\\
A2390-arc   & 1,9,10  &9.00$\pm$0.11 & 0.60$\pm$0.40 &  7.3$\pm$2.5 &  7.9$\pm$6.1\\
M0416-sys28 &1,4,9,10 &8.72$\pm$0.6 & 0.14$\pm$0.18 &  2.0$\pm$0.7 &  1.8$\pm$1.0\\
A2667-sys1  &1,4,9,10 &9.04$\pm$0.04 & 0.53$\pm$0.23 & 15.7$\pm$3.7 & 9.8$\pm$4.2\\
M1206-sys1	&1,4,9,10 &8.91$\pm$0.06 & 0.74$\pm$0.33 &107.3$\pm$30.7& 85.1$\pm$55.5\\
A521-sys1	&1,4,9,10 &9.05$\pm$0.08 & 2.86$\pm$0.50 &17.4$\pm$8.3& 30.2$\pm$31.8\\

\hline
\end{tabular}
\end{table}

\subsection{Stellar Mass}
\label{sec:mass}

We fit both spectra and photometry using the FSPS (Flexible Stellar Population Synthesis, \citealt{Conroy2009}) models implemented within a Bayesian framework in the {\sc Prospector}\footnote{ {\sc Prospector} publicly available at:  \url{https://github.com/bd-j/prospector}} code (Johnson et al., in prep). The only exception is A521-sys1, for which only the spectrum was fit and used to normalise the spectra, since only one \emph{HST} filter was available. {\sc Prospector} allows us to fit models produced with FSPS 'on the fly', i.e., the fit are not based on a precomputed grid of models but are instead computed for each exact set of parameters tested. We assume a decaying star formation history and choose the Padova isochrones \citep{Marigo2007,Marigo2008} with the \citealt{Chabrier2003} initial mass function and the MILES stellar library \citep{Sanchez-Blazques2007,Falcon-Barroso2011} to generate the models for all fits. We also adopt the \citealt{Charlot2000} dust attenuation here, with a fixed index of $-1.3$ for the star-forming regions, applied only to the star-light originating in stars younger than 10 Myrs, and an index of $-0.7$ for the global dust screen as suggested by \citealt{Charlot2000}, applied to all starlight equally. 

In total, 6 physical parameters were fit: stellar mass, the e-folding time of the star formation history, stellar population age, stellar metallicity and optical depth of both birth clouds and interstellar medium (ISM) (see Table~\ref{tab:prospector} for details and priors). Beside these physical parameters, a 3rd order polynomial is also fitted (similarly to what is done with {\sc pPXF}). For the sake of simplicity on an already quite complex fit, we decided to mask the emission lines during these fits, and do not attempt to fit them during the continuum fit with {\sc PROSPECTOR} and do not include nebular emission in the models. 

The fits for each galaxy can be found in Appendix~\ref{app:prospector}. The best values and errors, estimated from the marginalised distributions as in the previous section, can be found in Table~\ref{tab:prospector}. The magnification errors were included at this point in the mass and SFR error estimates. Overall, the best fits provide a good match to both the spectral and photometric data. Most of the derived model parameters are well constrained, with the exception of the attenuation, which in some instances have large tails extending to high values. In principle, the value of $\tau_{BC}$ obtained here should agree with the attenuation derived from the emission lines. In practice, since no FUV data is available, nor were the emission lines fitted with {\sc prospector}, this value cannot be constrained and will be degenerate with $\tau_{ISM}$. The obtained stellar masses vary from $\sim$1\,$\times\,10^{10}$ to $\sim$9\,$\times\,10^{10}$ M$_\odot$, with a mean value of 3.65\,$\times\,10^{10}$ M$_\odot$, and with an e-folding time that varies from 3 to 9 Gyrs and dominant ages from 3 to 7 Gyrs. 

\begin{table*}
\centering
\tabcolsep=0.2cm
\caption[SED fitting results]{SED fitting best results and their error (16th and 84th percentiles) obtained with {\sc Prospector} and including magnification corrections. M$_\star$ corresponds to the current mass of the galaxy; $\tau$ is the e-folding time of the star formation history and $age$ the age of the composite stellar population. We estimate the average SFR (SFR$_{\rm SED}$) as the ratio of the \textit{formed} mass (higher than the current mass, which does not include gas that was recycled into the ISM) and the age of the galaxy (the $age$ parameter). $\tau_{V, BC}$ is the attenuation factor for the birth clouds, applied only to starlight coming from stars with less than 10 Myrs, and $\tau_{V, ISM}$  is the global attenuation factor of the ISM and is applied to all starlight, respectively. The prior ranges of each parameter are shown in the first row. With the exception of $\tau$, where a logarithmic prior was used, all other priors are flat. The maximum allowed age is set by the age of the Universe at the redshift of each galaxy.}
\label{tab:prospector}
{\renewcommand{\arraystretch}{1.3}%
\begin{tabular}{|lccccccc|} 
\hline
Object      		& M$_\star$ 		               		& $\tau$                        	& $age$                    		&SFR$_{\rm SED}$               		& Z$_\star$                     & $\tau_{V, BC}$ 		& $\tau_{V, ISM}$  \\ 
	   		&  $(10^{10}$ M$_\odot)$ &  (Gyr) 		& (Gyr) 			&  (M$_\odot$ yr$^{-1}$)  		& ($\log$ Z/Z$_\odot$)   &                         		 & \\
\hline\hline
Range		& [0.5 , 10]  &  [0.01 , 10] & [0.2 ,4-8] & -- & [-1, 1] & [0, 4] & [0, 4] \\ 
AS1063-arc 		& 8.74$^{+0.39}_{-0.44}$	& 8.86$^{+0.27}_{-0.78}$ 	& 4.74$^{+0.46}_{-0.37}$ 	& 29.59$^{+3.12}_{-4.11}$ 		& $-0.48^{+0.04}_{-0.02}$	& 0.00$^{+0.01}_{-0.00}$ 	& 0.91$^{+0.05}_{-0.06}$\\
A370-sys1 	   & 2.49$^{+0.04}_{-0.04}$     & 7.10$^{+0.40}_{-1.47}$ 	& 7.27$^{+0.02}_{-0.24}$ 	& 5.68$^{+0.20}_{-0.18}$		& $-0.49^{+0.01}_{-0.01}$ 	& 0.00$^{+0.00}_{-0.00}$	& 0.65$^{+0.02}_{-0.07}$\\
A2390-arc      & 2.39$^{+0.50}_{-0.30}$		& 3.06$^{+1.65}_{-0.61}$ 	& 3.28$^{+0.44}_{-0.32}$ 	& 11.58$^{+2.86}_{-1.93}$ 		& $-0.41^{+0.05}_{0.13}$ 	& 3.45$^{+0.41}_{-0.47}$ 	& 0.40$^{+0.16}_{-0.10}$ \\
M0416-sys28 & 0.95$^{+0.25}_{-0.25}$		& 6.16$^{+0.67}_{-0.47}$	& 6.46$^{+0.03}_{-0.07}$	& 2.47$^{+0.65}_{-0.64}$		& -0.98$^{+0.03}_{-0.01}$ 	& 0.21$^{+0.03}_{-0.03}$ 	& 0.02$^{+0.03}_{-0.01}$ \\
A2667-sys1    	& 1.37$^{+0.21}_{-0.19}$ 	& 9.57$^{+0.32}_{-0.82}$ 	& 3.72$^{+0.07}_{-0.09}$ 	& 6.36$^{+1.14}_{-0.93}$ 		& $-0.98^{+0.01}_{-0.01}$	& 0.86$^{+0.03}_{-0.03}$ 	& 0.00$^{+0.01}_{-0.00}$\\
M1206-sys1 	& 7.87$^{+0.49}_{-0.47}$ 	& 2.07$^{+0.96}_{-1.31}$ 	& 4.38$^{+2.62}_{-2.29}$ 	& 39.91$^{+2.80}_{-2.69}$		& 0.07$^{+0.19}_{-0.19}$ 	& 0.24$^{+0.06}_{-0.16}$ 	& 0.26$^{+0.10}_{-0.08}$ \\
A521-sys1 	& 3.21$^{+0.77}_{-0.64}$ 	& 9.63$^{+0.30}_{-4.73}$ 	& 6.00$^{+0.00}_{-0.01}$ 	& 8.81$^{+2.05}_{-1.82}$		& -0.50$^{+0.01}_{-0.01}$ 	& 0.00$^{+0.01}_{-0.00}$ 	& 1.27$^{+0.07}_{-0.19}$ \\
M1149-sys1 	& 1.76$^{+0.11}_{-0.11}$ 	& 7.10$^{+0.40}_{-1.47}$ 	& 7.27$^{+0.02}_{-0.24}$ 	& 18.83$^{+2.41}_{-2.27}$		& $-1.00^{+0.01}_{-0.01}$ 	& 0.00$^{+0.01}_{-0.00}$ 	& 0.65$^{+0.02}_{-0.07}$\\
\hline
\end{tabular}
}
\end{table*}

\begin{figure}
\includegraphics[width=0.50\textwidth]{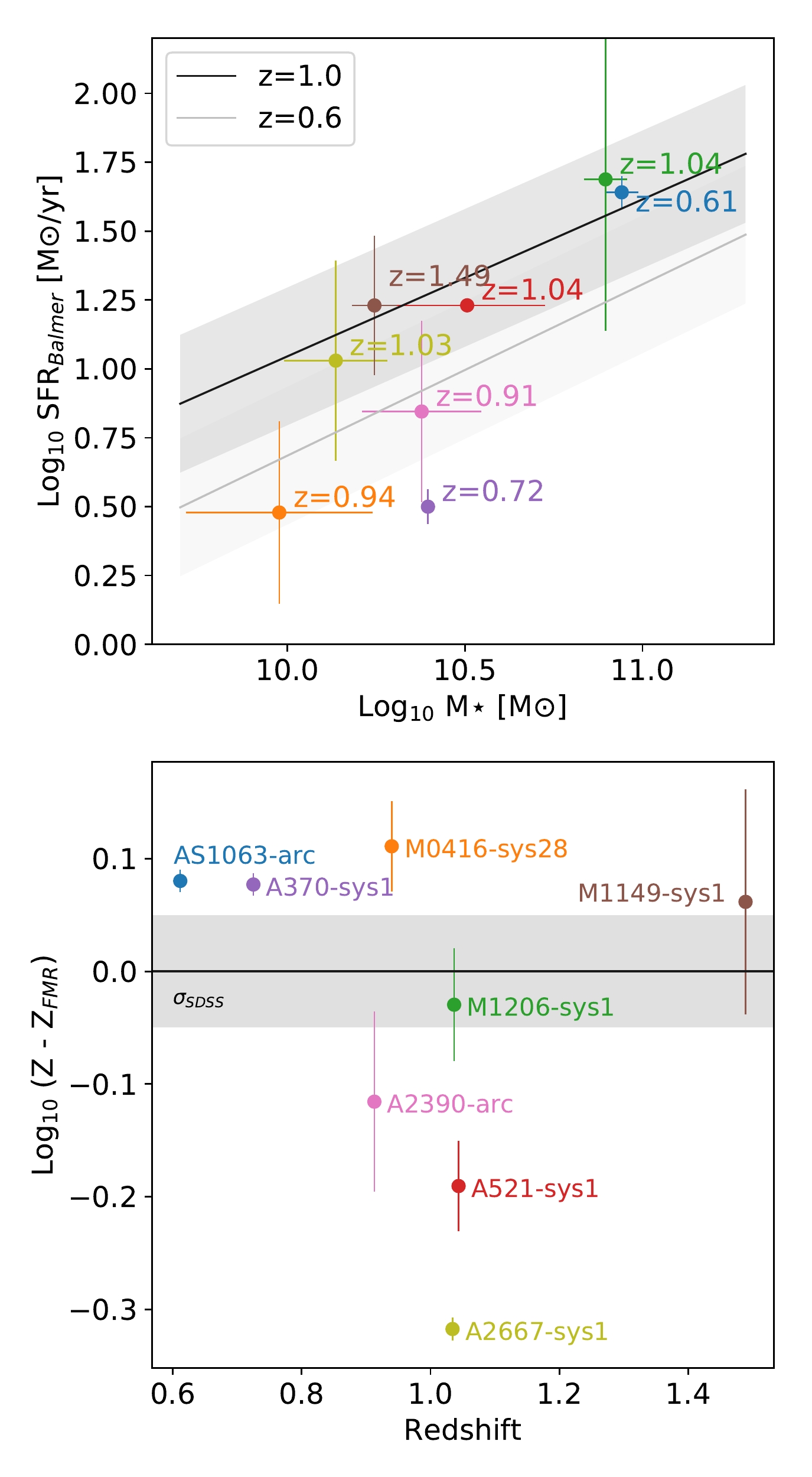}
\caption{Top panel: Gravitational arcs sample plotted against the main sequence. We plot in solid grey lines the parametrisation of \citealt{Whitaker2012} for this relation at $z=0.6$ and $z=1$, with the dispersion in grey. Bottom panel: Distance to the Fundamental Metallicity Relation (FMR) from \citealt{Mannucci2010} (with the 1$\sigma$ range of the SDSS sample form the same work shown in grey). Most galaxies lie close to this fundamental relation. The same colour scheme is used in both plots to identify objects. Both relations considered, these galaxies are typical star-forming galaxies for $z\sim$1.}
\label{fig:main_sequence}
\end{figure}

\subsection{Comparison with other samples}
\label{subsec:context}

We can now use the physical parameters previously derived both from emission lines and photometry to place this sample in context. For M1149-sys1, only the [\oii] line is present in the MUSE wavelength range, so it is not possible to constrain metallicity and SFR from these data. Two metallicity estimates exist for this galaxy in the literature. \citealt{Yuan2011} derive a metallicity of $12+\log$(O/H) = $8.36\pm0.04$ using the [\nii] / H$\alpha$ ratio, while \citealt{Wang2017} measure $12+\log$(O/H) = $8.70\pm0.10$  using \emph{HST} grism spectra. \citealt{Wang2017} use M08 calibrations, including line ratios with [\oii], [\oiii] and H$\beta$ emission lines, and calculate metallicity using a Bayesian inference framework, comparable to what is done here. We make use of both their metallicity and SFR ($16.99\pm4.3$ M$_\odot$/yr) values derived from emission lines  throughout this work.

Firstly, we compare the sample with the mass-SFR relation (top panel of Fig.~\ref{fig:main_sequence}). Within errors, all the galaxies lie close to what is measured in main sequence star-forming galaxies. The only exception is A370-sys1 that has a slightly lower star formation rate than expected from its derived mass. We then compare this sample with the fundamental metallicity relation \citep{Lara2010,Mannucci2010}, that  defines a tight relation between stellar mass, gas metallicity and SFR of normal star-forming galaxies (bottom panel of Fig.~\ref{fig:main_sequence}). Five out of the seven arcs -- AS1063-arc, A370-sys1, A2390-arc, M0416-sys28 and M1206-sys1 --  differ by up to 0.1 dex from this fundamental relation, which is higher than the dispersion derived for the local SDSS sample of about 0.05 dex, but still within the 0.5 dex dispersion observed for higher redshift samples \citep{Mannucci2010}. We conclude that all the galaxies studied in this work are normal star-forming galaxies and are good representatives of the population of disc galaxies at $z\sim1$.

We can estimate the gas mass of these galaxies by using the Kennicutt-Schmidt law \citep{KS_law} to convert SFR surface density to gas surface density. We take the global (dust and magnification corrected) SFR derived from the Balmer lines of the integrated spectrum and we calculate the SFR density by dividing it by the photometry aperture area (the area to which the spectra were normalised). Note that the area is calculated in the source plane. The SFR is then converted to gas surface density, and thereby also the total gas mass. These are listed in Table~\ref{tab:KS_law_gas}. The star formation rate densities, gas masses (M$_{\rm g,\,KS}$) and gas fractions (M$_{\rm g,\,KS}$/(M$_{\rm g,\,KS}$ + M$_\star$)) are listed in Table~\ref{tab:KS_law_gas}. These galaxies have low gas fractions that range from $\sim$0.1 to 0.3, despite the star-forming clumps clearly visible in most of them (particularly AS1063-arc, A370-sys1 and M1206-sys1). These values are consistent with the values obtained from the DYNAMO survey, a sample of comparable local galaxies that covers a stellar masses of 10$^{9}$ to 10$^{11}$ M$_\odot$ and star formation rates of $0.2-100$ M$_\odot\,$yr$^{-1}$, selected to be analogues to $z\sim$3 galaxies \citep{Green2014}. The gas fraction in our galaxies are also similar to what \citealt{White2017} found from making CO measurements and inverting the KS law.


\begin{table}
\caption{Star formation rate surface density ($\Sigma_{\rm SFR}$), gas surface density ($\Sigma_{\rm g}$) and gas mass (M$_{ \rm g,\,KS}$) estimated through the Kennicutt-Schmidt law and the respective gas fractions ($f_{\rm g,\,KS}$) calculated as M$_{g,\,KS}$/(M$_{\rm g,\,KS}$ + M$_\star$)}
\label{tab:KS_law_gas}
\centering
\tabcolsep=0.09cm
\begin{tabular}{|lcrcc|} 
\hline
Object		& $\Sigma_{\rm SFR}$					 &	$\Sigma_{\rm g}$			&  M$_{ \rm g,\,KS}$			&$f_{\rm g,\,KS}$\\
		& (M$_\odot\,$yr$^{-1}\,$kpc$^{-2}$) &	(M$_\odot\,$pc$^{-2}$)	& (10$^{10}$ M$_\odot$) &			\\
\hline
\hline
AS1063-arc	& 0.22$\pm$0.01 				& 124.66$\pm$5.25 			& 2.53$\pm$0.11		& 0.22$\pm$0.03\\
A370-sys1	& 0.03$\pm$0.01 				& 33.87$\pm$1.58			&0.31$\pm$0.01		& 0.11$\pm$0.01\\
A2390-arc	& 0.04$\pm$0.01 				& 107.72$\pm$26.98			&0.42$\pm$0.10		& 0.15$\pm$0.13\\
M0416-sys28 & 0.02$\pm$0.01 				& 21.49$\pm$5.63			&0.35$\pm$0.09		& 0.27$\pm$0.15\\
A2667-sys1	& 0.23$\pm$0.09 				& 139.24$\pm$38.30 			&0.58$\pm$0.16		& 0.30$\pm$0.09\\
M1206-sys1  & 0.70$\pm$0.39 				&246.00$\pm$93.41			&2.16$\pm$0.83		& 0.22$\pm$0.07\\
A521-sys1   & 0.03$\pm$0.01 				& 29.78$\pm$10.03 			& 1.74$\pm$0.59 	& 0.35$\pm$0.12 \\
\hline
\end{tabular}
\end{table}


\section{Resolved Properties}
\label{sec:resolved} 


\subsection{Observed Velocity Maps}
\label{subsec:obs_vel}

The velocity maps (or kinematic maps) were derived measuring the [\oii] emission lines present in all galaxies. The fits were performed with a slightly modified version of the {\sc camel}\footnote{{\sc camel} is available at \url{https://bitbucket.org/bepinat/camel.git}} code \citep{Epinat2010}, where we added the option of fitting binned data. {\sc camel} fits emission lines with 1D Gaussian models, having as free parameters the redshift, the Gaussian FWHM and the flux of each emission line. In order to more robustly probe the outer parts of the galaxies, we bin the data using the Voronoi binning method\footnote{Voronoi binning code: \url{http://www-astro.physics.ox.ac.uk/~mxc/software}} presented in \citealt{Cappellari2003}. We bin on the signal to noise of the [\oii] pseudo narrow bands used in Sect.~\ref{subsec:spec_extraction}. We impose a signal to noise of 5 per bin in all galaxies, which we find to be a good compromise between high spatial sampling and robust spectral fits. The velocity dispersion was allowed to vary between 0 and 250\kms. The 2D observed velocity maps of the 8 galaxies are presented in Fig.~\ref{fig:2d_kin}. These maps were inspected and the velocities values were corrected (adding or subtracting a constant value in the entire field) so that the velocity at the morphological centre (as given by the 2D exponential fit to the source plane image, see Table~\ref{tab:morph}) corresponds to the kinematic centre and has velocity zero.


\subsection{Fitting Velocity Maps}
\label{subsec:fit_vel}

We fit the 2D velocity maps with the three following kinematic models: an arctangent model \citep{Courteau+97}, an isothermal sphere model \citep{Spano2008} and an exponential disc model \citep{Freeman1970}. The first is an empirical model commonly used to fit local galaxies. The isothermal sphere model is a good approximation of the kinematics of a dark matter dominated galaxy, under the assumption that the dark matter halo has an isothermal profile. The exponential disc assumes a smooth exponentially decaying distribution of baryonic matter (and no dark matter) as seen in most disc galaxies (see \citealt{Epinat2010} for a summary of these models and analytical expression implemented in this work).

We fit these models directly to the 2D velocity maps produced by {\sc camel} (i.e. in image plane). Note that we do not fit the velocity dispersion maps. To account for lensing distortions, we produce high resolution displacement maps using the optimised cluster mass models and the {\sc Lenstool} software. These maps describe the geometrical transformation from source to image plane by predicting the spatial shifts (or displacement) that have to be applied to each source plane pixel to be projected to the image plane image.

The first step of the fit is to create a 2D kinematic map in source (unlensed) plane. We apply the lensing distortion to this kinematic model, using the displacement maps. We then convolve the lensed model velocity map with the seeing, using a Moffat kernel, flux-weighted by the [\oii] flux image from {\sc camel}. Finally, we apply the same binning to the models as the observations before comparing the modelled and the observed velocity fields. It is this final product that is compared with the data. 

Each model has seven free parameters, that are explored using a MCMC sampler:

\begin{itemize}
	\item X,Y position of the kinematic centre in image plane. 
	\item V$_{\rm sys}$ systemic velocity.
	\item Inclination of the galaxy (\textit{inc}).
	\item Major axis kinematic angle in \textit{source} plane (\textit{PA}). 
	\item Maximum Velocity ($V_{\rm max}$).
	\item Transition radius ($r_t$).
\end{itemize}

\begin{table}
\centering
\caption{2D kinematic fit priors and results for the three kinematic models tested: arctangent ("Atan"), exponential disc ("Exp") and isothermal sphere ("Iso"). Fitted parameters are the inclination ("Inc"), the position angle (PA), the maximum velocity (V$_{max}$) and the transition radius (r$_t$). The centre of the models was aligned with the morphological centre measured in Sect~\ref{subsec:lens_models}. For AS1063, the inclination was fixed, since the seeing of the observations ($\sim$1") did not allow to constrain this quantity. The position angle is given in source plane. The logarithmic evidence for each model is listed in the last column and the lowest evidence model is highlighted in bold in the first column.}
\label{tab:emcee_results}
\tabcolsep=0.022cm
\begin{tabular}{|lccccc|} 
\hline
& Inc.  &  PA   & $V_{\rm max}$            & $r_t$   & ln evidence\\ 
& (deg)   &  (deg)        &  (km$\,$s$^{-1}$)   & (kpc)   &   \\
\hline\hline
\multicolumn{6}{c}{AS1063-arc} \\
\hline
Priors    &  52     & [-53,-13]    & [50,500]  & [0.1,15]    &                \\ 
\textbf{Atan}&  -		& $-27\pm1$ 	& 296$\pm$2 & 3.0$\pm$0.2 & $-8314$\\ 
Exp 	  &  - 		& $-28\pm1$ 	& 266$\pm$1 & 8.7$\pm$0.1 & $-9012$\\
Iso       &  - 		& $-28\pm1$ 	& 262$\pm$1 & 8.9$\pm$0.1 & $-8620$\\
\hline
\multicolumn{6}{c}{A370-sys1} \\
\hline
Priors    &  [21,61]    & [$-12$,$-52$]    & [50,500]  & [0.1,15]    &                \\ 
Atan      &  53$\pm$1	& $-29\pm1$ 	& 239$\pm$2 & 2.0$\pm$0.2 & $-26786$\\
\textbf{Exp}&  54$\pm$1	& $-30\pm3$ 	& 207$\pm$5 & 5.1$\pm$0.1 & $-26607$\\ 
Iso       &  55$\pm$1	& $-31\pm1$ 	& 203$\pm$1 & 5.1$\pm$0.1 & $-26620$\\ 
\hline
\multicolumn{6}{c}{A2390-arc} \\
\hline
Priors    &  [31,71]    & [$-107$,$-57$]    & [50,500]  & [0.1,15]    &                \\ 
Atan      &  57$\pm$1	& $-101\pm1$ 	& 221$\pm$3 & 0.4$\pm$0.2 & $-7272$\\
Exp       &  58$\pm$1	& $-102\pm1$ 	& 229$\pm$3 & 6.2$\pm$0.2 & $-7254$\\ 
\textbf{Iso}&  58$\pm$2	& $-102\pm1$ 	& 225$\pm$2 & 5.0$\pm$0.2 & $-7230$\\ 
\hline
\multicolumn{6}{c}{M0416-sys28} \\
\hline
Priors    &  [37,77]    & [52,92]    & [50,500]  & [0.1,15]    &                \\ 
\textbf{Atan}&  51$\pm$2	& 67$\pm$7 	& 157$\pm$7  & 15.0$\pm$3.0 & $-1347$\\ 
Exp       &  76$\pm$9	& 58$\pm$4 	& 130$\pm$11 & 14.6$\pm$0.4 & $-1725$\\ 
Iso       &  53$\pm$10	& 63$\pm$5 	& 157$\pm$19 & 14.8$\pm$0.2 & $-1591$\\ 
\hline
\multicolumn{6}{c}{A2667-sys1} \\
\hline
Priors    &  [24,64]    & [$-184$,$-104$]    & [50,500]  & [0.1,15]    &                \\ 
\textbf{Atan}&  25$\pm$2	& $-117\pm1$ 	& 165$\pm$10 & 0.8$\pm$0.2 & $-6109$\\ 
Exp       &  25$\pm$2	& $-108\pm1$ 	& 175$\pm$10 & 4.5$\pm$0.2 & $-6364$\\ 
Iso       &  25$\pm$2	& $-108\pm1$ 	& 162$\pm$7  & 3.5$\pm$0.2 & $-6262$\\ 
\hline
\multicolumn{6}{c}{M1206-sys1} \\
\hline
Priors    &  [44,84]    & [$-10$,30]    & [50,500]  &[0.1,15]    &                \\ 
\textbf{Atan}&  70$\pm$1	& 24$\pm$1 	& 225$\pm$1    & 3.2$\pm$0.2 & $-16498$\\ 
Exp       &  60$\pm$10	& 20$\pm$8 	& 203$\pm$60   & 8.7$\pm$3.2 & $-21542$\\ 
Iso       &  60$\pm$3	& 20$\pm$1 	& 199$\pm$42   & 8.4$\pm$0.1 & $-20931$\\ 
\hline
\multicolumn{6}{c}{A521-sys1} \\
\hline
Priors    &  [31,81]    & [207,277]    & [50,500]  &[0.1,15]    &                \\ 
Atan      &  79$\pm$1	& 252$\pm$2 	& 120$\pm$6    & 0.4$\pm$0.2 & $-33413$\\ 
\textbf{Exp}&  72$\pm$1	& 256$\pm$2 	& 130$\pm$5    & 4.5$\pm$0.3 & $-33309$\\ 
Iso       &  75$\pm$2	& 257$\pm$3 	& 172$\pm$37    & 1.1$\pm$0.7 & $-33381$\\ 
\hline
\multicolumn{6}{c}{M1149-sys1} \\
\hline
Priors    &  [31,71]    & [$-73$,5]    & [50,500]  &[0.1,15]    &                \\ 
Atan      &  36$\pm$3	& $-32\pm1$ 	& 104$\pm$6    & 0.2$\pm$0.2 & $-12806$\\ 
\textbf{Exp}&  35$\pm$3	& $-32\pm1$ 	& 119$\pm$7    & 3.7$\pm$0.1 & $-12562$\\ 
Iso       &  36$\pm$3	& $-32\pm1$ 	& 114$\pm$7    & 2.4$\pm$0.1 & $-12708$\\ 
\hline
\end{tabular}
\end{table}

The first three parameters (X,Y,V$_{\rm sys}$) are mainly introduced to correct for small inaccuracies in the lensing correction (V$_{\rm sys}$ is used to keep the v=0 \kms at the centre even if the centre changes slightly.). We assume that the kinematic centre is the same as the morphology centre (measured in the reconstructed image), and correct the observed velocity fields in order to have zero velocity at this position. To correct any possible errors when passing from the position in source plane to image plane, during the fit we allow the kinematic centre to vary within a 2$\times$2 MUSE pixel box around the original position and let the velocity adjust up to $\pm$100 \kms. The corrections for all galaxies were small with displacements of 1 pixel and velocity adjustments of up to $\pm$30 \kms. 

The maximum velocity and transition radius ($V_{\rm max},r_t$) correspond to the maximum velocity (and the radius at which it occurs) for the exponential disc and isothermal sphere models. The arctangent model velocity does not reach a maximum at a finite radius, so the transition radius in the arctangent model corresponds to the radius at which 70\% of the asymptotic velocity ($V_{\rm max}$) value is reached, and it is therefore generally lower for the arctangent model than for the other two kinematic models.

\begin{landscape}%
\begin{figure}
\includegraphics[width=0.7\textheight]{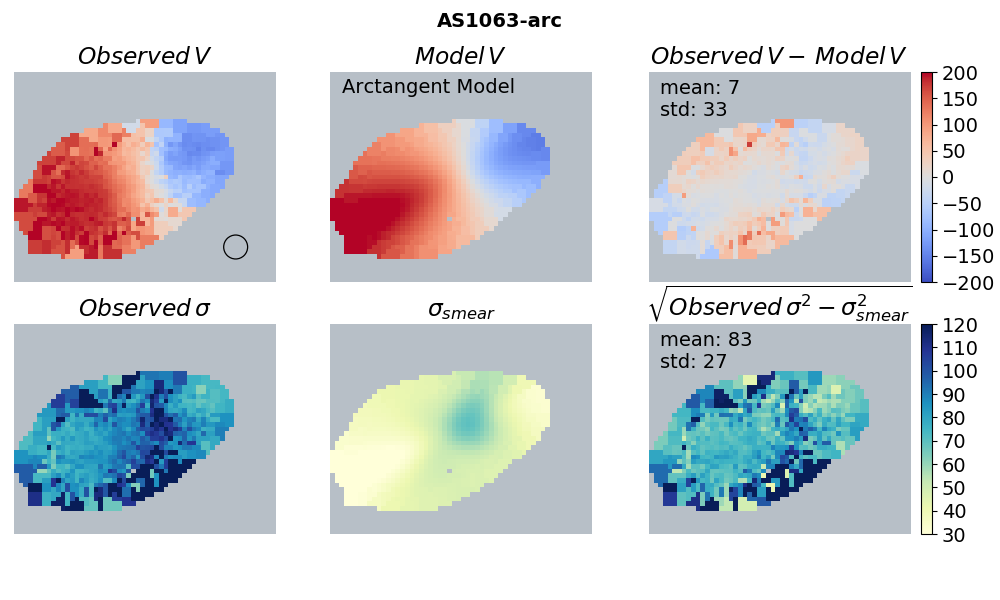}
\includegraphics[width=0.7\textheight]{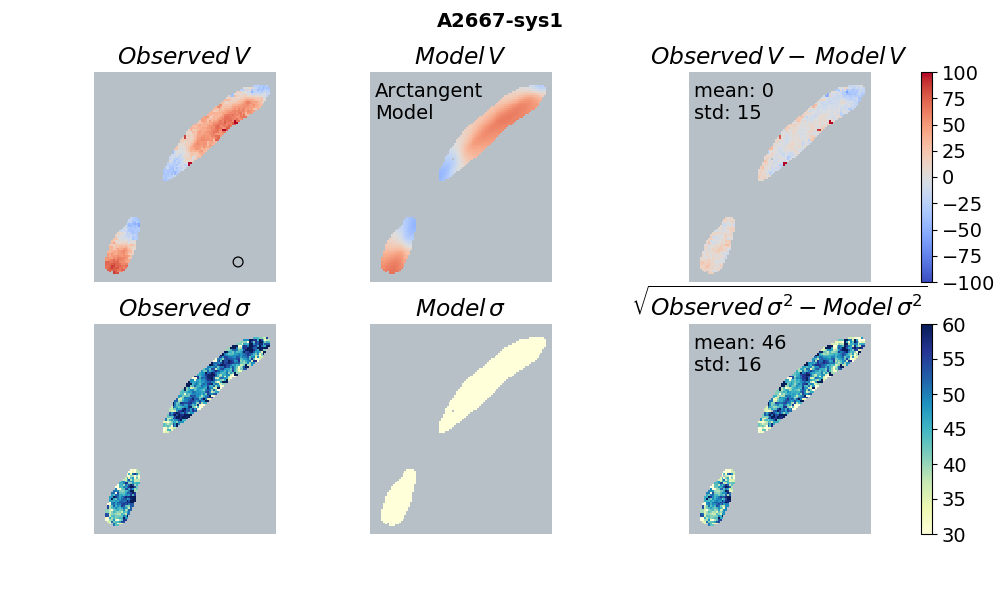}\\
\includegraphics[width=0.7\textheight]{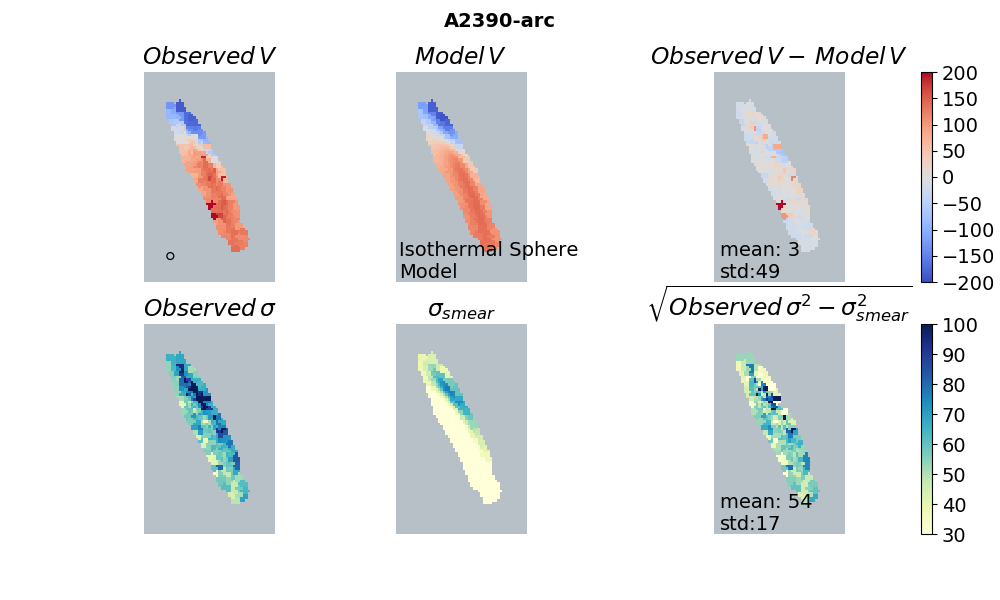}
\includegraphics[width=0.7\textheight]{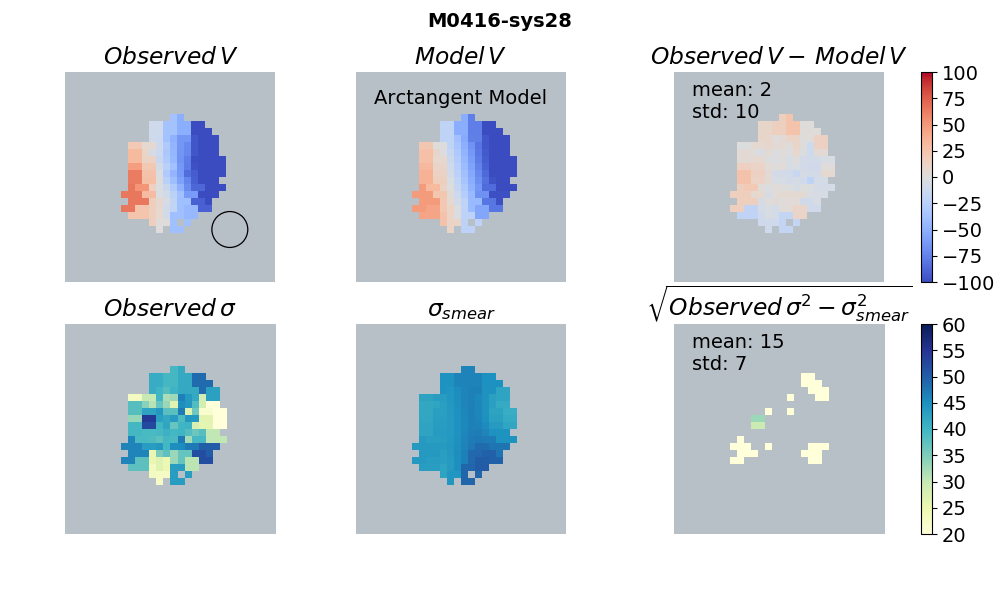}\\
\caption{Observed and model velocity maps of the 8 gravitational arcs. For each galaxy, the 3 first upper panels display the velocity maps: the observed velocity derived from [\oii] emission, the best model (with the model name) and the residuals. On the first panel, a black circle gives the seeing size for that observation. The 3 lower panels are the equivalent for the velocity dispersion. Both colour bars are in in units of \kms.} We have rejected and masked (in grey) values with unreliable beam-smearing corrections ($\sigma_{\textrm{smear}}$>$\sigma_{\textrm{obs}}$).
\label{fig:2d_kin}
\end{figure}
\end{landscape}

\begin{landscape}%
\begin{figure}
\includegraphics[width=0.7\textwidth]{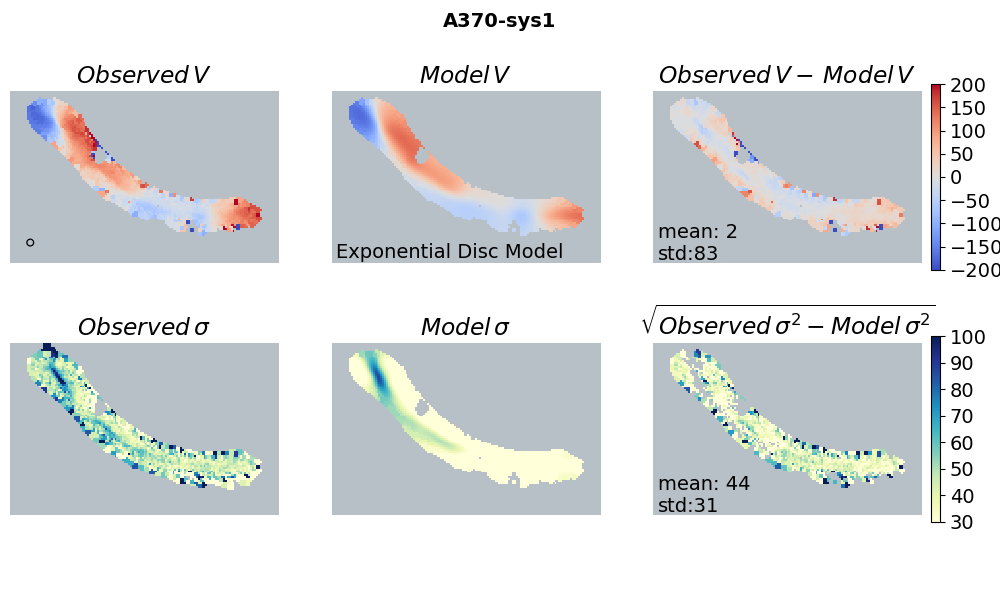}
\includegraphics[width=0.7\textwidth]{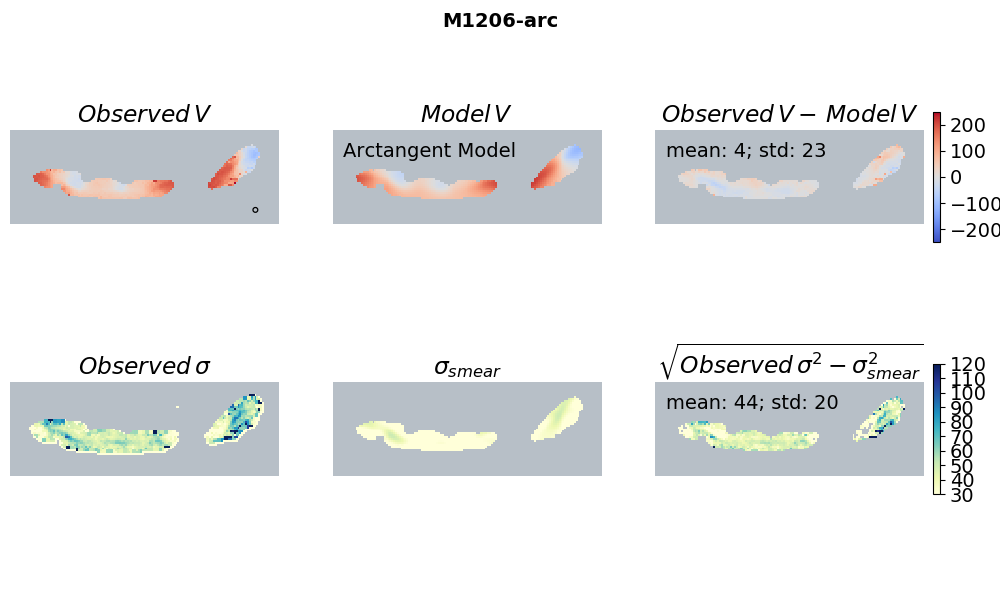}\\
\includegraphics[width=0.7\textwidth]{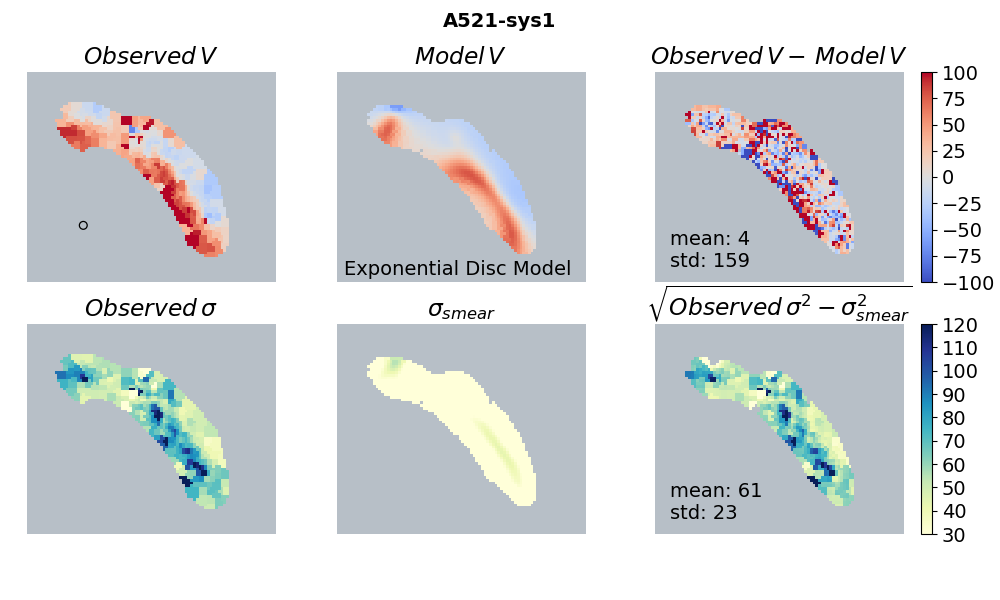}
\includegraphics[width=0.67\textwidth]{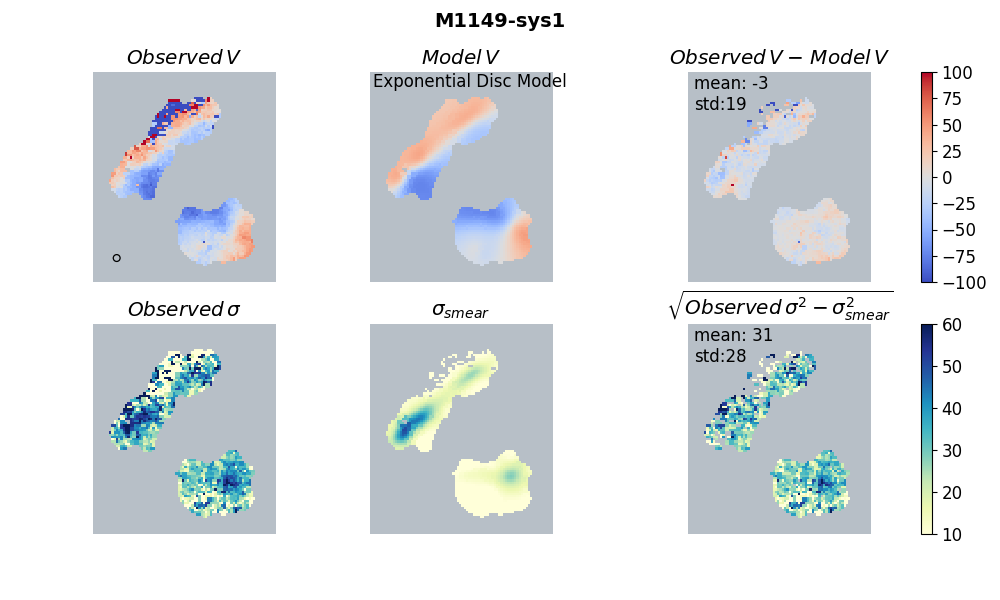}
\contcaption{}
\end{figure}
\end{landscape}

We use the {\sc emcee} \citep{Foreman-Mackey2013} MCMC (Monte Carlo Markov Chain) implementation, with 100 walkers and let the chain evolve for 1000 steps, after an initial burn-in phase of 100 steps. The chains are inspected to confirm convergence. Also here, we adopt a Gaussian likelihood function:

\begin{equation}
\label{eq:lnprob_kin}
\ln\,p = - \frac{1}{2} \sum_r \left[ \frac{(Model(inc,PA,V_{max},r_t) - Vel)^2}{\sigma^2} + \ln(2\pi\sigma^2) \right]
\end{equation}

where $Model(inc,PA,V_{max},r_t)$ is one of the kinematic models (arctangent, isothermal sphere or exponential disc) with inclination $inc$, position angle $PA$, maximum velocity $V_{max}$ and transition radius $r_t$, after being lensed, convolved with the seeing and binned. The model is compared with the observed velocity field ($Vel$) and weighed by the errors, $\sigma$. For these, we only consider the errors of the measured velocity, as given by {\sc camel}, corresponding to the formal errors of the $\chi^2$ minimisation of each bin. We assume flat priors to all the parameters being fit.

We choose a wide prior for the maximum velocity and transition radius that is the same for all galaxies. The inclination was given a wide prior of $\pm$20 degrees around the inclination derived from the morphological fit. The only exception is AS1063-arc, observed with the worst seeing in this sample ($\sim$1"), where the inclination was unconstrained by the kinematic data and was set to the value measured in the \emph{HST} image. The position angle was also given a wide prior ($\pm$20 degrees) around the position angle measured in the exponential fit to the \emph{HST} images. Finally, we present the Bayesian evidence of each model, a relative measure of how well a model performs, as calculated by {\sc emcee}.

\subsection{Results}
\label{sec:properties_all} 

The results for each model are presented in Table~\ref{tab:emcee_results}, where the best value of each parameter corresponds to the median of the marginalised distributions. Also here, the half distance between the 16th and 84th percentiles was taken as the error for each parameter. From the Bayesian evidences values,  the preferred kinematic model in this sample is the arctangent model, with galaxies AS1063-arc, A2667-sys1, M0416-sys28 and M1206-sys1 being better fit by this model. (However, notice that the fit of M0416-sys28 is not constrained, with the transition radius hitting the upper limit allowed). The remaining galaxies, A370-sys1, A521-sys1 and M1149-sys1 are better fit by an exponential disc model and only A2390-arc by an isothermal sphere. And although there is a clearly preferred model for each galaxy, none of the models is globally preferred. The position angles (PA), that are expected to be the same among the three models, agree quite well for each galaxy. The inclination shows a larger spread, but still displays a reasonable agreement. The 2D maps for the best model with the parameters of Table~\ref{tab:emcee_results} are plotted in the middle panel of Fig.~\ref{fig:2d_kin}.

As previously stated, intrinsic velocity dispersions are difficult to estimate due to the beam-smearing effect. Here we benefit from high magnification of the lensing and the good seeing of the MUSE observations that reduce this problem, but nevertheless we still correct each pixel using the 2D velocity kinematics. From the best model for each galaxy, we calculate the expected velocity dispersion created by the beam-smearing effect ($\sigma_{\textrm{smear}}$) applying equation A24 from \citealt{Epinat2012}. At the [\oii] wavelengths for this sample, MUSE has a LSF with FWHM of $\sim$40-80 km$\,s^{-1}$ (as from \citealt{Bacon2017} fit of the UDF mosaic data). We subtract in quadrature the LSF value at the [\oii] wavelength of each galaxy. The intrinsic velocity dispersion, corrected for instrument broadening and velocity smear, is then given by: $\sigma = \sqrt{\sigma_{\textrm{obs}}^2 - \sigma_{\textrm{inst}}^2 - \sigma_{\textrm{smear}}^2 }$. We plot these 2D maps in the lower right panel in Fig.~\ref{fig:2d_kin}.


All galaxies are rotation-dominated, with regular 2D velocities and  $V/\sigma$ ratios between $\sim$2 and $\sim$10, estimated from the maximum velocity $V_{\rm max}$ and the mean of the $\sigma$ maps (see Table~\ref{tab:kin_sigma}). These values are in agreement with $V/\sigma$ ratios found for other high-$z$ samples such as the KMOS$^{3D}$ kinematics survey \citep{Wisnioski2015}, with $V/\sigma$ between $\sim$3-11 for galaxies at $z\sim1$, the MUSE sample of low mass galaxies \citep{Contini2016}, with the range $\sim$1-5 and the KLASS survey of lensed galaxies at $0.7<z<2.3$ \citep{Mason2017}, with $V/\sigma\sim$8 at $z\sim1$. 

\begin{table}
\centering
\caption{Kinematic properties of the sample. V$_{\rm max}$ is the maximum velocity of the best model (repeated from Table~\ref{tab:emcee_results}); $\sigma$ is the mean and standard deviation of intrinsic velocity dispersion (corrected for instrumental broadening and beam-smearing) measured in the 2D maps of Fig.~\ref{fig:sample}. We also calculate the Gini coefficient of the intrinsic velocity dispersion maps (G$_\sigma$) and of the [\oii] narrow band images (G$_{[\oii]}$), as a comparison. The velocity dispersions are high but quite homogeneous.}
\label{tab:kin_sigma}
\tabcolsep=0.13cm
\begin{tabular}{|lccccc|} 
\hline
Object			&	V$_{\rm max}$		& $\sigma$ & V$_{\rm max}$/$\sigma$  & G$_{\sigma}$ & G$_\oii$\\
			&  (\kms) 		& (\kms)   &			 &    			& 		  \\
\hline
\hline
AS1063-arc	& 296$\pm$2 	& 84$\pm$33 & 3.5$\pm$1.4 & 0.16 & 0.82\\
A370-sys1	& 207$\pm$7 	& 44$\pm$31 & 4.7$\pm$3.3 & 0.29 & 0.90\\
A2390-arc	& 225$\pm$2 	& 54$\pm$17 & 4.2$\pm$1.3 & 0.16 & 0.88\\
M0416-sys28	& 157$\pm$7 	& 15$\pm$7 & 10.5$\pm$4.9 & 0.25 & 0.86\\
A2667-sys1	& 165$\pm$10  	& 46$\pm$20 & 3.6$\pm$1.6 & 0.14 & 0.93\\
M1206-sys1	& 225$\pm$1 	& 44$\pm$31 & 5.1$\pm$3.6 & 0.20 & 0.83\\
A521-sys1	& 130$\pm$5 	& 63$\pm$23 & 2.1$\pm$0.8 & 0.26 & 0.87\\
M1149-sys1	& 119$\pm$7 	& 31$\pm$28 & 3.8$\pm$3.5 & 0.30 & 0.83\\
\hline
mean		& 191		   & 48 	   & 4.7		 & 0.23 & 0.87\\
\hline
\end{tabular}
\end{table}

\begin{figure}
\includegraphics[width=0.50\textwidth]{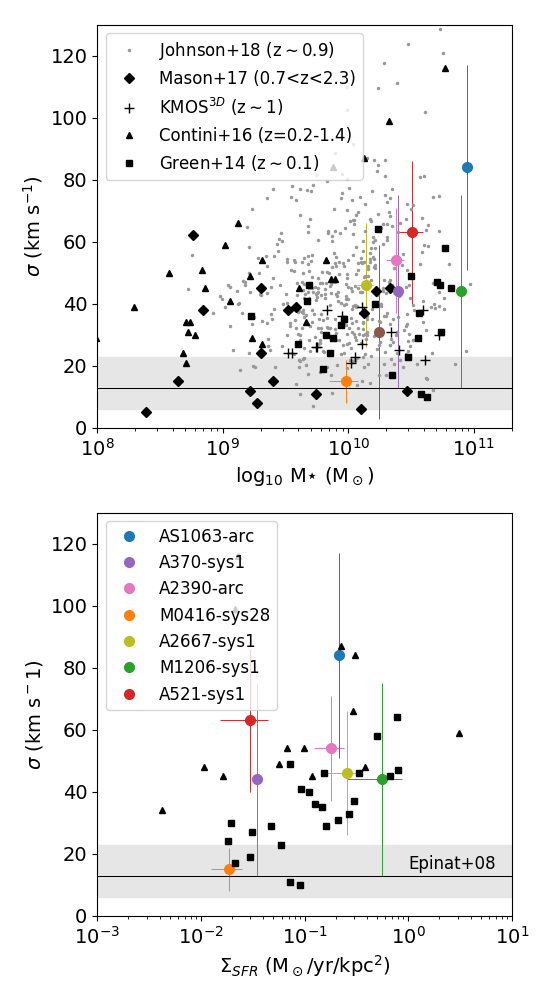}
\caption{Comparison of the velocity dispersion of the gravitational arcs with other samples. Velocity dispersion against stellar mass on the top panel and velocity dispersion against star formation surface density on the bottom (M1149-sys1 is omitted from this plot since we do not derive its SFR). The mean value for the local sample of rotating galaxies from \citealt{Epinat2008} in plotted as a black line, with scatter masked by the grey area; the local ($z=0.1$) DYNAMO sample of \citealt{Green2014} in black squares (we plot only the rotating disc galaxies); the $z\sim0.2-1.4$ sample from \citealt{Contini2016} in black triangles; a sub-set of the $z=1$ KMOS$^{3D}$ galaxies, with velocity dispersions derived by \citealt{DiTeodoro2016} as black crosses (only in the top panel); the lensed galaxies from the KLASS sample \citep{Mason2017} are plotted as black diamonds (we have excluded possible mergers and galaxies with only upper limits). The KMOS Redshift One Spectroscopic Survey Sample at z$\sim$1 from \citealt{Johnson2017} is also plotted in grey dots. The gravitational arcs sample velocity dispersions are in good agreement with all these samples.}
\label{fig:local_sample}
\end{figure}


\section{Discussion}
\label{sec:discussion} 

In this section, we discuss the implications of the kinematic fit results and the morphology of the velocity dispersion maps. We also explore the connection between star formation and turbulence in these typical $z\sim$1 discs. A summary of the individual properties of the galaxies, and comparison with previous studies of these objects, can be found in Appendix~\ref{sec:properties_each}.

\subsection{Best Kinematic Model}
\label{subsec:best_kin}

Recent work by \citealt{Genzel2017} and \citealt{Lang2017} found that both the individual velocity curves of massive galaxies and the stacked velocity curve of 101 star-forming galaxies in the range $0.6<z<2.6$, display a significant drop in the velocity curve beyond the turnover radius. This would imply a higher baryonic mass fraction in high-$z$ galaxies than previously though. We briefly investigate this issue.

As stated before, four out of eight galaxies are better fit by the arctangent model, an empirical model. The exponential disc model, which describes the rotation velocity of a gas disc, has lower Bayesian evidences on three of the galaxies and the isothermal sphere model, an approximation to what is expected for an isothermal dark matter halo, on only one. Most importantly, all models are remarkably similar in the inner parts of the galaxies. For most galaxies, the three models (with the best parameter estimations) are visually indistinguishable up to 1 R$_e$ (e.g. A370-sys1 and A2667-sys1), or even 2 R$_e$ (e.g. AS1063-arc). 

In none of the galaxies analysed here do we see a drop in the velocity curve at large radii, which is confirmed by the fact that the arctangent model (the only one that does not decrease velocity with radius) is the most preferred model. However, with the exception of AS1063-arc and A2667-sys1, our data does not extend out as far as the stacked sample of \citealt{Lang2017}, so we cannot rule out the possibility of a drop in the velocity beyond 2 effective radii. We also cation the reader that our analysis is too simplistic to explore this subject, since a combination of the two components -- baryons and dark matter -- is expected, and a joint fit should be done in order to gain some more insight on this matter, as done in \citealt{Lang2017}.

\begin{figure*}
\includegraphics[width=0.24\textwidth]{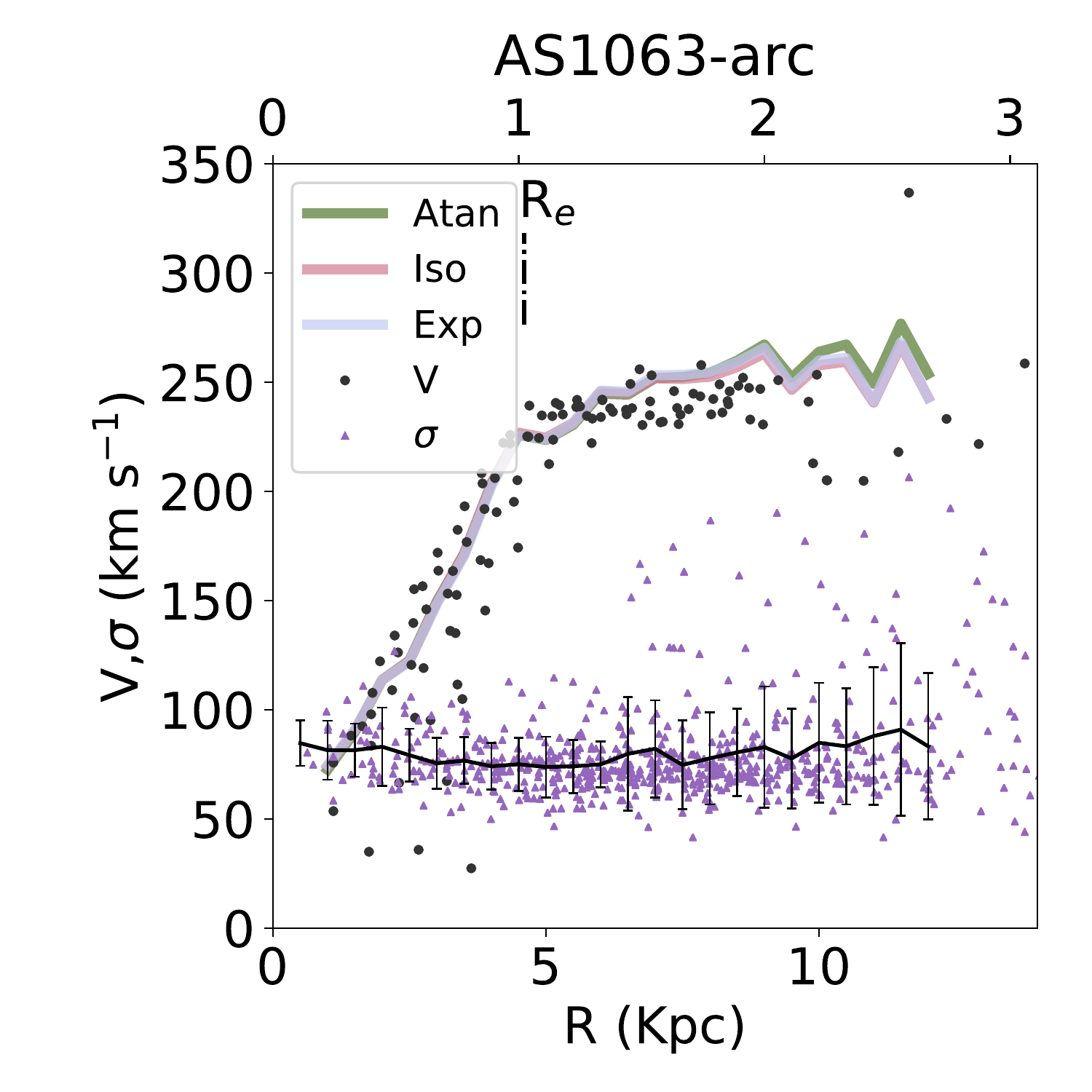}
\includegraphics[width=0.24\textwidth]{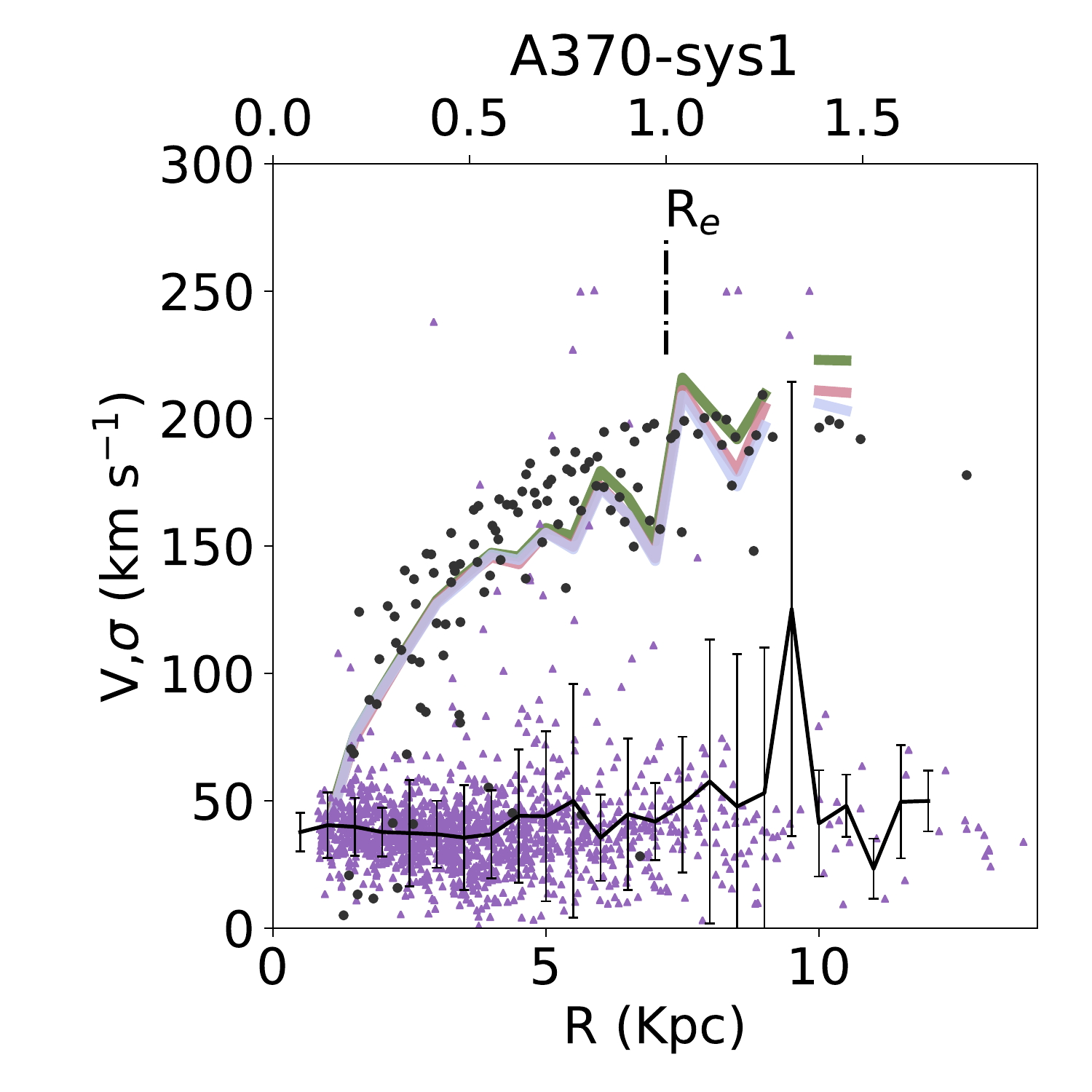}
\includegraphics[width=0.24\textwidth]{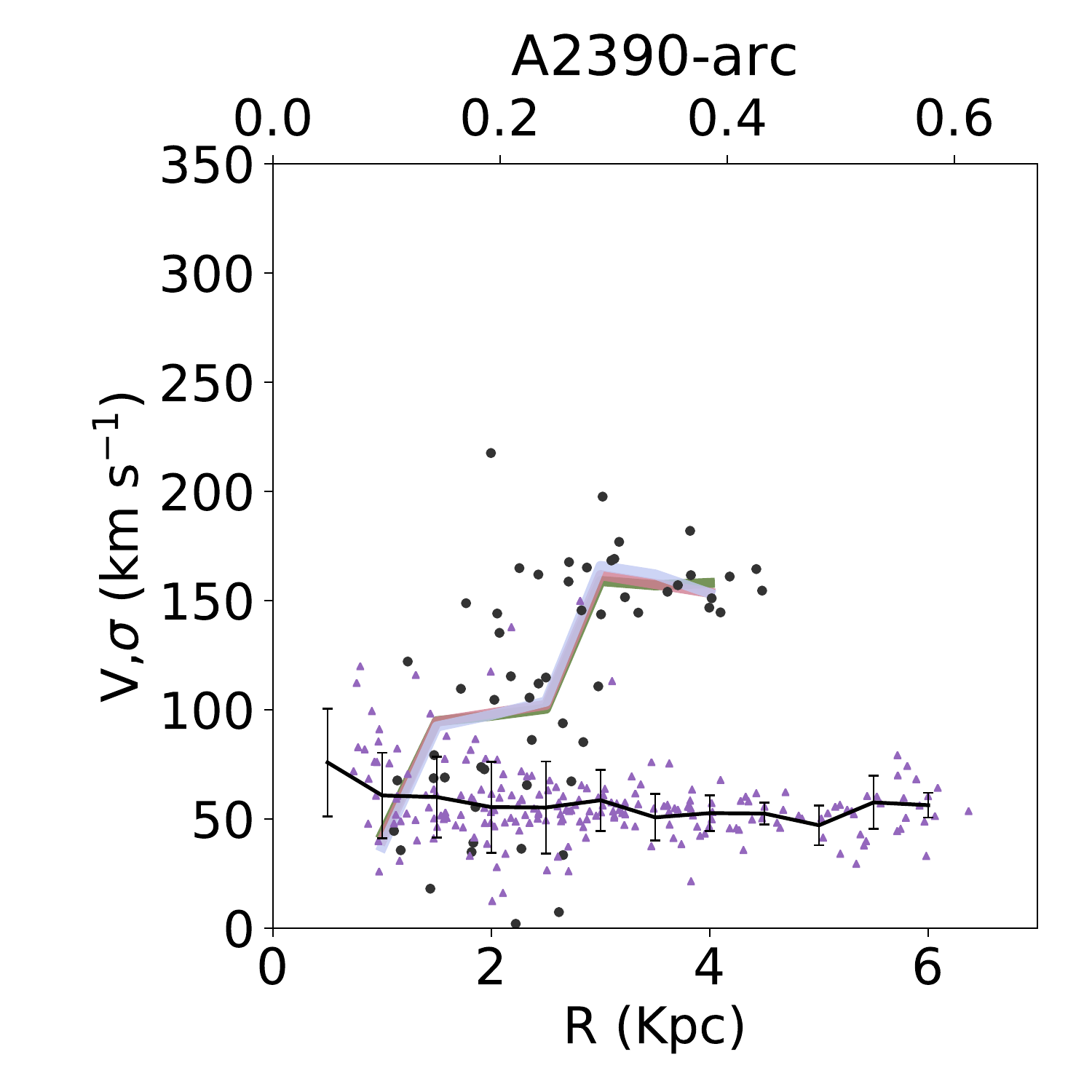}
\includegraphics[width=0.24\textwidth]{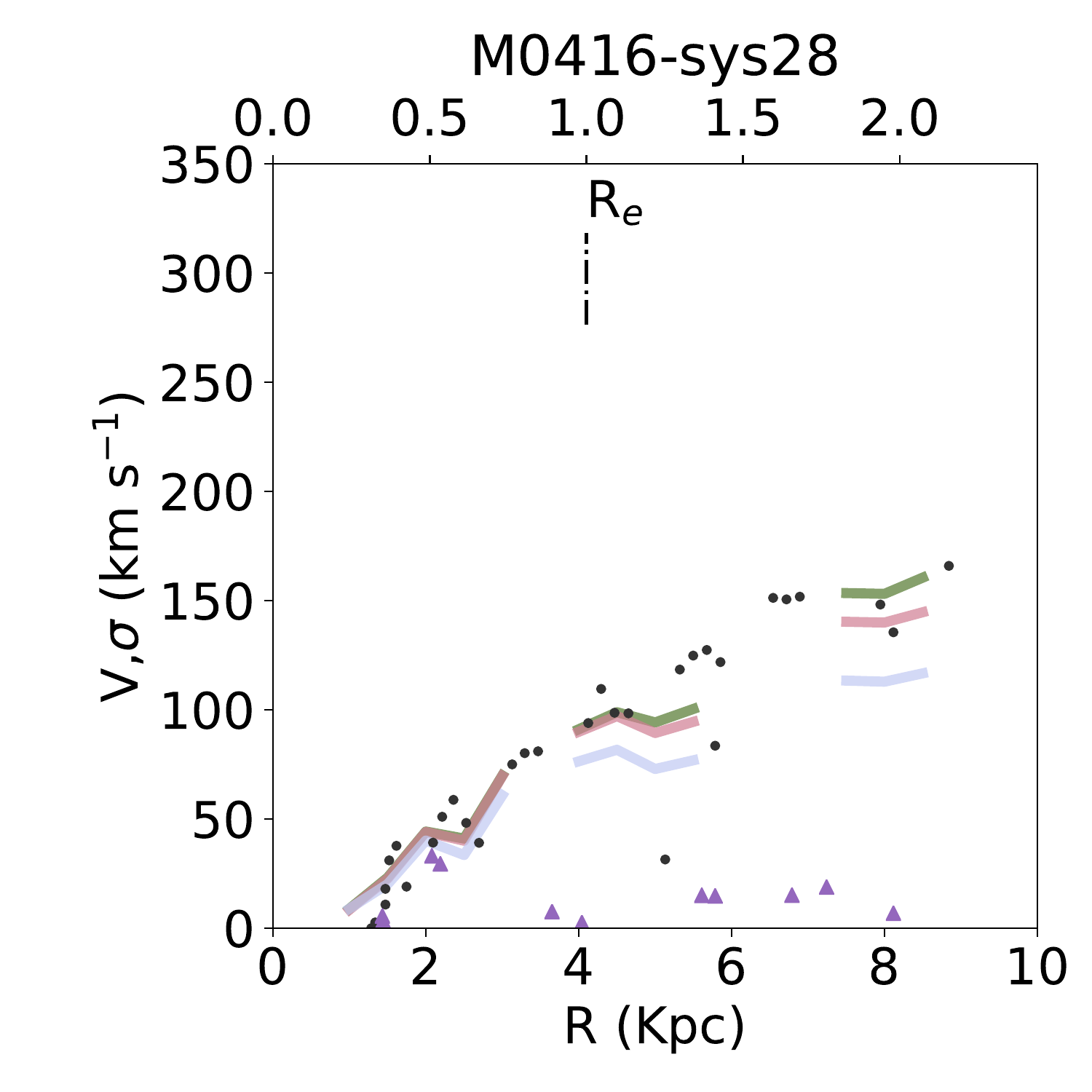}\\
\includegraphics[width=0.24\textwidth]{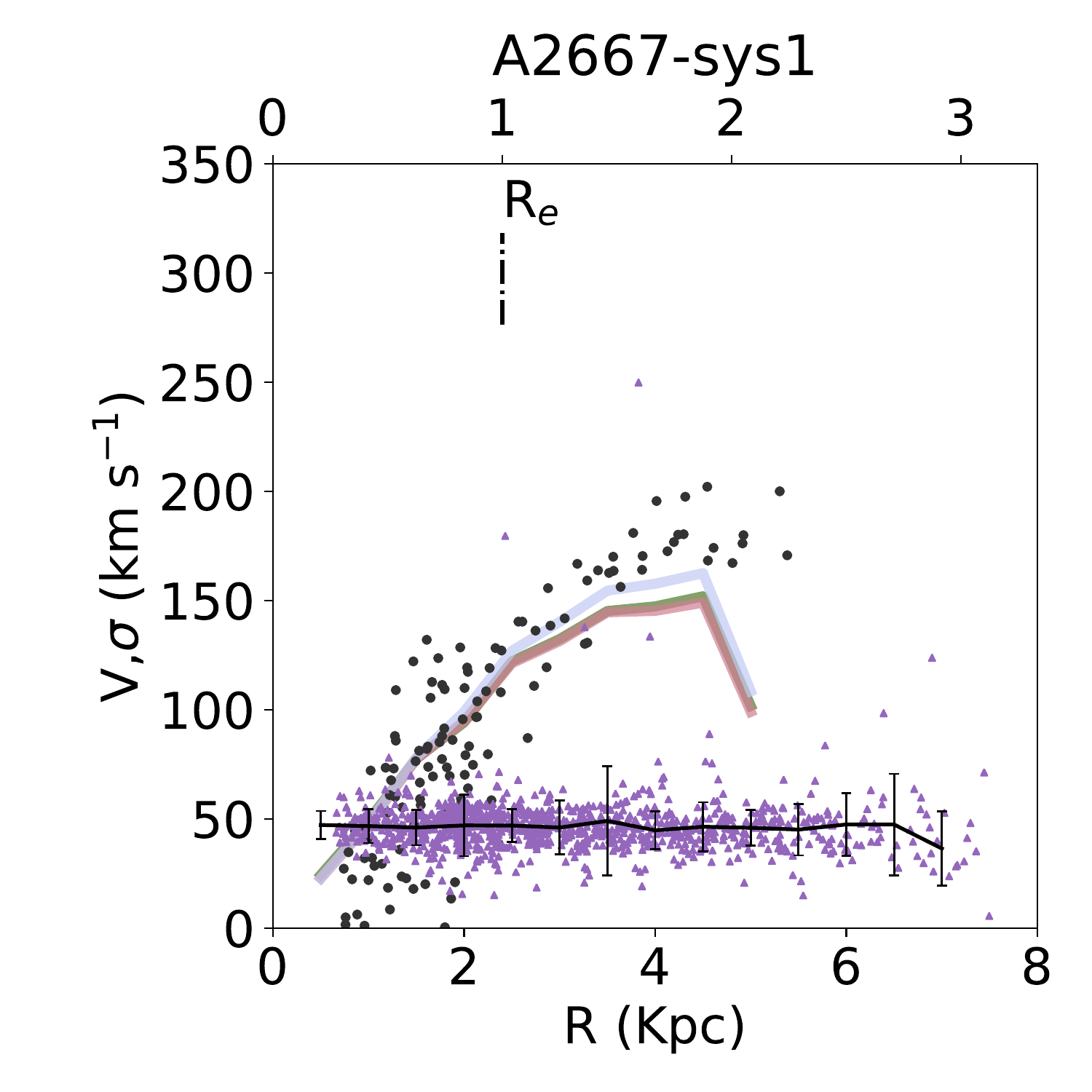}
\includegraphics[width=0.24\textwidth]{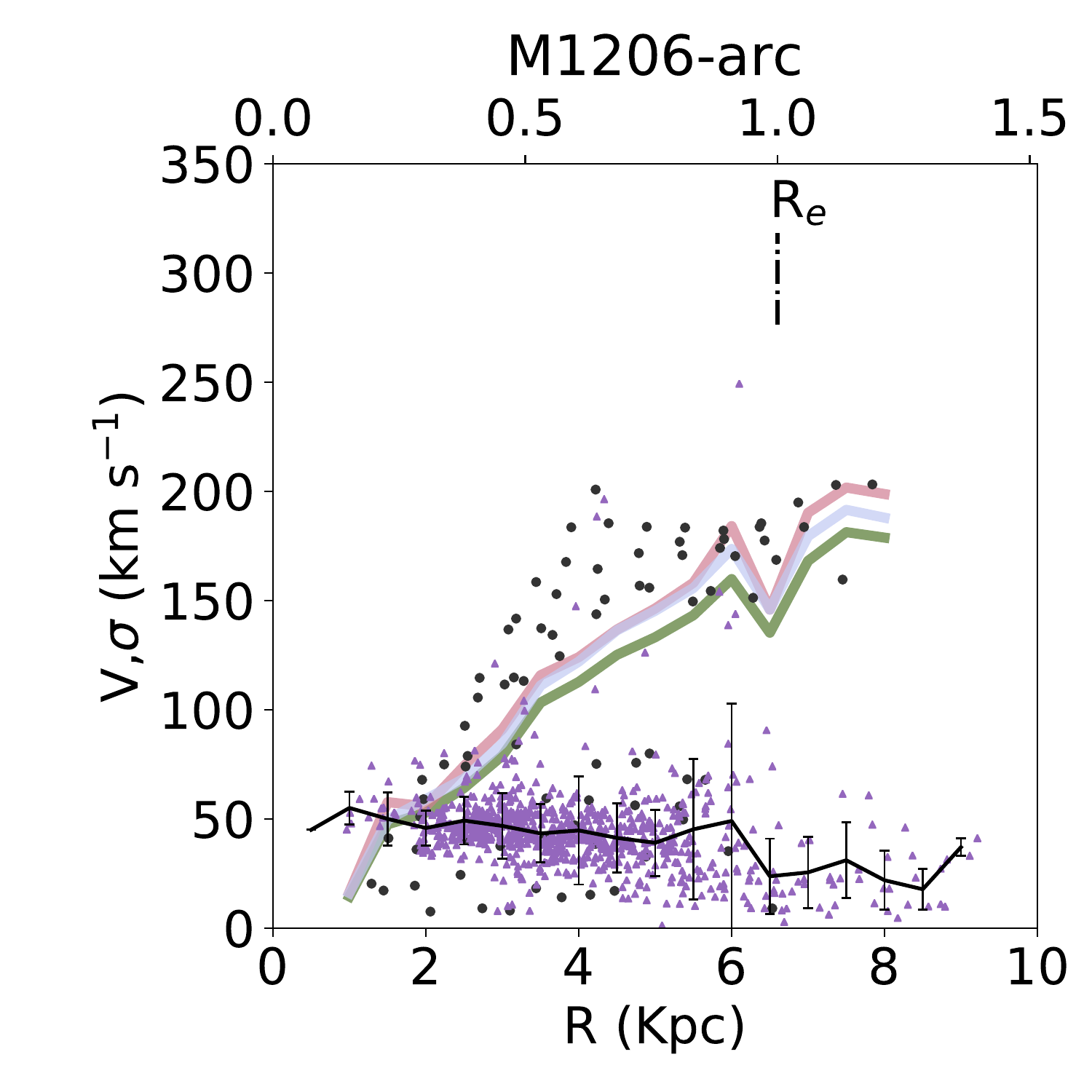}
\includegraphics[width=0.24\textwidth]{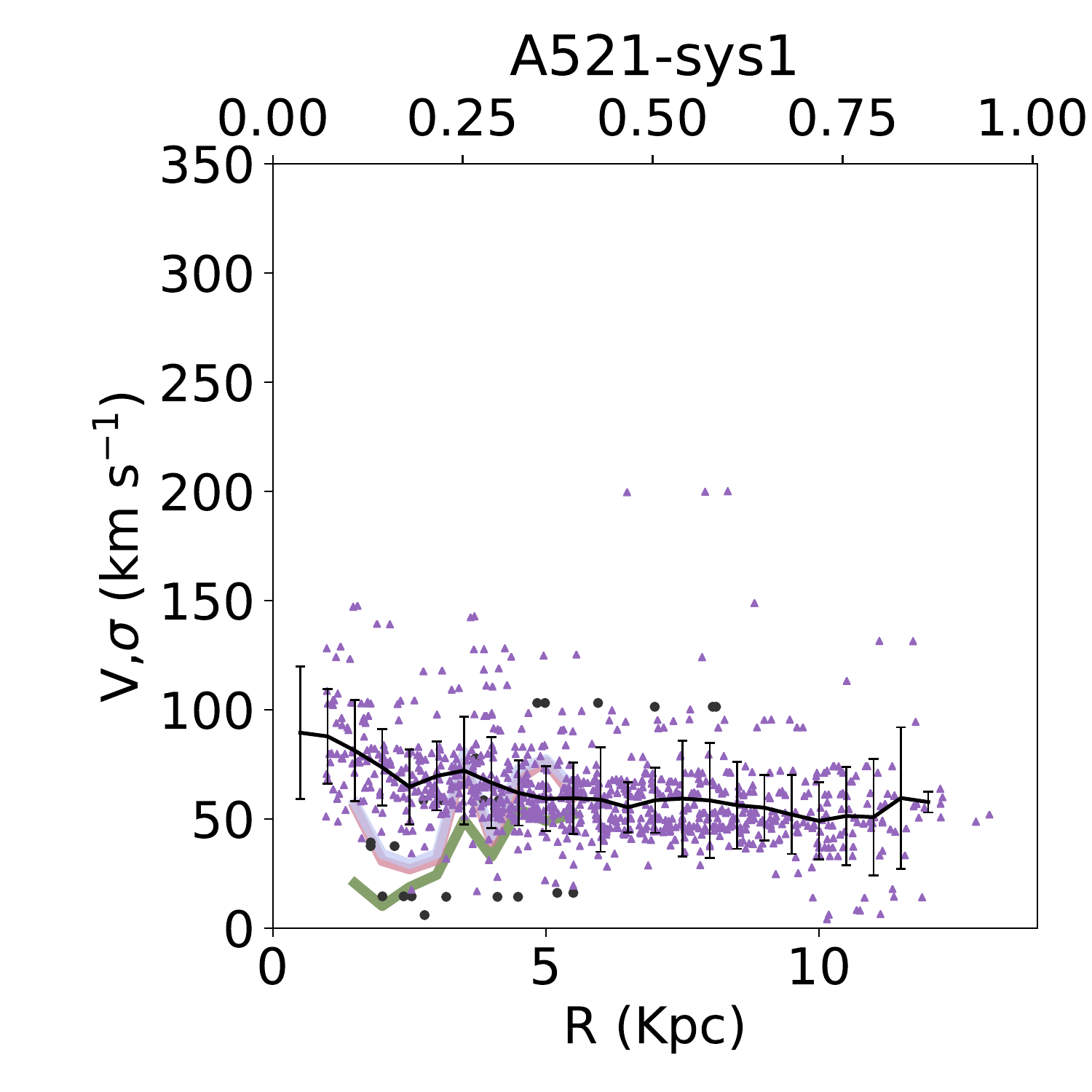}
\includegraphics[width=0.24\textwidth]{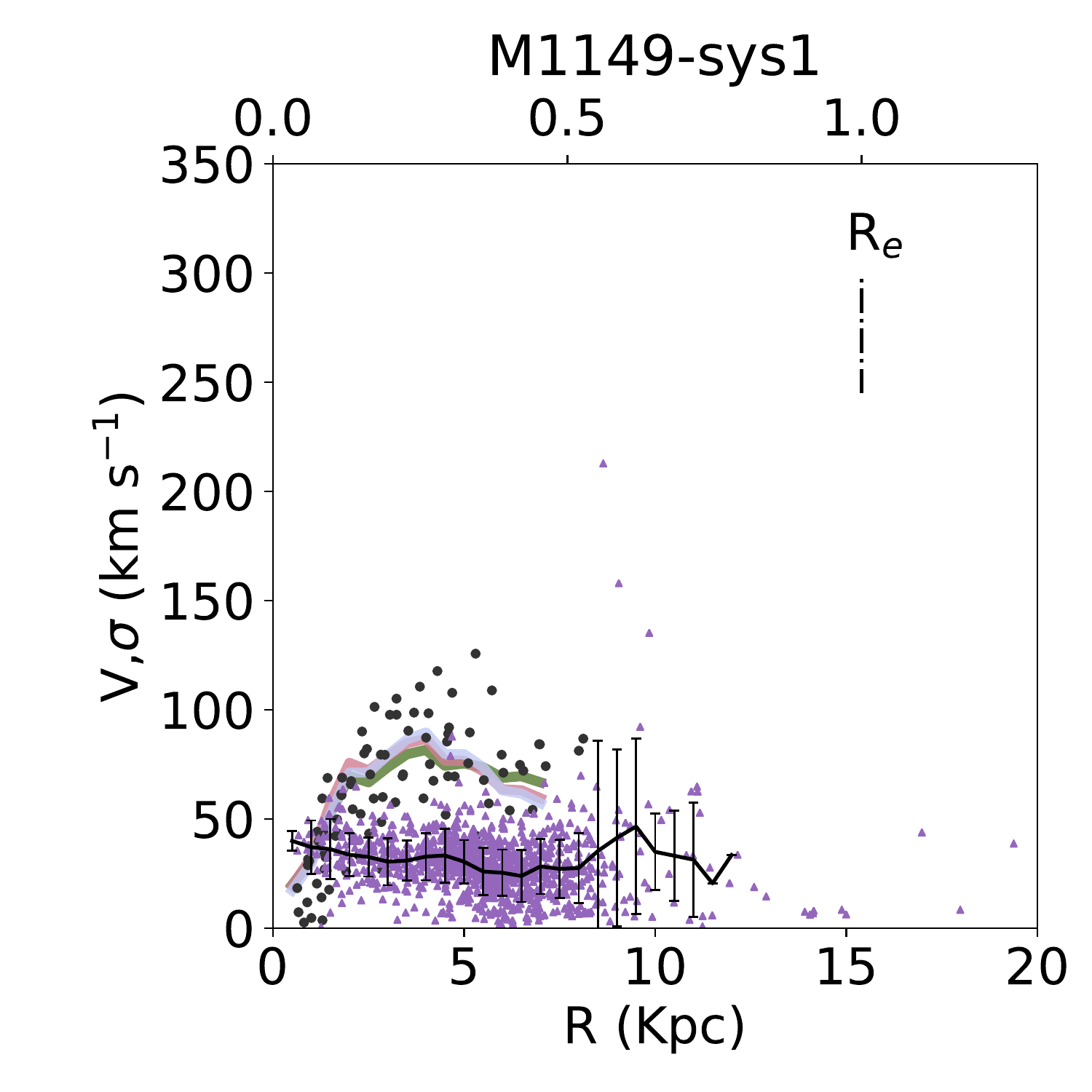}
\caption{Radial velocity and velocity dispersion in source plane for individual bins. Velocity was extracted along the major kinematic axis with an aperture of 0.5" both for the data (back circles) and for the models (thick coloured lines) and is inclination corrected. Velocity dispersion corresponds to the intrinsic velocity dispersion (beam-smearing corrected) and was extracted in the full galaxy. A vertical bar marks the effective radius (R$_{e}$).}
\label{fig:1d_vel}
\end{figure*}

\subsection{Velocity Dispersion Morphology}
\label{subsec:resolved_disp}

We now turn our attention to the velocity dispersion. The mean intrinsic velocity dispersion of these disc galaxies ranges from 15 to 84 \kms, with a mean value of 48\kms. We plot these values in Fig.~\ref{fig:local_sample} and compare them with other samples. The mean velocity dispersions are comparable to the global values derived for the KMOS$^{3D}$, the MUSE samples for $0.2<z<1.4$ galaxies, and the lensed sample from the KLASS survey (\citealt{Wisnioski2015}, \citealt{Contini2016} and \citealt{Mason2017}, respectively). Our velocity dispersions are also in agreement with the values derived for the DYNAMO sample of local analogues \citealt{Green2014,Olivia2017}. When compared to the GHASP survey \citep{Epinat2010}, a sample of nearby late-type galaxies, these lensed galaxies have velocity dispersions higher by a factor of 1.4 to 4.2. 

Several studies have investigated the kinematics of lensed galaxies \citep[e.g][]{Jones2010,Livermore2015,Leethochawalit2016,Mason2017}, mostly focusing on the velocity fields of these galaxies and not so much in the resolved velocity dispersions. The sample presented in the current work was selected to be extremely extended and with high physical resolution and the high signal to noise of our data results in relatively small formal errors per bin in the 2D velocity dispersion maps. These are derived with {\sc camel} and, for AS1063-arc, A370-sys1, A2390-arc, M0416-sys28, A2667-sys1 and M1206-sys1, have a median value that varies between 5 and 8 \kms (but can be as high as 50 \kms). The remaining galaxies, A521-sys1 and M1149-sys1, have higher median errors, 24 and 16 \kms, respectively. Another source of uncertainty is the LSF correction, that we perform using expression 7 of \citealt{Bacon2017}. For our lowest redshift galaxy, the LSF dispersion is about 1.1\AA\, (see figure 15 of \citealt{Bacon2017}), which translates in a dispersion of $\sim$50\kms. For the higher redshift galaxies this value is lower. Neither of these sources of uncertainty directly depends on the position of the bin in the velocity map. On the other hand, the beam-smearing correction does depend on the position of the bin in the 2D map. Calculate the errors associated with this correction is computationally expensive (since a reasonable number of velocity dispersion maps would have to be produced), so we estimate the uncertainties associated with the beam-smearing correction in a simplified way. We produce 16 velocity dispersion models with combinations of perturbed $V_max$, $r_t$, $PA$ and the inclination. Since this is a simplified estimation of the error, we perturbed these parameters by $\pm$20 \kms, $\pm$1 kpc and $\pm$5 degrees, values that are considerably higher than our typical errors. Comparing these perturbed models with the best model for each galaxy, we obtain median differences between 5 and 10 \kms. Overall, our major source of error is the LSF estimation. Nevertheless, \citealt{Bacon2017} measure the LSF to be largely constant in the MUSE FOV, with an average standard deviation of $\sim0.05$\AA, therefore the effects on the study of the velocity dispersion structure are expected to be small.  

The observed, $\sigma_{\rm smear}$ and intrinsic velocity dispersion 2D maps of the 8 galaxies analysed in this work can be seen in the lower panels of Fig.~\ref{fig:2d_kin}. Both the observed and the $\sigma_{\rm smear}$ (the dispersion expected due to observational effects) maps have a higher velocity dispersion at the centre of the galaxies (exceptions are A2667-sys1 and M0416-sys28). The intrinsic 2D velocity dispersion maps reveal no significant 2D structure at our precision limit, with the dispersion of the $\sigma$ maps varying between 7 and 33 \kms (see Fig.~\ref{fig:2d_kin}). Our intrinsic velocity dispersion maps are broadly consistent with a constant velocity dispersion value throughout the entire galaxy.

\begin{figure*}
\includegraphics[width=\textwidth]{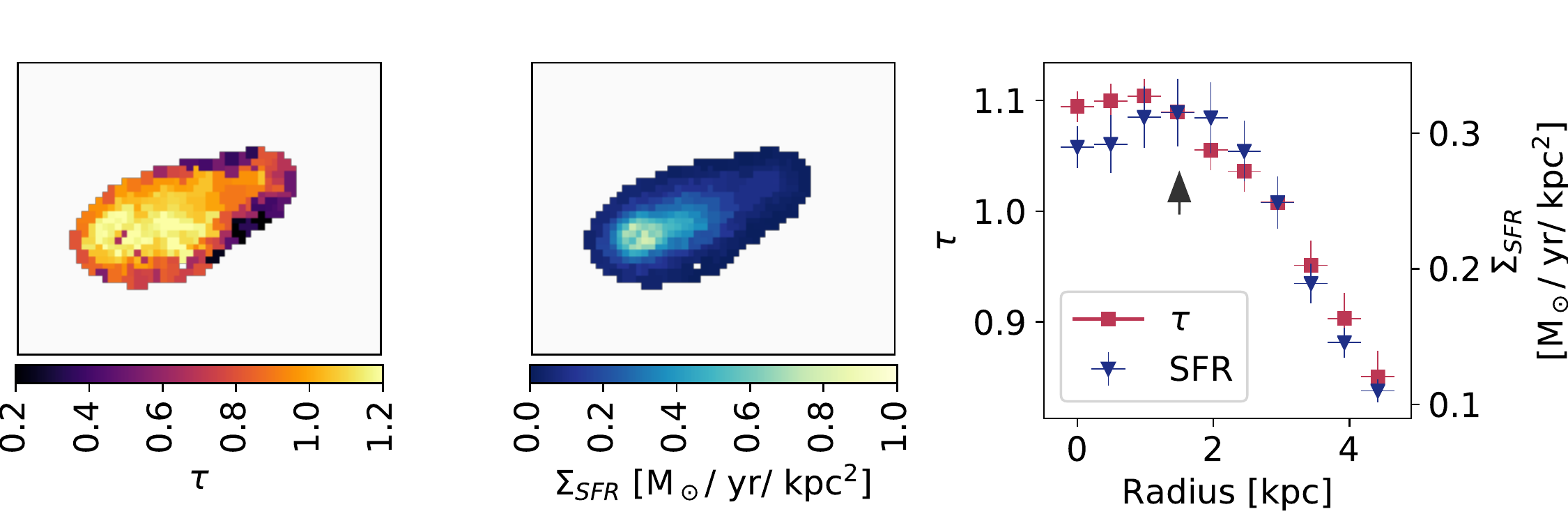}
\includegraphics[width=\textwidth]{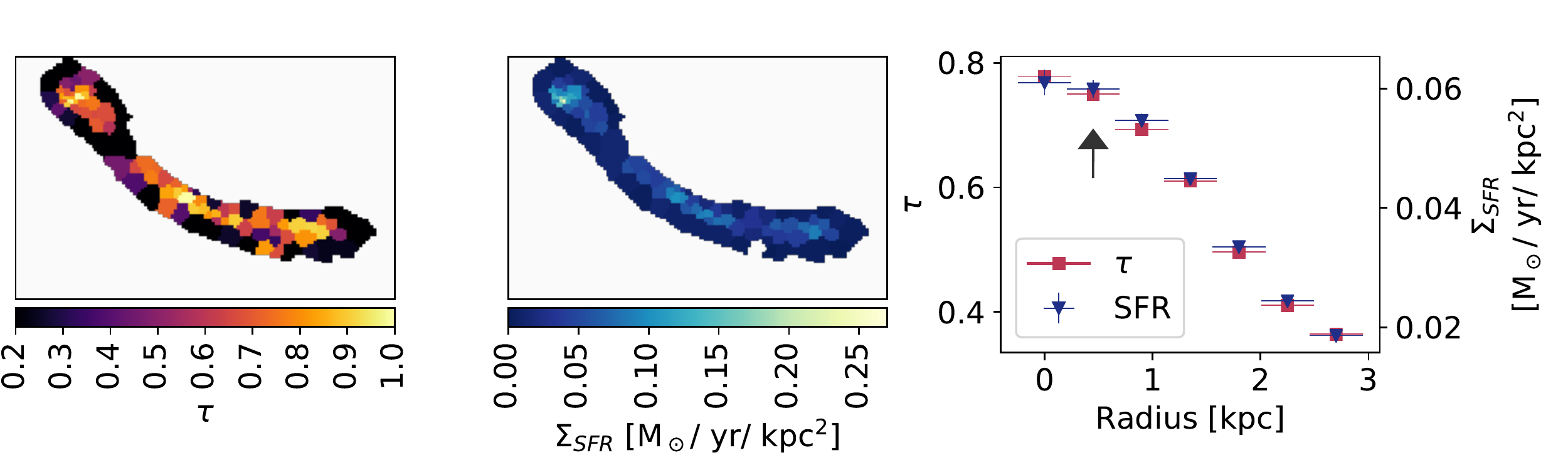}
\caption{Resolved extinction and SFR density maps for AS1063-arc (top) and A370 (bottom). In each row, from left to right: attenuation ($\tau$) map, star formation rate surface density, from H$\beta$ (units of [M$_\odot\,$yr$^{-1}\,$kpc$^{-2}$]) and 1D radial profiles of both. These were obtained by averaging the values of extinction and SFR within distorted elliptical annuli of increasing radius (see Fig.\ref{fig:sample}). These distorted elliptical annuli correspond to circular apertures in source plane.   from the centre of the galaxy. The position of the strongest star forming clumps are marked  with a black arrow. }
\label{fig:resolved_sfr}
\end{figure*}

In order to better quantify how homogeneous the velocity dispersion maps are we also measure their Gini coefficient \citep{Gini1912}. This is a measure of inequality in a data set and can vary from 1, absolute inequality, to 0, absolute equality (where, in this case, all the pixels have the same value). We obtain values between 0.14 and 0.30 with a mean value of 0.23 (see Table~\ref{tab:kin_sigma}). This suggests a fairly homogeneous distribution of the values of the velocity dispersion and it is significantly lower than the values obtained doing the same exercise with the [\oii] pseudo-narrow bands, that map the star-forming clumps, where we measure Gini coefficients between 0.82 and 0.90. The relatively homogeneous velocity dispersion maps, compared with clumpy [\oii] and UV morphologies, suggest that the star-forming clumps do not strongly disturb the local velocity dispersion, i.e., the clumps do not seem to locally increase turbulence. We caution however that we did this exercise based on individual pixel values, not deriving velocity dispersions per clump. Recent work analysing the DYNAMO sample by \citealt{Olivia2017} found statistically significant structure in 6 of these local analogue galaxies with increased velocity dispersion possibly associated with clumps. These structures generally show an increase of the order of $\sim$20\kms. This is very close to the precision we can reach with our data, and it is unlikely that we would detect this, even if this structure were present. 

\subsection{Turbulence and star formation}
\label{subsec:sigma_sfr}

As stated in the introduction, the reasons why high redshift discs are more clumpy and turbulent, and how these two properties relate to each other, are still under discussion. While some works point to the higher gas fractions of these younger discs as the reason for their clumpy morphologies and turbulence (e.g. \citealt{Genzel2011,Livermore2015}), others found a correlation between SFR and higher turbulence (e.g. \citealt{Green2010,Swinbank2009}). We investigate this in the two lowest redshift galaxies of this sample, where the resolved extinction and extinction corrected SFR maps can be robustly derived. We will present and discuss the metallicity maps of these galaxies in a separate paper (Patr\'icio et al., in prep.) and focus here on the SFR density. 

To produce the star formation density ($\Sigma_{\rm SFR}$) dust corrected maps, we first subtract the continuum from the observed cube, since this could bias the measurement of the Balmer lines. This is done by fitting the local continuum with {\sc pPXF}. The data were spatially binned using H$\gamma$ pseudo-narrow bands, which guarantees that both H$\gamma$ and H$\beta$ can be reliably measured in each bin. The line fluxes are fit in the continuum subtracted cube using {\sc camel}, as done for the observed velocity maps. We calculate the extinction and SFR from the emission lines in each bin as done in the integrated spectrum (see Section~\ref{subsec:integrated_properties}), with the exception that only the $[\oiii]\lambda\,5007$/H$\beta$,  $[\oiii]\lambda\,5007$/$[\oii]\lambda\,3727$, $[\oiii]\lambda\,5007$/$[\oiii]\lambda\,4959$ and H$\gamma$/H$\beta$ ratios were used and that the parameter space explored this time was smaller (extinction between 0 and 2.5) based on the results obtained in the integrated spectra. The resulting maps can be seen in Fig.~\ref{fig:resolved_sfr}. Since velocity dispersion and SFR density were derived with a different binning, in the following analysis we only consider the binning in $\Sigma_{SFR}$ and obtain the corresponding velocity dispersion by binning the intrinsic velocity dispersion maps to the same grid.

\begin{figure}
\includegraphics[width=0.50\textwidth]{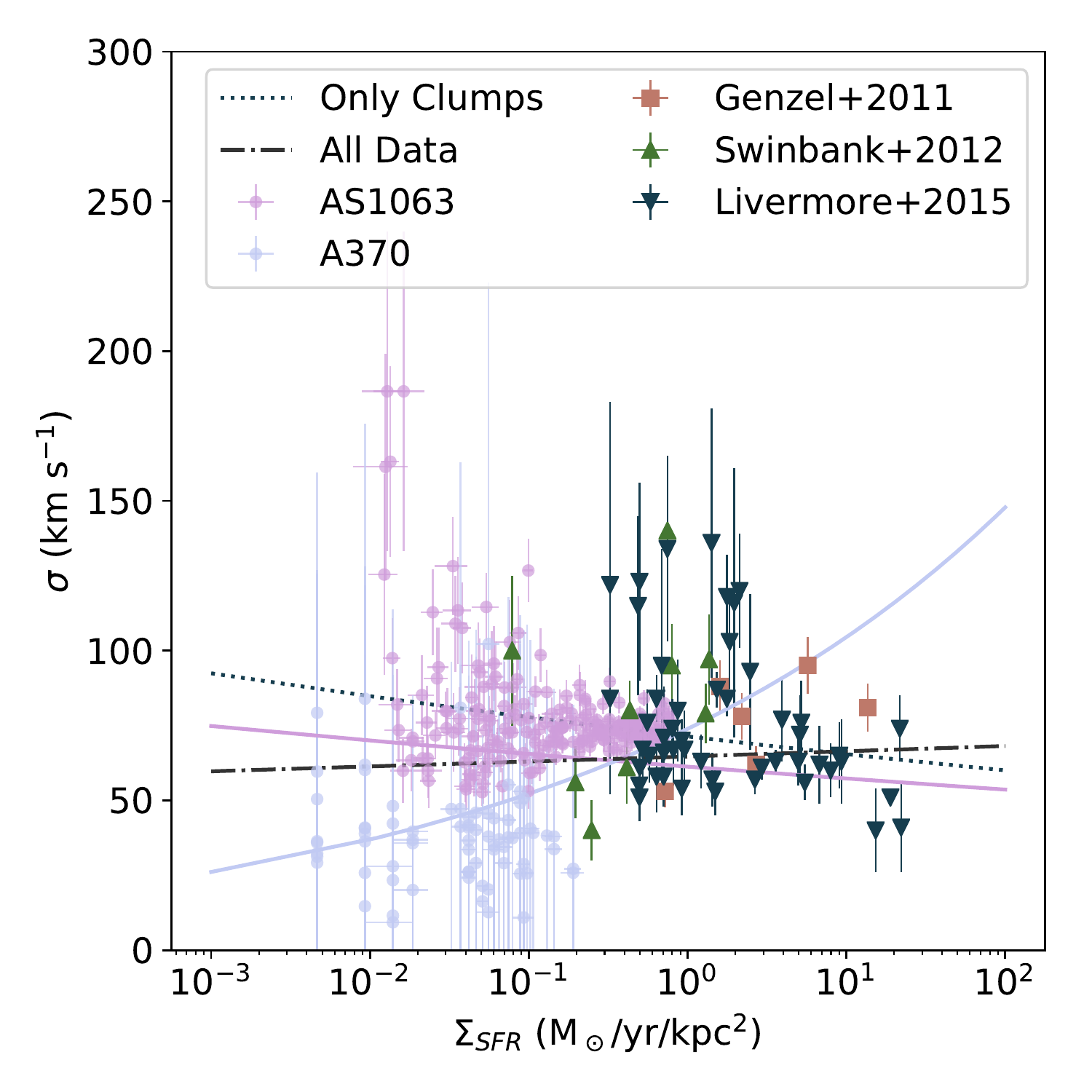}
\caption{Velocity dispersion vs. SFR density for several sub-galactic samples: in purple and blue circles, the results from AS1063 and A370, per Voronoi bin; in brown squares, the \citealt{Genzel2011} giant star-forming clumps at $z\sim$2; in green triangles the sub-kiloparsec clumps at $1<z<4$ of \citealt{Livermore2015} and in dark triangles the clumps from \citealt{Swinbank2012a}. The lines show the results of the fit. Purple and blue full lines are fits to AS1063-arc and A370-sys1, respectively, dotted line the fit only to the \citealt{Livermore2015}, \citealt{Swinbank2012a} and \citealt{Genzel2011}, and dashed dotted line fit to all data.}
\label{fig:sigma_clumps}
\end{figure}

There are different results in the literature on the relation between star-formation rate density and velocity dispersion. While \citealt{Genzel2011} found little correlation between the two in their sample of massive giant clumps at $z\sim$2, \citealt{Swinbank2012a}, using a sample of galaxies observed with Adaptive Optics, found it was well described by the relation $\sigma \propto \Sigma_{\rm SFR}^{1/n} + k$. Since clump selection depends heavily on the criteria and data used (e.g. \citealt{Dessauges-Zavadsky2017}) we use the full 2D information of these two galaxies and plot in Fig.~\ref{fig:sigma_clumps}, for each Voronoi bin, the star formation density and the respective velocity dispersion. We also gather data on the velocity dispersion and star formation density of the high-$z$ clumps available in the literature: 5 giant star forming clumps at $z\sim2$ from \citealt{Genzel2011}, from the $z\sim1$ lensed galaxies clumps from \citealt{Livermore2015} and from the clumps of nine $0.8<z<2.23$ galaxies from the HiZELS survey \citep{Swinbank2012a}. for the last two, the $\Sigma_{\rm SFR}$ were computed from the SFR and the radii of the clumps. 

We first investigate if these two quantities, $\Sigma_{\rm SFR}$ and $\sigma$ are correlated by computing the Spearman rank correlation ($\rho$) between the two (see Table~\ref{tab:sigma_sfr_fit}). Since the data on the star-forming clumps gathered in the literature can be quite different from the individual bins studied here, we do this analysis separately. 

The full data set on properties of clumps from \citealt{Genzel2011}, \citealt{Swinbank2012a} and \citealt{Livermore2015} shows very little correlation between $\Sigma_{\rm SFR}$ and $\sigma$ , with $\rho$=-0.12. Taking only the data of AS1063-arc we find a Spearman correlation of $-0.21$ and for A370-sys1 a correlation of 0.20, once more both very weak. 

We proceed to fit both these quantities with a physically motivated relation, following \citealt{Lehnert2009} and \citealt{Swinbank2012a} analysis. Assuming a marginally stable disc, with a Toomre parameter\footnote{$Q=\sigma\kappa / \pi G\Sigma_{\textrm{gas}}$, where $\sigma$ is the velocity dispersion; $\kappa$ the epicyclic frequency; $G$ the gravitational constant and $\Sigma_{\textrm{gas}}$ the gas surface density)} equal to unity, and the Kennicutt-Schmidt relation, we would expect that the velocity dispersion and the star formation density are related by:

\begin{equation}
\label{eq:sigma_sfr}
\sigma_0 = \epsilon \Sigma_{\rm SFR} ^{\gamma}
\end{equation}

Also here we analyse the data separately, fitting data only from individual star-forming clumps from the literature, only from AS1063-arc or A370-sys1, and finally all of it. The fit results and their reduced $\chi^2$ are listed in Table~\ref{tab:sigma_sfr_fit}. We generally find very low $\gamma$ values, all below 0.1 except A370-sys1 (0.307$\pm$0.041), far from the $\sim$0.74 found by \citealt{Swinbank2012a} and implied by the local Kennicutt-Schmidt law. \citealt{Lehnert2009} also present pixel-per-pixel data on 11 $z\sim$2 star-forming galaxies observed with AO and find that $\gamma$=0.33 is a good match to their data, which is in agreement with what we see for A370-sys1, but not to AS1063-arc or the individual clumps. 

\begin{table}
\caption{Relation between $\sigma_0$ and $\Sigma_{\rm SFR}$ for different subsets of data in Fig.~\ref{fig:sigma_clumps}: \textit{Only Clumps} for individual clumps from \citealt{Genzel2011}, \citealt{Swinbank2012a} and \citealt{Livermore2015}; \textit{AS1063-arc} and \textit{A370-sys1} for the bin-per-bin values derived here and \textit{All Data} for all the above. We list the Spearman rank correlation ($\rho$) and the $\epsilon$ and $\gamma$ values derived from the fit and the reduced $\chi^2$ of the fit. We also report the fit values from \citealt{Genzel2011}. Overall, $\sigma_0$ and $\Sigma_{\rm SFR}$ are weakly correlated.}
\label{tab:sigma_sfr_fit}
\centering
\tabcolsep=0.13cm
\begin{tabular}{|lcccc|} 
\hline
Data			& $\rho$	&$\epsilon$	&	$\gamma$ 		& $\chi^2/dof$ \\
\hline
\hline
\citealt{Genzel2011}	& -- 		&66 		&  0.1$\pm$0.04 	& -- \\
Only Clumps		& $-0.12$		&68$\pm$2	& $-0.044\pm0.023$ 	& 2.83	\\
AS1063-arc 	    & $-0.21$		&61$\pm$1	&  0.004$\pm$0.012 	& 6.64	\\
A370-sys1	    & 0.18		&106$\pm$15	&  0.314$\pm$0.042	& 0.20	\\
All Data		& 0.24		&62$\pm$1	&  0.006$\pm$0.008 	& 4.56  \\
\hline
\end{tabular}
\end{table}

\begin{figure*}
\includegraphics[width=\textwidth]{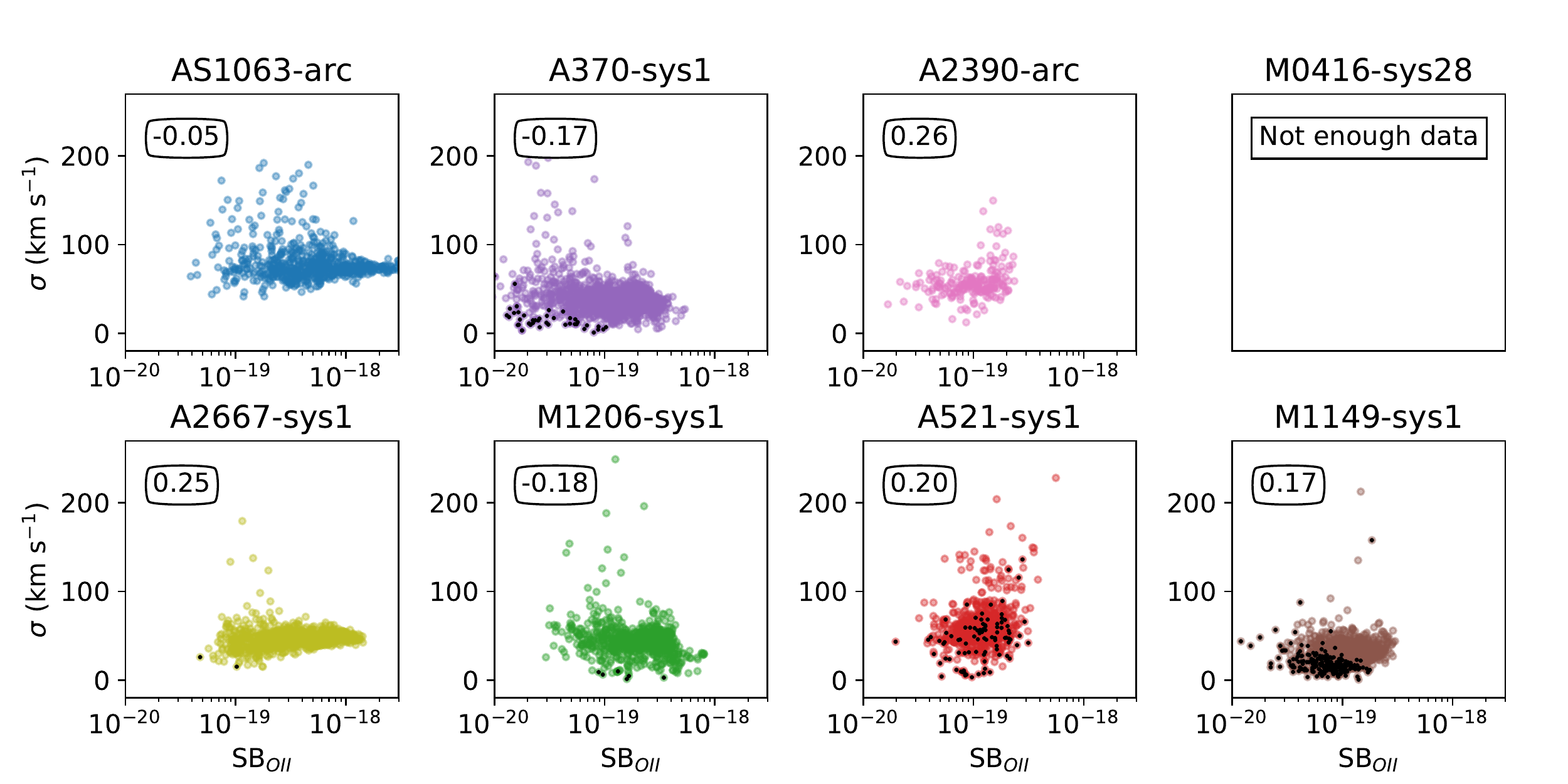}
\caption{[\oii] surface brightness and velocity dispersion for the full sample except M0416-sys28 where not enough data was available. The Spearman coefficients of these two quantities are plot in the upper left corner of each panel. Unreliable points, (in black circles in the plots, see text for details) were not included in the calculation of the Spearman coefficients. There is no correlation between [\oii] surface brightness and velocity dispersion in the sample.}
\label{fig:oii_vs_sigma}
\end{figure*}

Because we made use of fully resolved (as opposed to clump) data of highly magnified objects, we reach lower star formation rate surface densities. Indeed, there is only a small overlap in $\Sigma_{\rm SFR}$ between AS1063-arc and A370-sys1 when compared with the clump sample gathered from the literature. Other resolved studies, such as \citealt{Lehnert2009} and \citealt{Swinbank2009}, generally reach star formation rate surface densities as low as 0.1 $M_{\odot}$\,yr$^{-1}$\,kpc$^{-2}$, while much of the our data sits below this limit (although see \citealt{Cresci2009} for other $\Sigma_{\rm SFR}<0.1$ $M_{\odot}$\,yr$^{-1}$\,kpc$^{-2}$ measurements of high-$z$ galaxies). However, for $\Sigma_{\rm SFR}<0.1$ $M_{\odot}$\,yr$^{-1}$\,kpc$^{-2}$, velocity dispersions are expected to almost not vary with star formation surface density, even in a stellar-feedback driven turbulence scenario \citep{Krumholz2017}, so it is not surprising not to find a correlation at low star-formation rates. We can nevertheless look at the expected velocity dispersions for different scenarios for the measured stellar formation rate densities. For $\Sigma_{\rm SFR}$ between 0.02 and 1 $M_{\odot}$\,yr$^{-1}$\,kpc$^{-2}$, where our data is more reliable, we find values of $\sigma$ of 20 to 100 \kms. Comparing with theoretical predictions from \citealt{Krumholz2017} (their figure 4) we see that these high velocity dispersions for low star formation density are only reached for models that include transport (i.e. gravitational instability) driven turbulence. Our data best compares with the transport+feedback and no-feedback predictions for local dwarf galaxies. These predictions are however done for integrated properties of the galaxies, and not resolved as presented in this work, which makes a direct comparison difficult.

We try to extend this study beyond these two galaxies, where we can correct each bin for extinction, to the other galaxies of the sample, but in a more simplistic way. Since we cannot derive resolved extinction maps for the remaining 6 objects, we test only if there is a correlation between the observed [\oii] surface brightness (that probes star formation and is unaffected by magnification since gravitational lensing conserves surface brightness) and the velocity dispersion. A similar analysis, using H$\alpha$ surface brightnesses, was presented in \citealt{Lehnert2009} that found a positive correlation between the two. We plot these two quantities and the Spearman coefficients in Fig.~\ref{fig:oii_vs_sigma}. To make sure that the Spearman coefficients were not biased by unreliable measurements we have removed data points for which i) formal errors were 0 \kms; ii) in which the velocity dispersion reached the maximum allowed value of 250 \kms; or iii) that have S/N lower than 2. These unreliable values are plotted in black squares in Fig.~\ref{fig:oii_vs_sigma}.  

This crude approach has limitations. Comparing the $\rho$ coefficients obtained for AS1063-arc and A370 using both methods, we see they they are extremely different. However, in neither analysis do we find a strong correlation between star-formation rates (or [\oii] surface brightness) and turbulence. All the 7 considered galaxies have $\rho\leq0.26$, denoting a very faint dependence between these two quantities (M0416-sys28 does not have enough data to perform this exercise). Bearing in mind the limitations of this test, we conclude that there is no strong evidence of local velocity dispersions being sustained by (low) star formation, which is corroborated by the more careful analysis of AS1063-arc and A370-sys1, and is in agreement with previous results from \citealt{Genzel2011} and \citealt{Livermore2015}, but in contrast with what was found by \citealt{Lehnert2009}, \citealt{Swinbank2012a} and \citealt{Lehnert2013}.


\subsection{Energy Budget}
\label{sec:energy}

In the previous section we argued that in this sample we do not see strong evidence that local SFR and velocity dispersion are correlated. There might be natural explanations for this: the sample probes low star formation rates, where it is not possible to distinguish if the turbulence is mostly driven by star formation. Another possible test to investigate the origin of the higher turbulence of high-$z$ discs is to study the energy necessary to sustain those high velocity dispersions and where it could come from. Here we use \citep{Krumholz2017} equations 1 and 5 to derive the energy associated with turbulence in a disc and star formation, respectively. We reproduce both expressions here:

\begin{equation}
\label{eq:energy_turb}
\left(\frac{dE}{dA}\right)_{\mathrm{turb}} = 3.1 \times 10^9 \frac{\Sigma_g}{[10\,\textrm{M}_{\odot}\,\textrm{pc}^{-2}]} \left(\frac{\sigma}{[10\,\textrm{km}\,\textrm{s}^{-1}]}\right)^2
\end{equation}

\begin{equation}
\label{eq:energy_sf}
\left(\frac{dE}{dA}\right)_{\textrm{SF}} = 3.1 \times 10^9 \frac{
\frac{\Sigma_g}{[10\,\textrm{M}_{\odot}\,\textrm{pc}^{-2}]}\,\frac{\sigma}{[10\,\textrm{km}\,\textrm{s}^{-1}]}\,\frac{r}{[10\,\textrm{kpc}]}\,}
{\left(\frac{V}{[200\,\textrm{km}\,\textrm{s}^{-1}]}\right) }
\end{equation}

where $\left(\frac{dE}{dA}\right)_{turb}$ is the energy per unit area in [{\rm erg}\,{\rm cm}$^{-2}$] associated with turbulence in a disc and $\left(\frac{dE}{dA}\right)_{\rm SF}$ the energy associated with star formation; $\Sigma_g$ is the gas surface density; $\sigma$ the velocity dispersion; $\Sigma_{\rm SFR}$ the star formation rate surface density; $r$ the radius and $V$ the rotation velocity. (The energy associated with star formation is calculated over a galactic dynamical time $t_{\rm dyn} = r\,V^{-1}$).

We calculate both these energies within the photometric aperture (to which the total spectra were normalised), making use of the gas densities from Table~\ref{tab:KS_law_gas}. The velocity in expression \ref{eq:energy_sf} was calculated for the photometric aperture radius using the best model and parameters for each galaxy. We warn that $\Sigma_g$ and SFR are not independent, since $\Sigma_g$ was derived using the inverse Kennicutt-Schmidt law. We report these values in Table~\ref{tab:energy_budget}.

\begin{table}
\caption{Energy associated with turbulence $\left(\frac{dE}{dA}\right)_{\textrm{turb}}$ and star formation $\left(\frac{dE}{dA}\right)_{\textrm{SF}}$ following \citealt{Krumholz2017} equations 1 and 4. The gas densities used are listed in Table~\ref{tab:KS_law_gas}. The radius corresponds to the radius of the circularised areas used to calculate the gas densities. The velocities ($V$) at that radius were calculated for the best kinematic model of each galaxy. M1149-sys1 was not analysed, since the SFR cannot be reliably calculated from MUSE data.}
\label{tab:energy_budget}
\centering
\tabcolsep=0.07cm
\begin{tabular}{|lccccc|} 
\hline
Object &   radius &    V    &   $\left(\frac{dE}{dA}\right)_{\textrm{SF}}$   &     $\left(\frac{dE}{dA}\right)_{\textrm{turb}}$ &  $\frac{E_{\textrm{turb}}}{E_{\textrm{SF}}}$  \\
	&	(kpc)  & (\kms)	& (erg\,cm$^{-2}$) & (erg\,cm$^{-2}$) &  \\
\hline\hline
AS1063-arc & 17 & 280 & 6.83e+12 &  2.48e+12 & 0.36\\
A370-sys1 & 11 & 162 & 6.37e+11 &  1.69e+11 & 0.26 \\
A2390-arc & 7 & 217 & 2.00e+12 &  8.40e+11 & 0.42 \\
M0416-sys28 & 15 & 105 & 1.76e+11 &  6.67e+09 & 0.04 \\
A2667-sys1 & 8 & 159 & 3.25e+12 &  6.97e+11 & 0.21 \\
M1206-sys1 & 11 & 205 & 7.94e+12 &  1.22e+12 & 0.15 \\
A521-sys1 & 30 & 55 & 6.00e+12 &  3.32e+11 & 0.06  \\
\hline
\end{tabular}
\end{table}

We find that in our sample, star formation can provide the necessary energy to sustain turbulence, requiring only 6\% to 46\% of the available energy. This is somehow at odds with what is found by \citealt{Krumholz2017} in their disc evolution model where at $z>0.5$ disc turbulence is dominated by transport-driven mechanisms (e.g. from gas accretion) and not stellar-feedback, especially for more massive galaxies, although both processes co-exist. Such a difference can come from the approximations made to calculate these energies, particularly in estimating the gas surface density.



\section{Summary and Conclusions}
\label{sec:conclusion} 

We discussed the physical properties of seven clumpy discs at $0.6<z<1.5$, that 
are highly magnified by strong lensing and can therefore be resolved to sub-kiloparsec scales. From high-quality MUSE data (with $<$0.7'' seeing) and abundant \emph{HST} archival data we derive the following:

\begin{itemize}
\item We derive stellar masses, global metallicities and extinctions from multi-band \emph{HST} photometry and multiple line diagnostics for the full sample and for the first time for 3 of the 8 objects (AS1063-arc, A370-sys1 and A521-sys1). Comparing these objects with the main-sequence of star-forming galaxies, we confirm that these are typical star-forming galaxies for the considered redshifts (Fig.~\ref{fig:main_sequence}).
\item The observed velocity fields, derived from the MUSE data cubes, all reveal smooth velocity gradients with a velocity dispersion peak at the centre of rotation, typical of rotating discs. These are rotation-dominated galaxies with V/$\sigma$ between $\sim$2 and $\sim$10, also typical of what was found by IFU surveys (Fig.\ref{fig:local_sample}). We do not see any evidence of a drop in the velocity curve within the sample, even for galaxies where 2 effective radii are reached (Fig.~\ref{fig:1d_vel}).
\item The observed velocity maps were fit with three different kinematic models (arctangent, isothermal sphere and exponential disc model), with a novel method that allows to fit several multiple images at the same time, accounting for lensing distortions. The arctangent model was a better fit to 4 galaxies of the sample, the exponential disc model to 3 and the isothermal sphere model to the remaining galaxy. However, all models deliver very similar results, particularly in the central part of the galaxies, which seems to indicate that the velocity is not probed far enough to reliably infer which model is the best (and therefore to verify if the velocity curves decay at larger radii). 
\item The intrinsic velocity dispersion maps (that we correct for beam-smearing effect) do not display any strong 2D structure, revealing fairly isotropic 2D maps with a mean Gini coefficient of 0.23 (compared with 0.87 for the [\oii] maps, Table~\ref{tab:kin_sigma}).
\item There is no strong evidence of increased (or decreased) velocity dispersion at the positions of the star-forming clumps. We also do not find a strong correlation between the local (resolved) SFR density and the velocity dispersion for the two galaxies where the dust attenuation corrected SFR maps could be derived in a resolved way (Fig.~\ref{fig:sigma_clumps}). Comparing the resolved results with theoretical predictions, we conclude that the high velocity dispersion found for relatively low star formation rate surface densities are more in agreement with scenarios where the turbulence is driven by gravitational instability \citep{Krumholz2017}.
\item However, from an energetic point of view, the observed star-formation is sufficient to explain the observed turbulence. 
\item We find that the star-forming clumps share the dynamical properties of the ionised gas surrounding them (rotation velocity and turbulence). In a future work, we will study the chemical composition of these galaxies and the clumps within them, to further test the in-situ clump formation hypothesis.
\end{itemize}

Finally, here we only focus on the ionised gas, traced by emission lines. The analysis of the properties of molecular gas, made from high-resolution ALMA data obtained for a number of these objects, will allow to probe the turbulence of the cold gas and to have a more complete view of the process of formation of star-forming clumps.


\section*{Acknowledgements}

We warmly thank Benjamin Johnson for helping using {\sc prospector} and updating the code to fit spectra and photometry simultaneously. We thank Jorge S\'anchez Almeida and Andrea Ferrara for insightful suggestions about clumps, turbulence and energy. We thank Mark Swinbank, Philippe Amram, Miroslava Dessauges and Mathieu Puech for useful comments on this work. We thank David Bina for reducing A2390 and A2667 MUSE data. We also thank the referee for a constructive report with useful suggestions.

VP, DC, JR, JM, DL and BC acknowledge support from the ERC starting grant 336736-CALENDS. TC, NB and BE acknowledge support from the ANR FOGHAR (ANR-13-BS05-0010-02) and the OCEVU Labex (ANR-11-LABX-0060). TC acknowledge support from and the A*MIDEX project (ANR-11-IDEX-0001-02) funded by the "Investissements d'avenir" French government program. KBS was supported by the BMBF grant 05A14BAC. JB acknowledges support by Funda{\c c}{\~a}o para a Ci{\^e}ncia e Tecnologia (FCT) through national funds (UID/FIS/04434/2013) by FEDER through COMPETE2020 (POCI-01-0145-FEDER-007672), and by FCT through Investigador FCT contract IF/01654/2014/CP1215/CT0003. JR, FB and RP also acknowledge support by the Programa de Cooperaci{\'{o}n Cient{\'{\i}}fica ECOS SUD Program C16U02. FB acknowledge support CONICYT grants Basal-CATA PFB-06/2007, 
FONDECYT Regular 1141218 and the Ministry of Economy, Development, and Tourism's Millennium Science Initiative through grant IC120009, awarded to The Millennium Institute of Astrophysics, MAS.

Based on observations made with ESO Telescopes at the La Silla Paranal Observatory under programme IDs 060.A-9345, 094.A-0115, 095.A-0181, 096.A-0710, 097.A-0269, 100.A-0249 and 294.A-5032. Also based on observations obtained with the NASA/ESA Hubble Space Telescope, retrieved from the Mikulski Archive for Space Telescopes (MAST) at the Space Telescope Science Institute (STScI). STScI is operated by the Association of Universities for Research in Astronomy, Inc. under NASA contract NAS 5-26555. Part of this work was supported by the ECOS SUD Program C16U02. This research made use of Astropy, a community-developed core Python package for Astronomy \citep{astropy} and of VizieR catalogue access tool, CDS, Strasbourg, France.




\bibliographystyle{mnras}
\bibliography{bibliography} 


\appendix

\begin{landscape}
\section{Sample Integrated Spectra}
\label{app:spectra}

Sample integrated spectra and {\sc pPXF} fit. Details about the {\sc pPXF} fit are given inside the inset box: reduced $\chi^2$ of the fit, number of templates used of a total of 448 and degree of multiplicative (\textit{mdegree}) and additive (\textit{degree}) polynomials added (see Section~\ref{subsec:emission_line_measurements} for details). M1149-sys1 is not shown since in the MUSE spectra only [\ion{O}{ii}] is seen (spectra and absorption line analysis in \citealt{Karman2016}).

\begin{figure}
\includegraphics[width=0.63\textwidth]{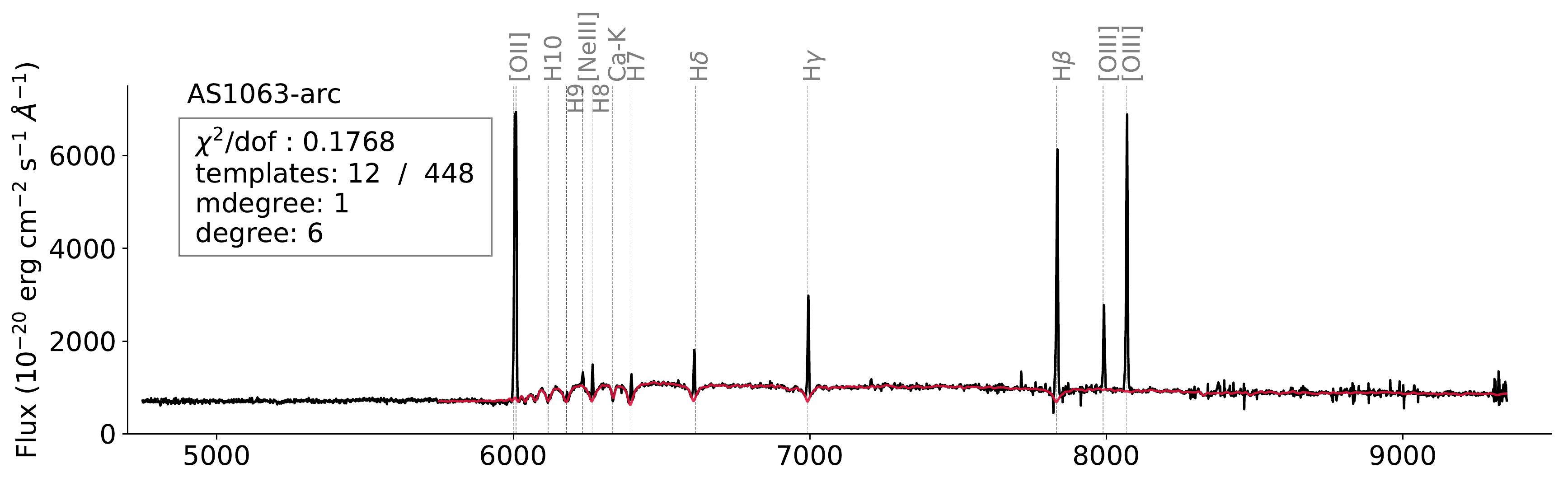}
\includegraphics[width=0.63\textwidth]{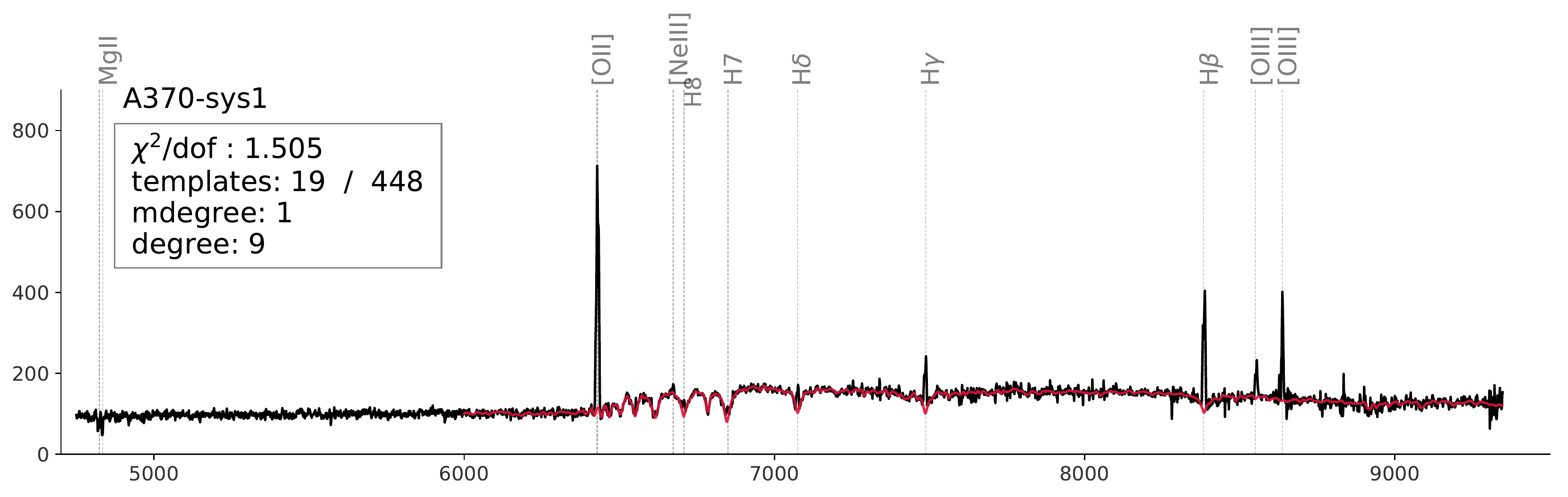}\\
\includegraphics[width=0.63\textwidth]{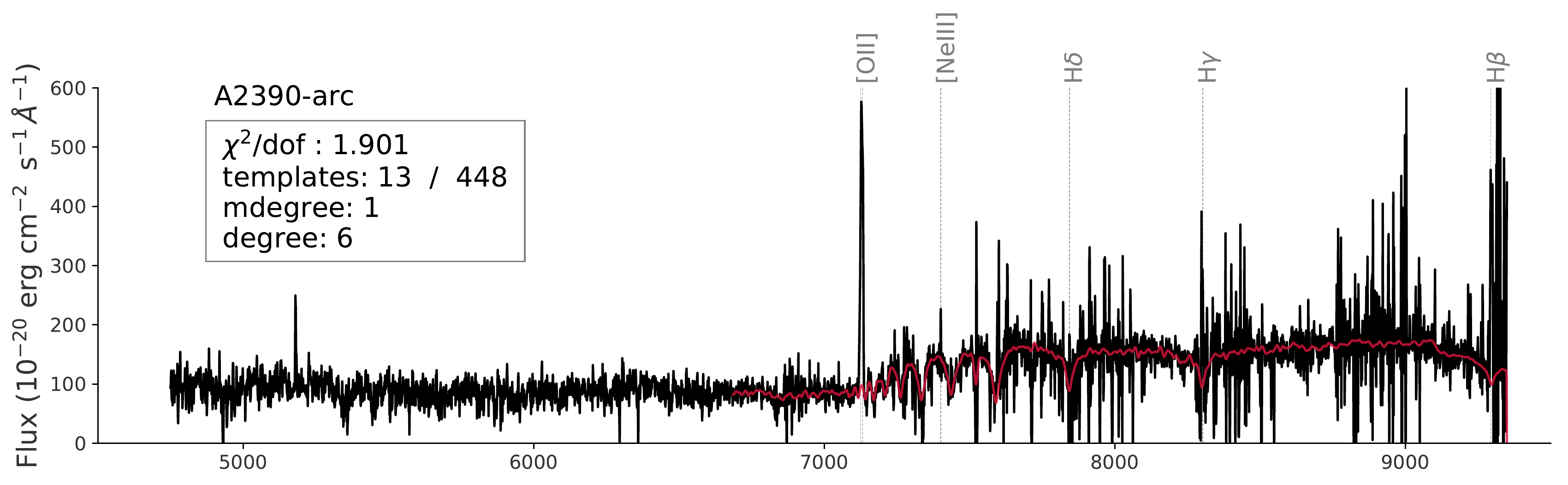}
\includegraphics[width=0.63\textwidth]{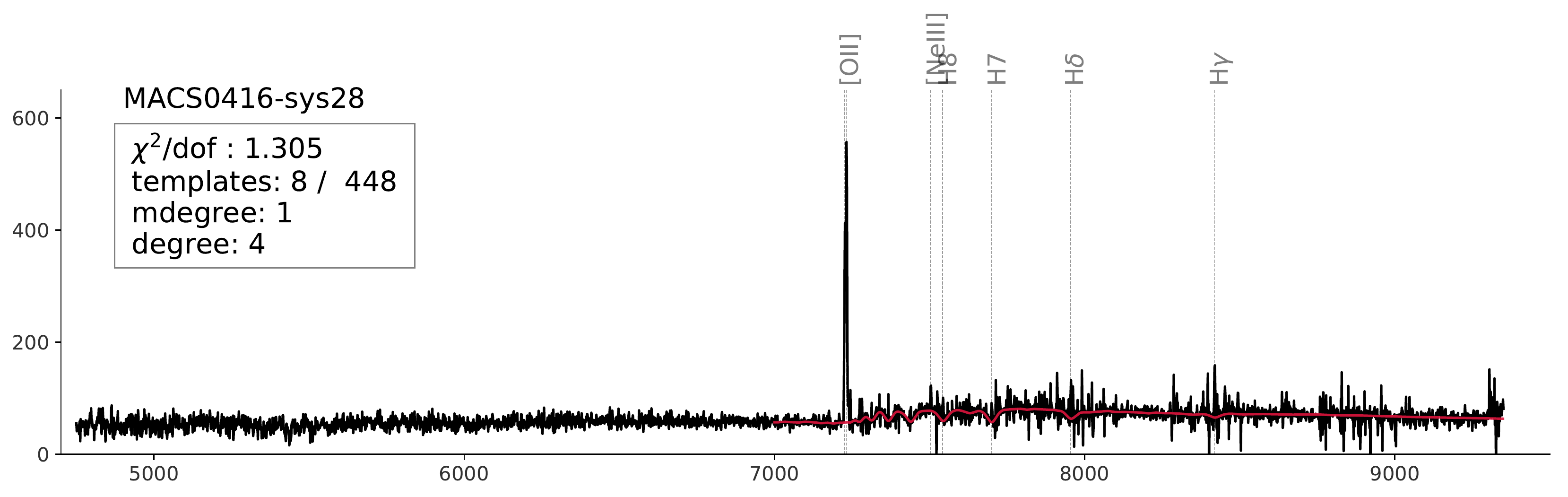}\\
\includegraphics[width=0.63\textwidth]{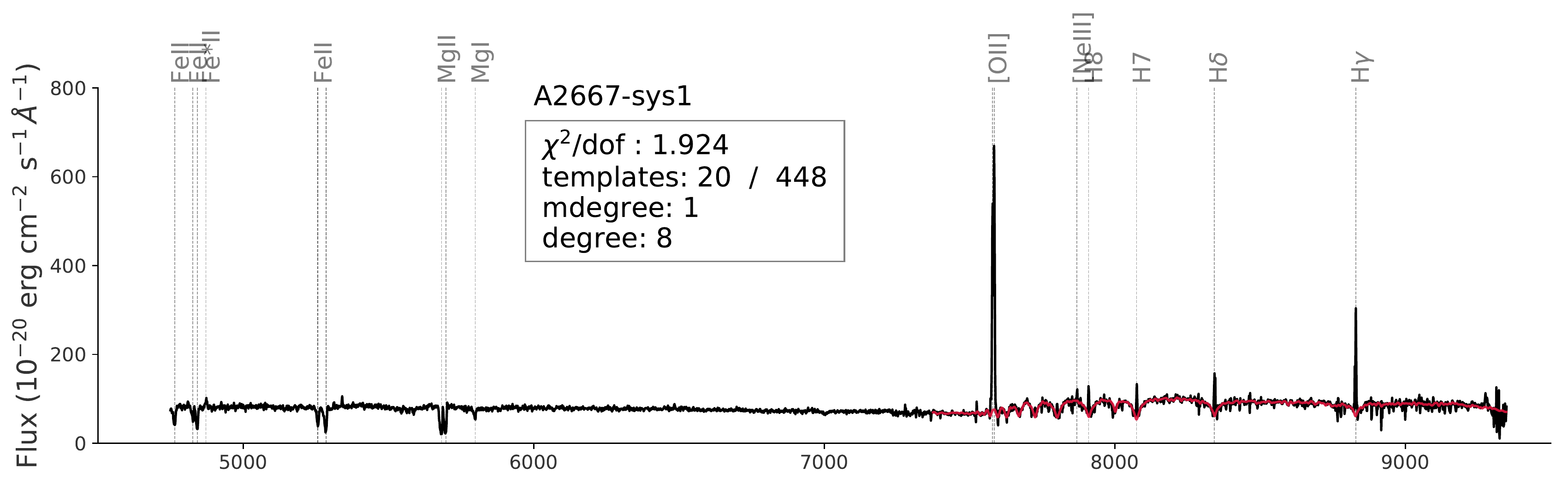}
\includegraphics[width=0.63\textwidth]{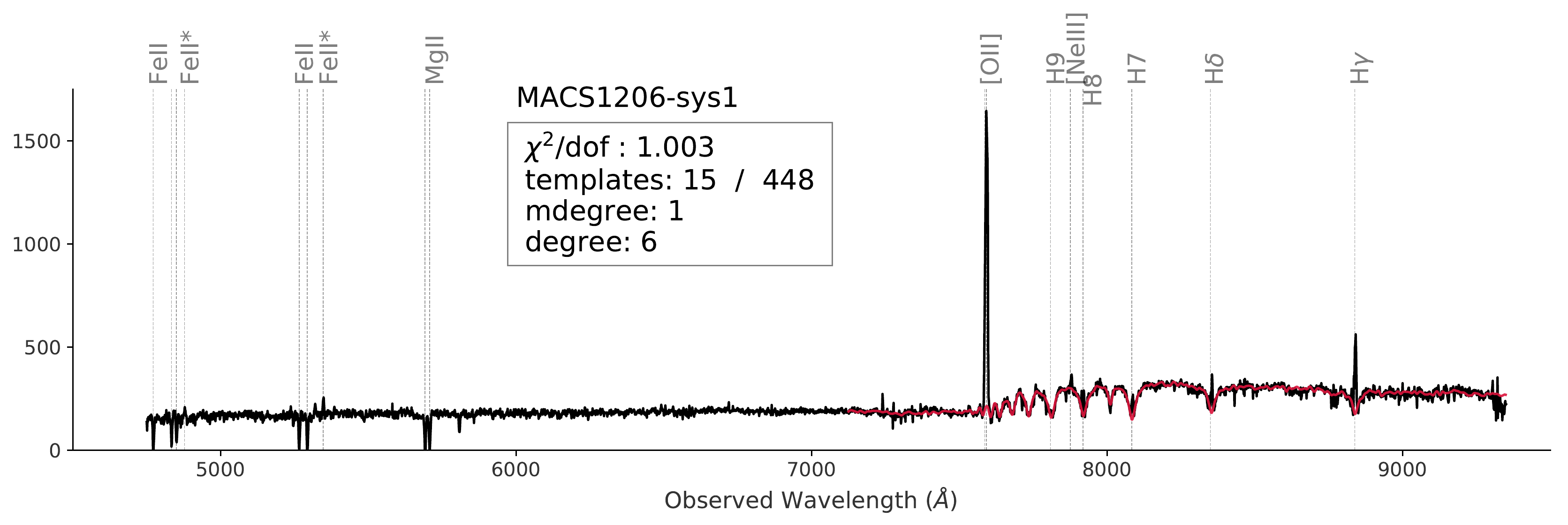}\\
\includegraphics[width=0.63\textwidth]{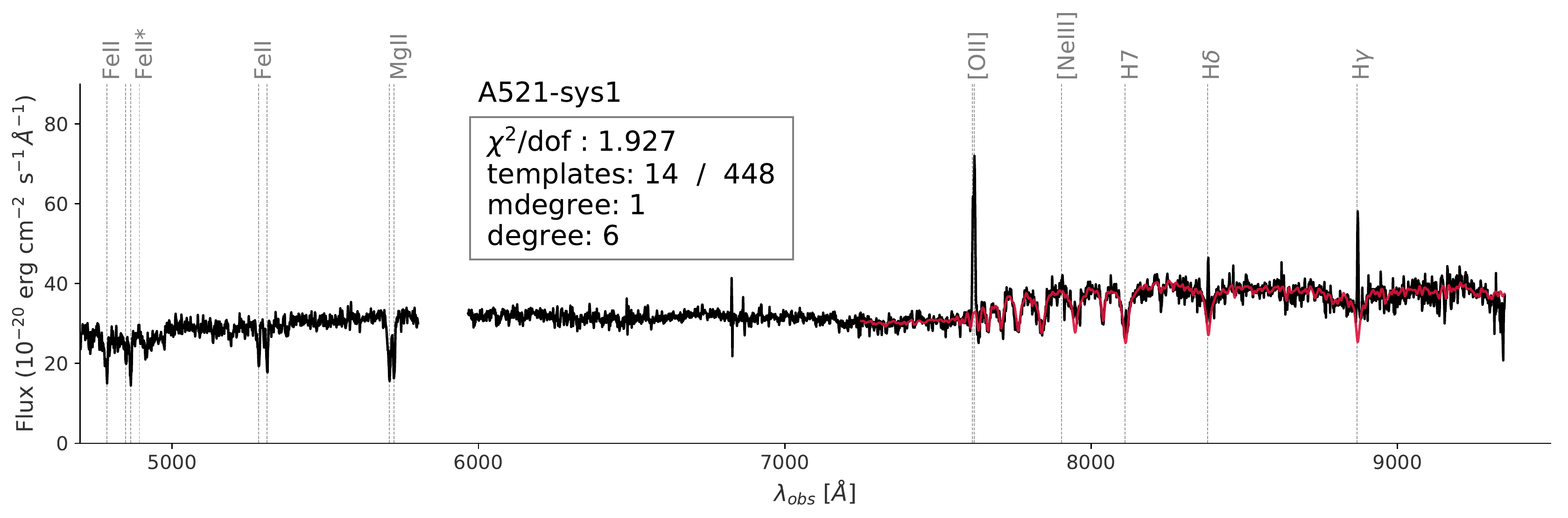}
\caption{Integrated MUSE spectra of the sample. Intrinsic flux, corrected for magnification, is shown in thick back and {\sc pPXF} fits to the continuum in red.}
\label{fig:sample_spectra}
\end{figure}
\end{landscape}

\begin{landscape}
\section{Photometry and Line Emission Fluxes}
\label{app:photometry}

Here we list the photometry derived from the publicly available \emph{HST} data (programmes listed in the Table) with respective errors and magnification errors. This data was used, together with the integrated spectra, to derive the mass and stellar population properties of this sample. We also list the observed emission lines fluxes, measured in the continuum subtracted spectra and corrected for magnification but not for attenuation, used to derive the metallicity and attenuation factors in Section~\ref{sec:mass}. 


\centering 
\begin{table}
\caption{AB magnitude of the available \emph{HST} bands for each object. All are \textit{amplification corrected} magnitudes. This factor $\mu$, calculated in the same aperture, is also listed in the table. Errors on the magnitudes include cluster member subtraction residuals and a systematic error of 0.03 mag. Magnification errors were not added to these, since they are achromatic, and are listed separately in the err$_\mu$ column. The data were gathered from several programmes. Programmes ID: (1) 14037; (2) 12458; (3) 11507; (4) 14038; (5) 10504; (6) 5352; (7) 12459; (8) 13496; (9) 10491; (10) 12069; (11) 10504; (12) 8882; (13) 13504; (14)  11312}
\label{tab:photom}
\tabcolsep=0.16cm
\begin{tabular}{|lccccccccccccccc|} 
		\hline 
Object & F435W & F475W & F606W & F625W & F775W & F814W & F850LP & F105W & F110W &F125W & F140W & F160W  & err$_\mu$ & $\mu_{\emph{HST}}$ & Ref.\\
        \hline\hline

AS1063-arc & 22.33$\pm$0.04 & -- & 21.36$\pm$0.03 & 21.14$\pm$0.03 & 20.66$\pm$0.03 & 20.58$\pm$0.03 & 20.42$\pm$0.03 & 20.11$\pm$0.03 & 20.04$\pm$0.03 & 19.99$\pm$0.03 & 19.82$\pm$0.03 & 19.64$\pm$0.03 & 0.02 & 3.97$\pm$0.06 & (1)(2)\\  

A370-sys1 & 24.46$\pm$0.05& 24.23$\pm$0.07& 23.58$\pm$0.04& 23.36$\pm$0.09 & -- & 22.64$\pm$0.03 & --& 22.16$\pm$0.03 & -- & 21.93$\pm$0.03& 21.83$\pm$0.03& 21.68$\pm$0.03 & 0.04 & 9.73$\pm$0.32 & (1)(4) \\
 
A2390-arc & -- & -- & 23.99$\pm$0.08$^*$ & -- &  -- & 22.68$\pm$0.07 & 23.71$\pm$0.56 & -- & 22.16$\pm$0.9 & 22.11$\pm$0.17 & --& 21.58$\pm$0.53 & 0.02 & 10.72$\pm$0.18 & (5)(6)\\

M0416-sys28 & 24.89$\pm$0.06& 24.78$\pm$0.05& 24.41$\pm$0.04& 24.52$\pm$0.05& 23.57$\pm$0.04& 23.41$\pm$0.04& 23.29$\pm$0.05& 23.12$\pm$0.04& 22.83$\pm$0.04& 23.0$\pm$0.04& 22.93$\pm$0.04& 22.86$\pm$0.04 & 0.28 & 5.8$\pm$1.5& (7)(8)\\

A2667-sys1 & -- & 24.21$\pm$0.04$^{**}$ & 24.04$\pm$0.04 & -- & -- & 23.25$\pm$0.04 & 22.7$\pm$0.07 & -- & 22.71$\pm$0.03 & -- & -- & 22.23$\pm$0.04 & 0.15 & 13.41$\pm$1.9 & (11)(12)\\

M1206-sys1  & 24.04$\pm$0.27 & 23.52$\pm$0.13 & 23.09$\pm$0.08 & 22.91$\pm$0.1 & 22.24$\pm$0.08 & 22.05$\pm$0.06 & 21.66$\pm$0.09 & 20.04$\pm$0.04 & 21.28$\pm$0.04 & 21.14$\pm$0.04 & 21.02$\pm$0.04 & 20.92$\pm$0.04 & 0.06 & 3.25$\pm$0.18 & (9)(10) \\ 

A521-sys1 & -- & -- & 24.74$\pm$0.07 & -- & -- & -- & -- & -- & -- & -- & -- & -- & 0.12 & 3.7$\pm$0.2 & (14)\\

M1149-sys1 & 24.23$\pm$0.03 & -- & 23.94$\pm$0.02 & -- & -- & 23.62$\pm$0.01 & -- & 23.07$\pm$0.01 & -- & 22.92$\pm$0.01 & 22.89$\pm$0.01 & 22.74$\pm$0.01 & 0.02 & 3.79$\pm$0.08 & (13)\\
\hline
\end{tabular}
* : F555W; **: F450W
\end{table}

\begin{table}
\caption[Emission line fluxes]{Emission line fluxes in units of 10$^{-18}$ erg\,cm$^{-2}$\,s$^{-1}$. These were measured in the continuum subtracted spectra and errors were derived from the MC technique described in section \ref{subsec:spec_extraction}. We note that the A2390-arc spectrum is more noisy and does not allow to accurately determine [\neiii] flux and fainter Balmer lines.}
\small
\label{tab:line_fluxes}
\tabcolsep=0.17cm
\begin{tabular}{|lcccccccccc|} 
\hline
Object      & [\oii]      &  [\oii]    & [\neiii]       & H8 & H7 & H$\delta$ & H$\gamma$ & H$\beta$ & [\oiii] & [\oiii] \\

$\lambda_{rest}$  & 3726.03 & 3728.82 & 3868.75 & 3889.05 & 3970.07 & 4101.74 & 4340.47& 4861.283 & 4958.91 & 5006.84 \\ 
\hline\hline
AS1063-arc & 288.00$\pm$8.00 & 283.00$\pm$8.00 & 16.00$\pm$6.00 & 40.00$\pm$5.00 & 33.00$\pm$4.00 & 61.00$\pm$4.00 & 126.00$\pm$4.00 & 315.00$\pm$6.00 & 98.00$\pm$7.00 & 327.00$\pm$6.00 \\
A370-sys1 & 28.29$\pm$0.54& 25.45$\pm$0.68& 1.19$\pm$0.26& 2.71$\pm$0.28& 2.44$\pm$0.25& 4.66$\pm$0.26& 9.30$\pm$0.30& 21.41$\pm$0.51& 5.73$\pm$0.27& 16.71$\pm$0.68 \\
A2390-arc & 19.19$\pm$1.78 & 30.13$\pm$1.83 & -- & -- & -- & 5.71$\pm$0.95 & 11.73$\pm$1.55 & 26.91$\pm$1.66 & -- & -- \\
A2667-sys1 & 19.58$\pm$0.14& 24.86$\pm$0.17& 0.58$\pm$0.23& 3.06$\pm$0.12& 5.57$\pm$0.39& 10.29$\pm$0.29 & -- & -- & -- & -- \\
M0416-sys28 & 14.87$\pm$0.38& 20.63$\pm$0.42& 1.85$\pm$0.31& 1.61$\pm$0.33& 1.38$\pm$0.39& 2.45$\pm$0.35& 3.91$\pm$0.38 & -- & -- & -- \\
M1206-sys1 & 66.09$\pm$2.97& 85.69$\pm$2.60& 3.95$\pm$1.45& 7.30$\pm$1.57& 7.81$\pm$0.92& 12.16$\pm$1.36& 27.38$\pm$1.14 & -- & -- & -- \\
A521-sys1 & 1.54$\pm$0.06& 1.90$\pm$0.06& 0.21$\pm$0.04& 0.44$\pm$0.05& 0.39$\pm$0.03& 0.98$\pm$0.07& 1.88$\pm$0.06 & -- & -- & --  \\
M1149-sys1 & 3.95$\pm$0.28 & 4.92$\pm$0.32 -- & -- & -- & -- & -- & -- & --& -- & --\\
\hline
\end{tabular}
\end{table}
\end{landscape}

\begin{landscape}
\section{SED fitting Results}
\label{app:prospector}

The photometry and integrated spectra of the galaxies was simultaneously fit using the SED fitting code {\sc PROSPECTOR} (see Section~\ref{sec:mass} for details). Emission lines were masked. Here we presented the comparison of the photometric and spectral data with the model predictions for both. The inset shows a zoom of the Balmer series.

\begin{figure}
\includegraphics[width=0.68\textwidth]{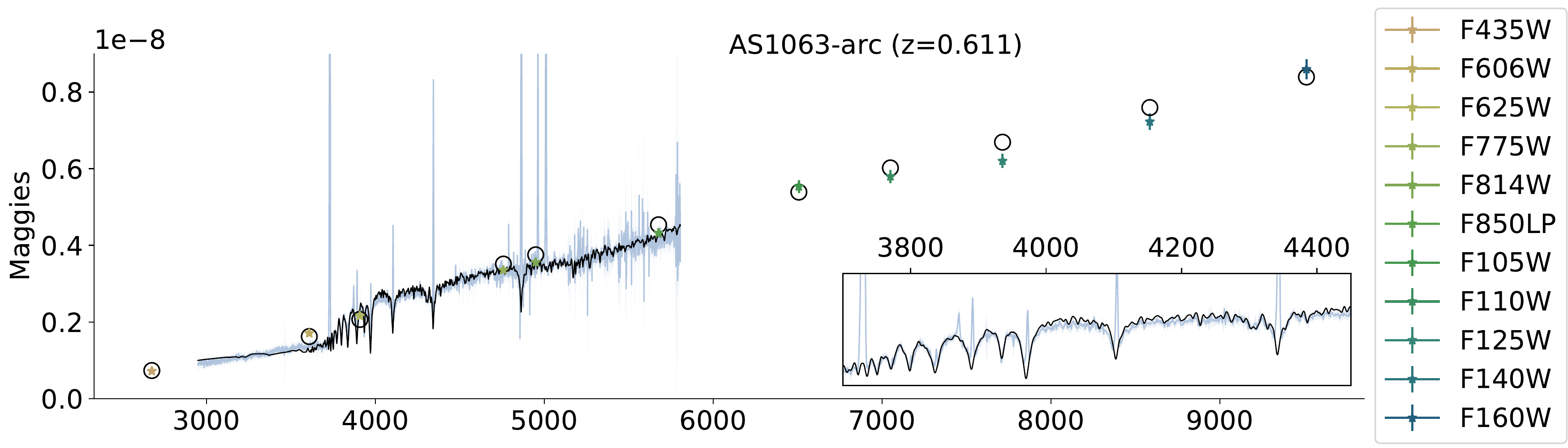}
\includegraphics[width=0.68\textwidth]{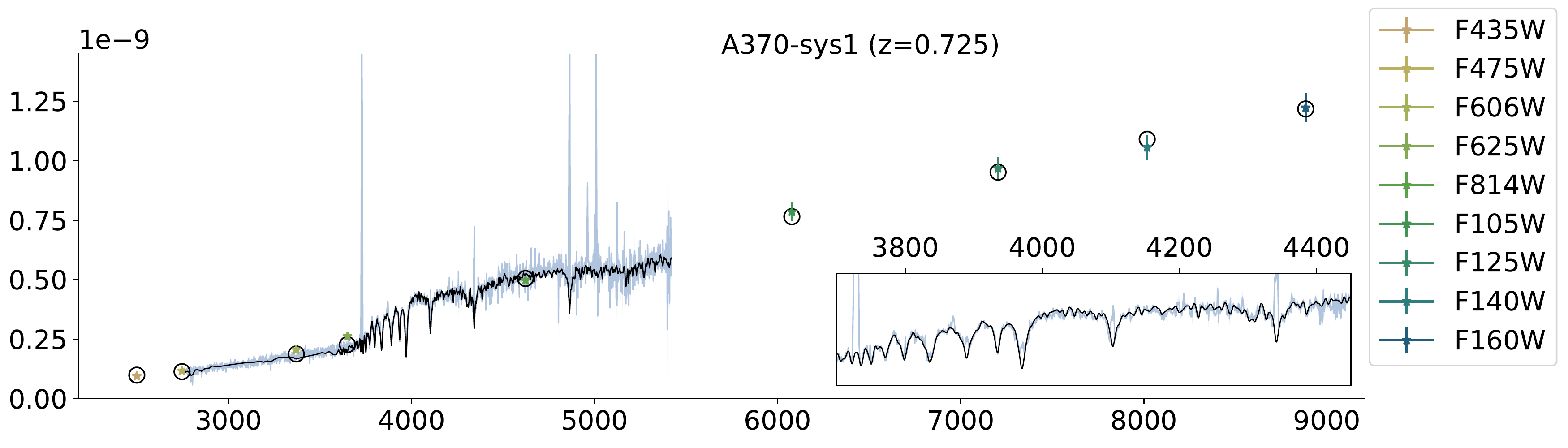}\\
\quad
\includegraphics[width=0.68\textwidth]{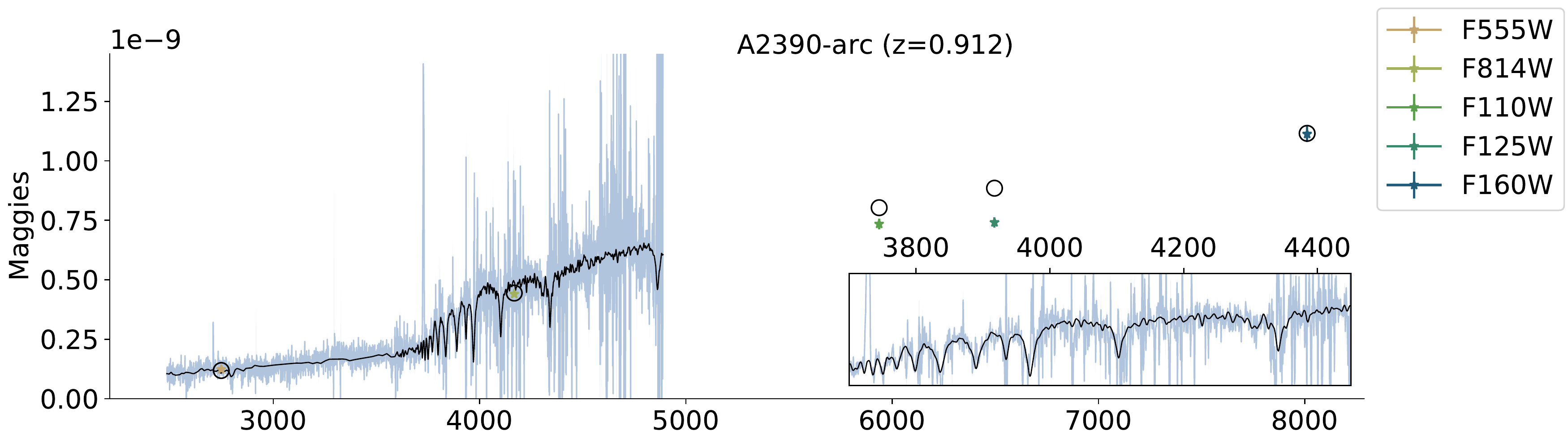}
\includegraphics[width=0.68\textwidth]{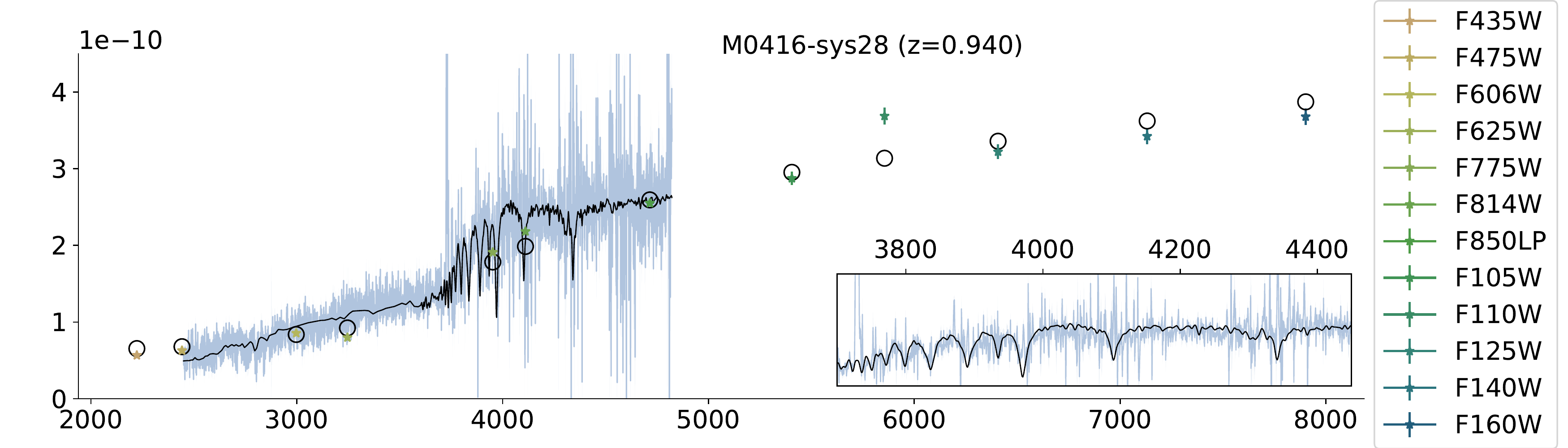}\\
\quad
\includegraphics[width=0.68\textwidth]{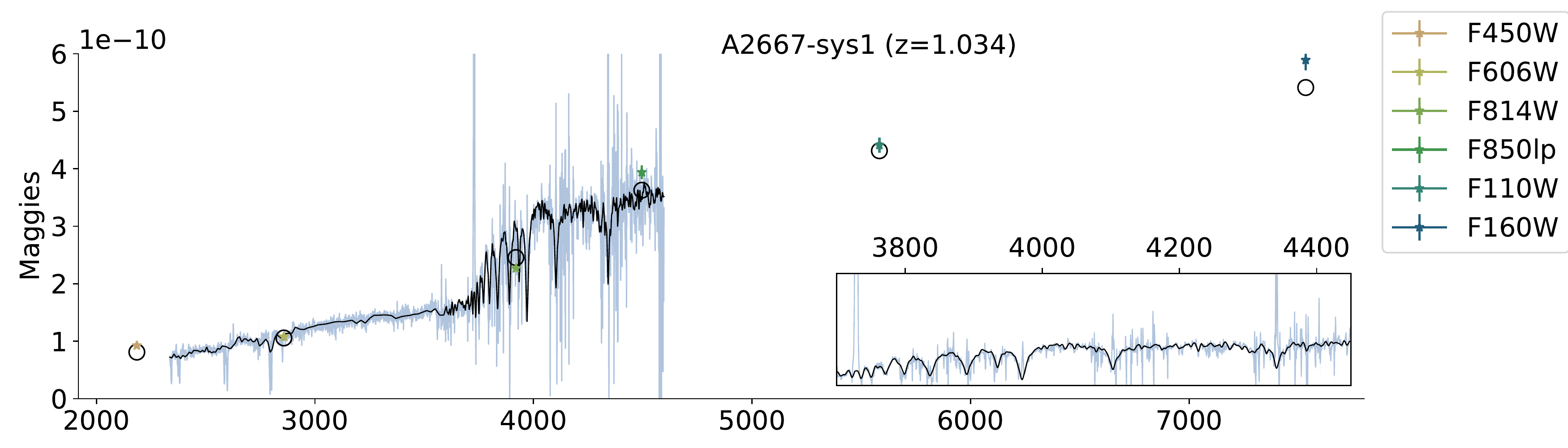}
\includegraphics[width=0.68\textwidth]{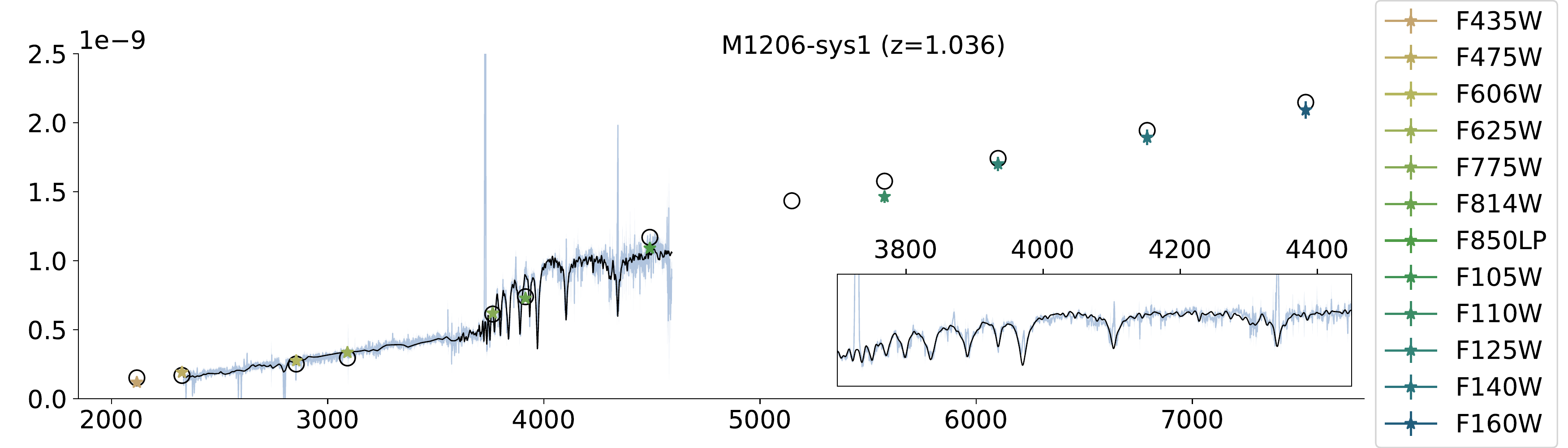}\\
\quad
\includegraphics[width=0.68\textwidth]{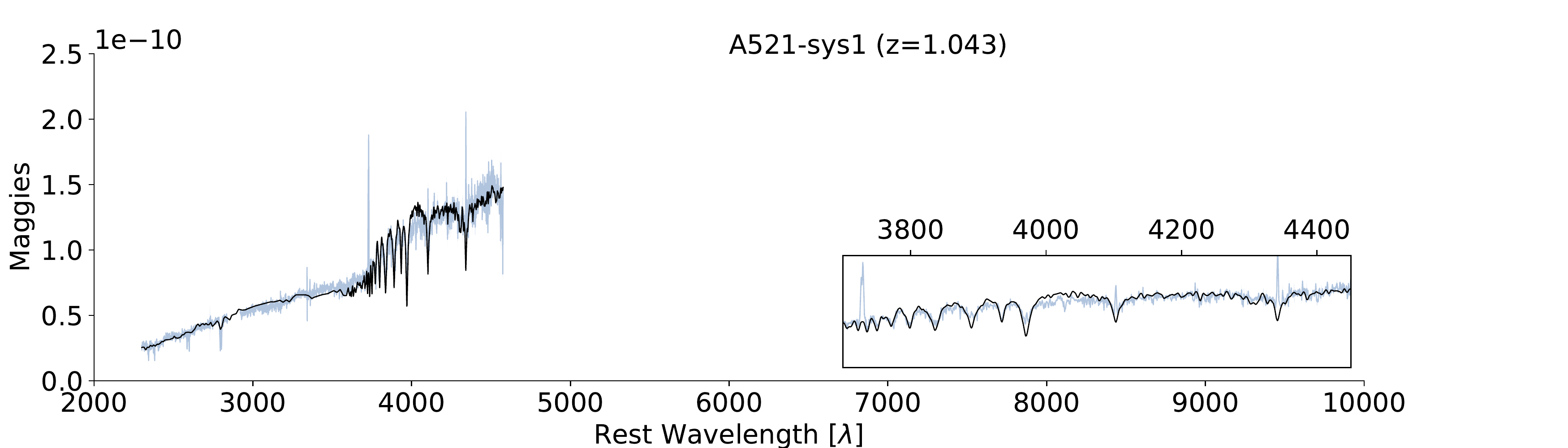}
\includegraphics[width=0.68\textwidth]{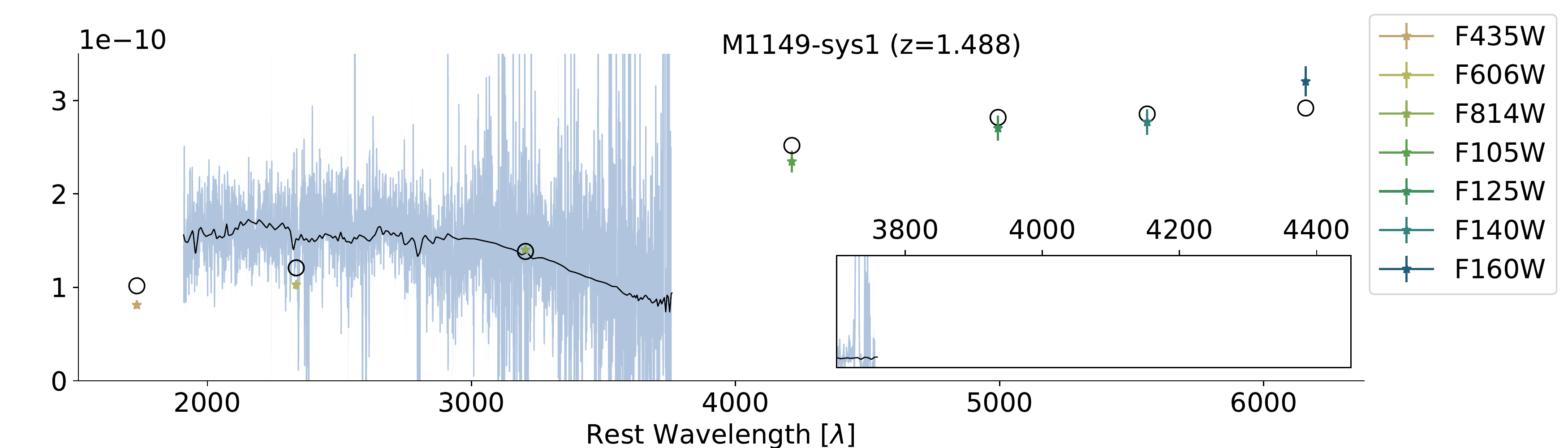}
\caption{SED Fits. Best model in black (both circles and line) and data in colour. Flux in unit is \textit{maggies}, linear flux units used in SDSS-III Collaboration (1 maggie = 3.631$\times$10$^{3}$ Jy). Wavelengths are in \AA\, and rest frame.}
\label{fig:sed_fits}
\end{figure}
\end{landscape}

\section{Metallicity and attenuation fits}
\label{app:met_and_ext}

\begin{figure}
\includegraphics[width=0.23\textwidth]{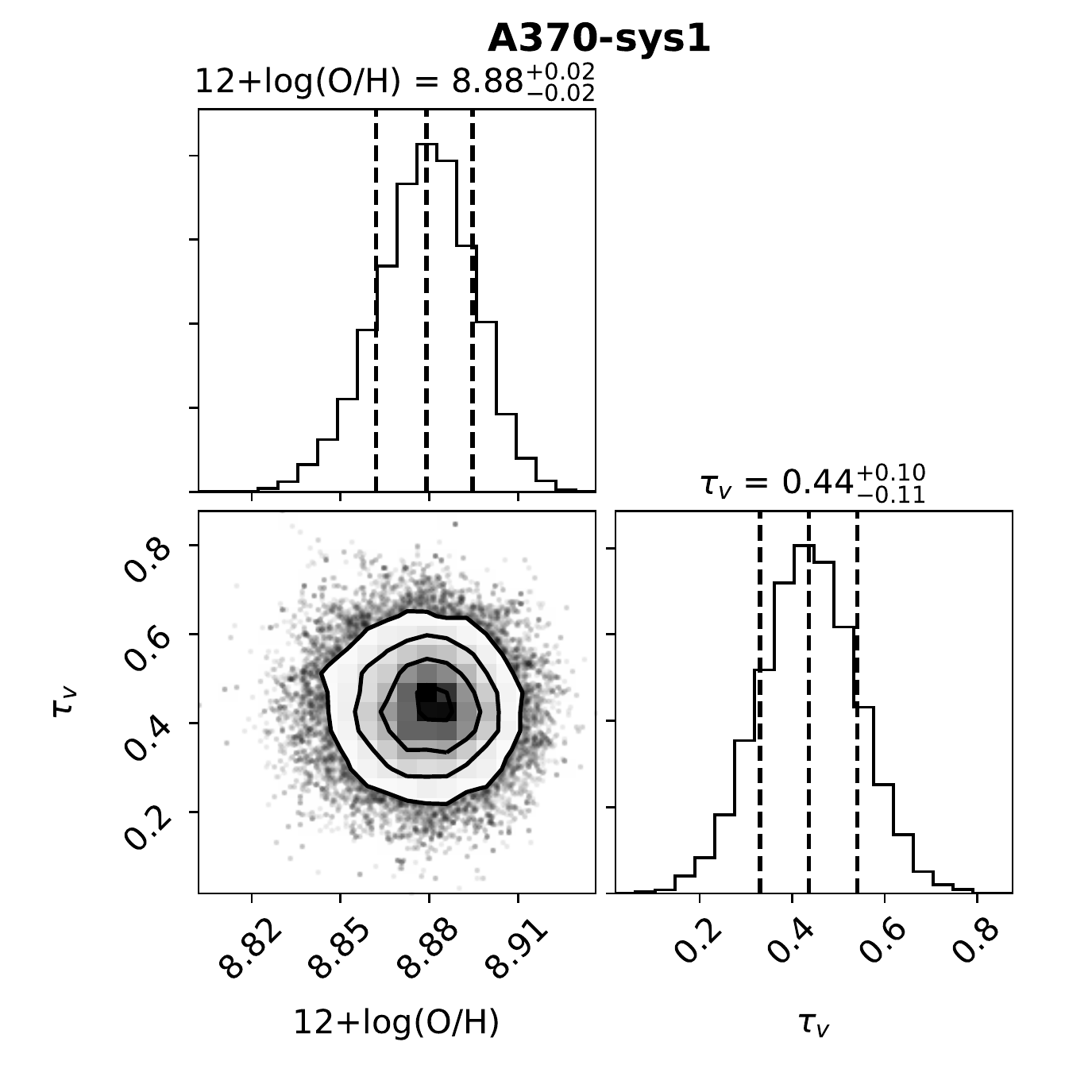}
\includegraphics[width=0.23\textwidth]{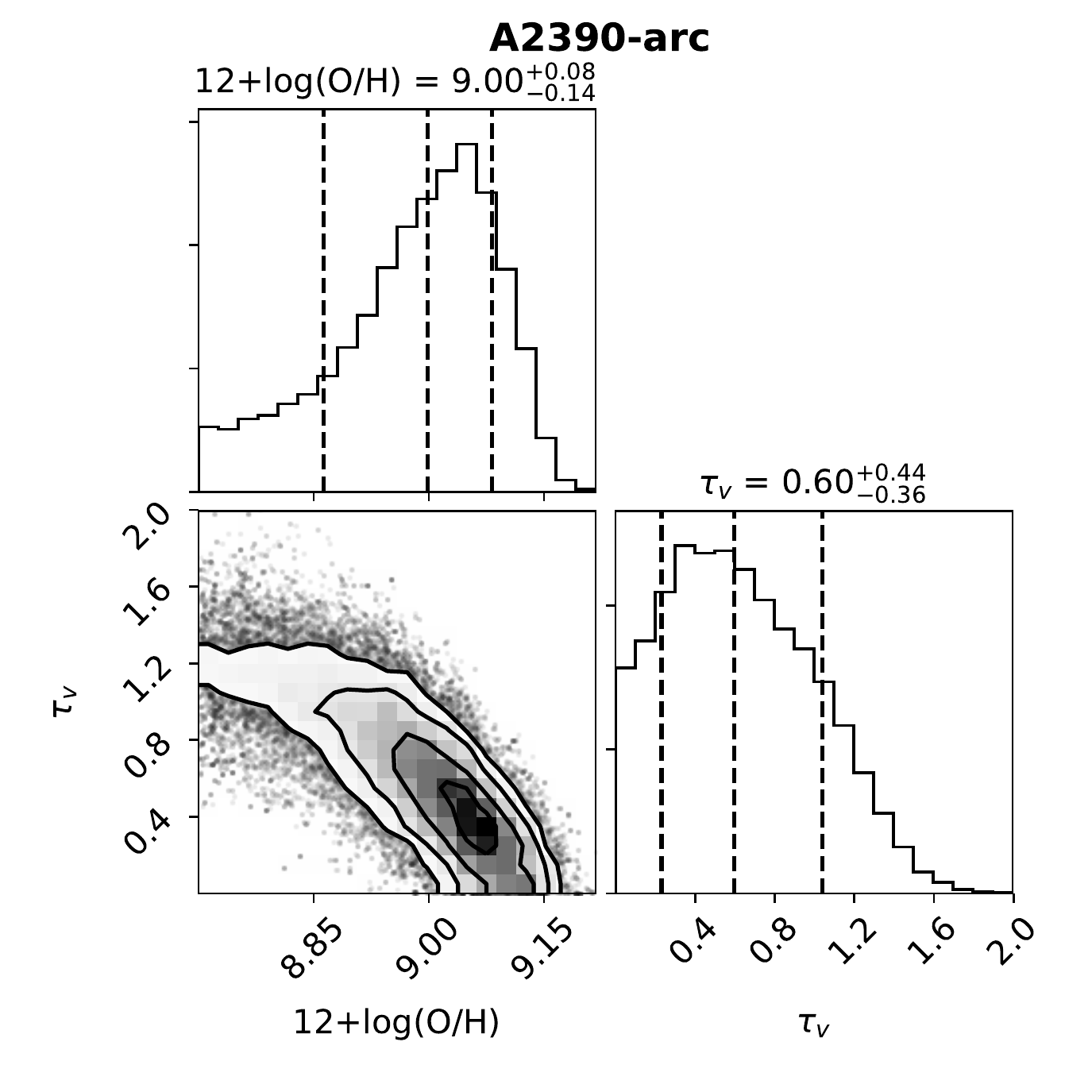}\\
\includegraphics[width=0.23\textwidth]{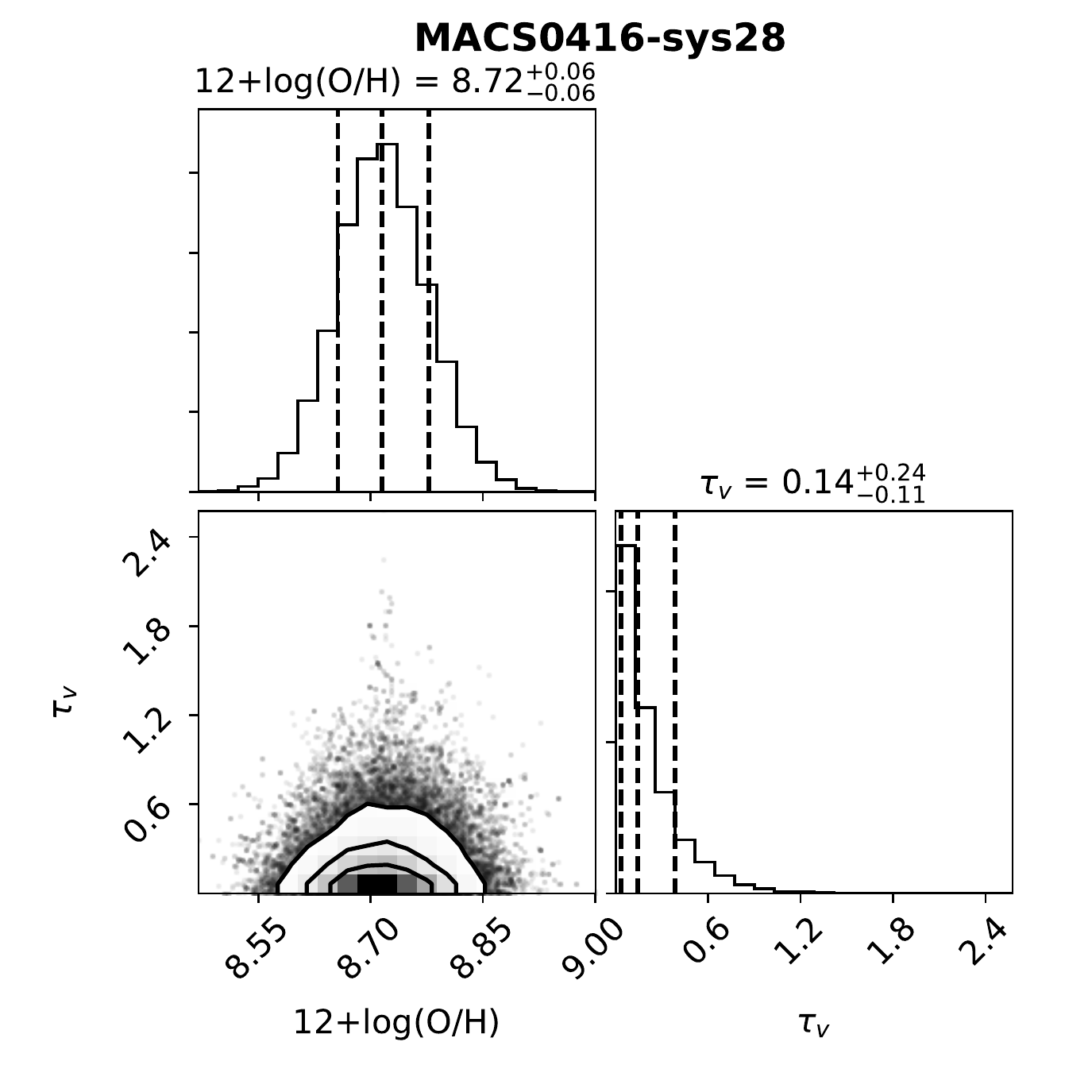}
\includegraphics[width=0.23\textwidth]{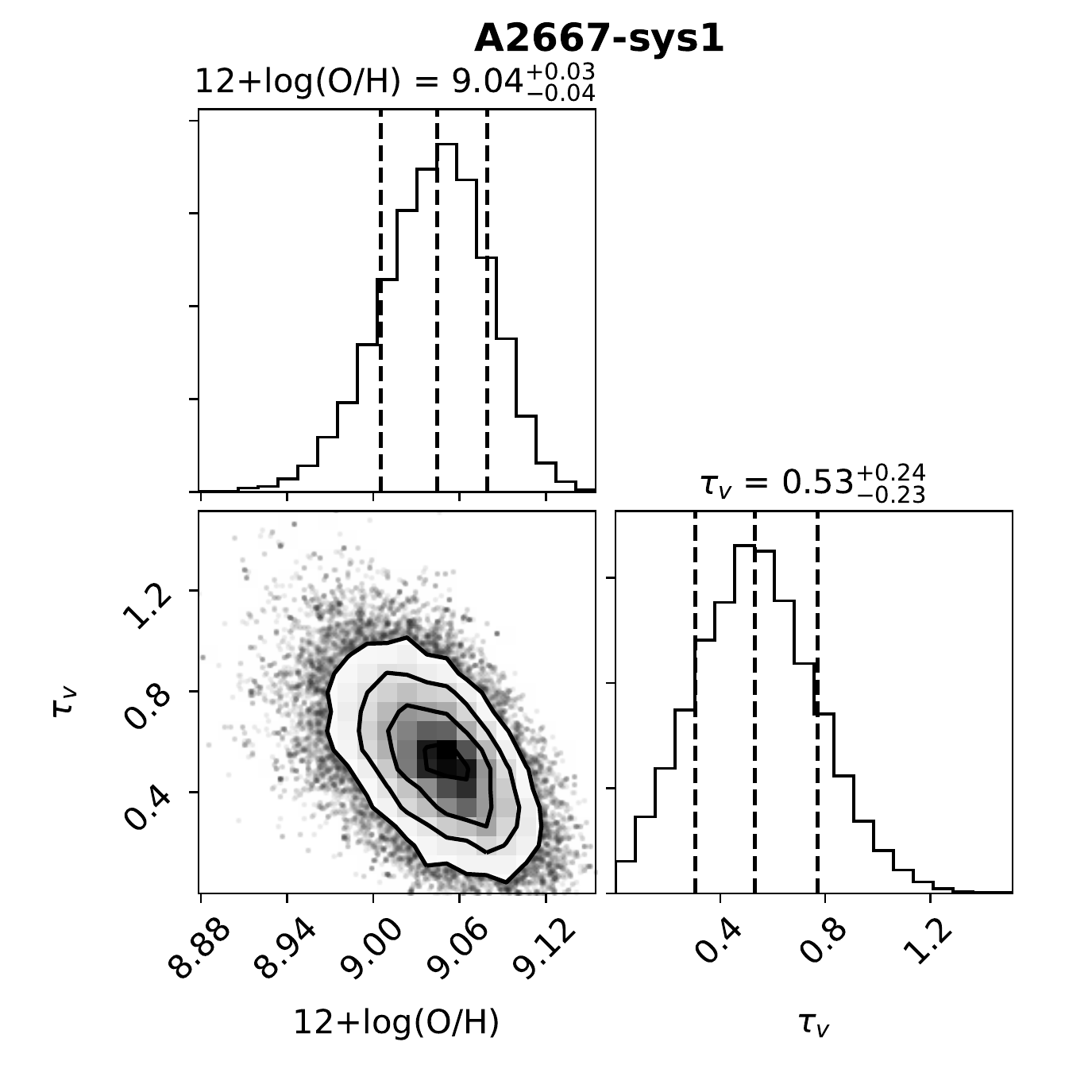}\\
\includegraphics[width=0.23\textwidth]{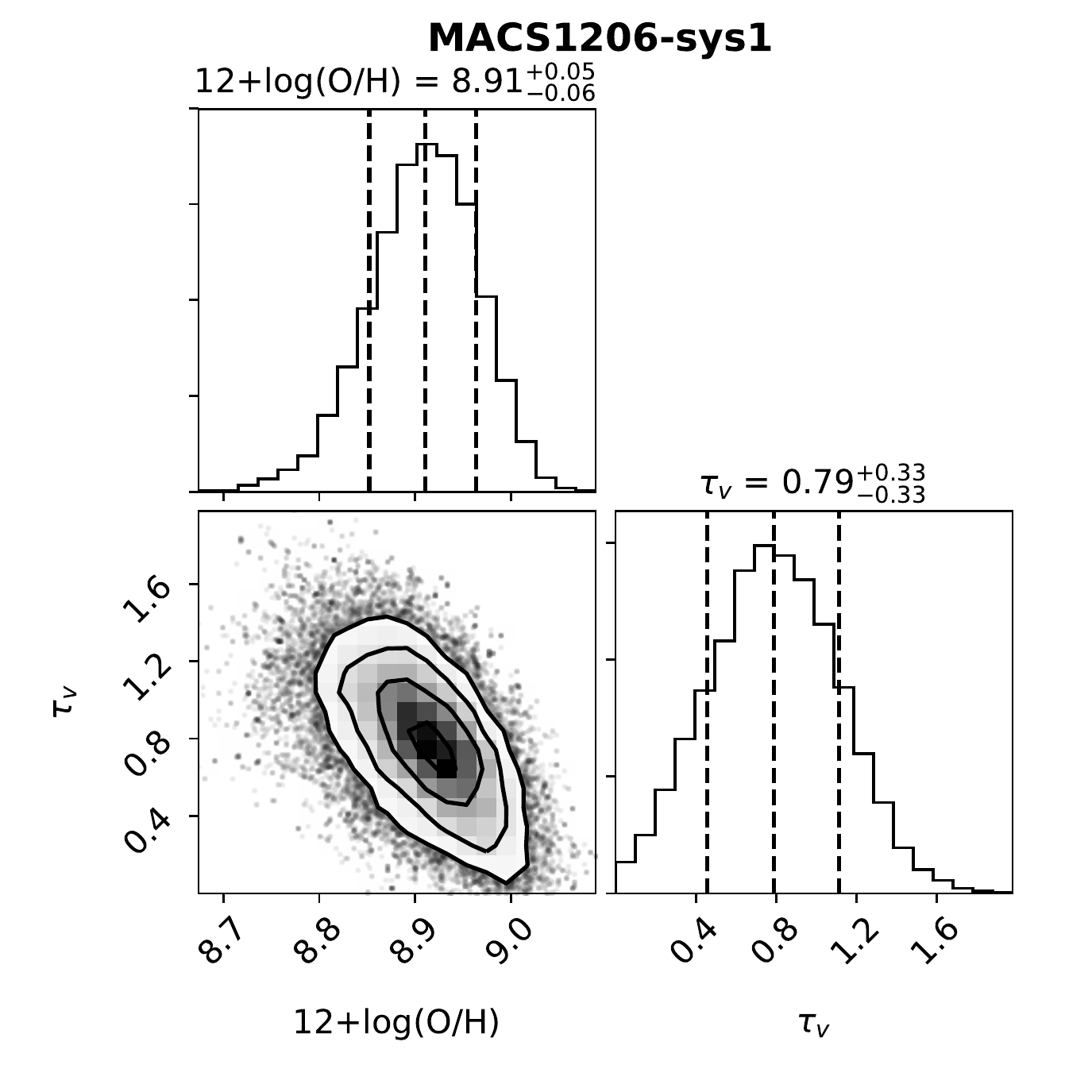}
\includegraphics[width=0.23\textwidth]{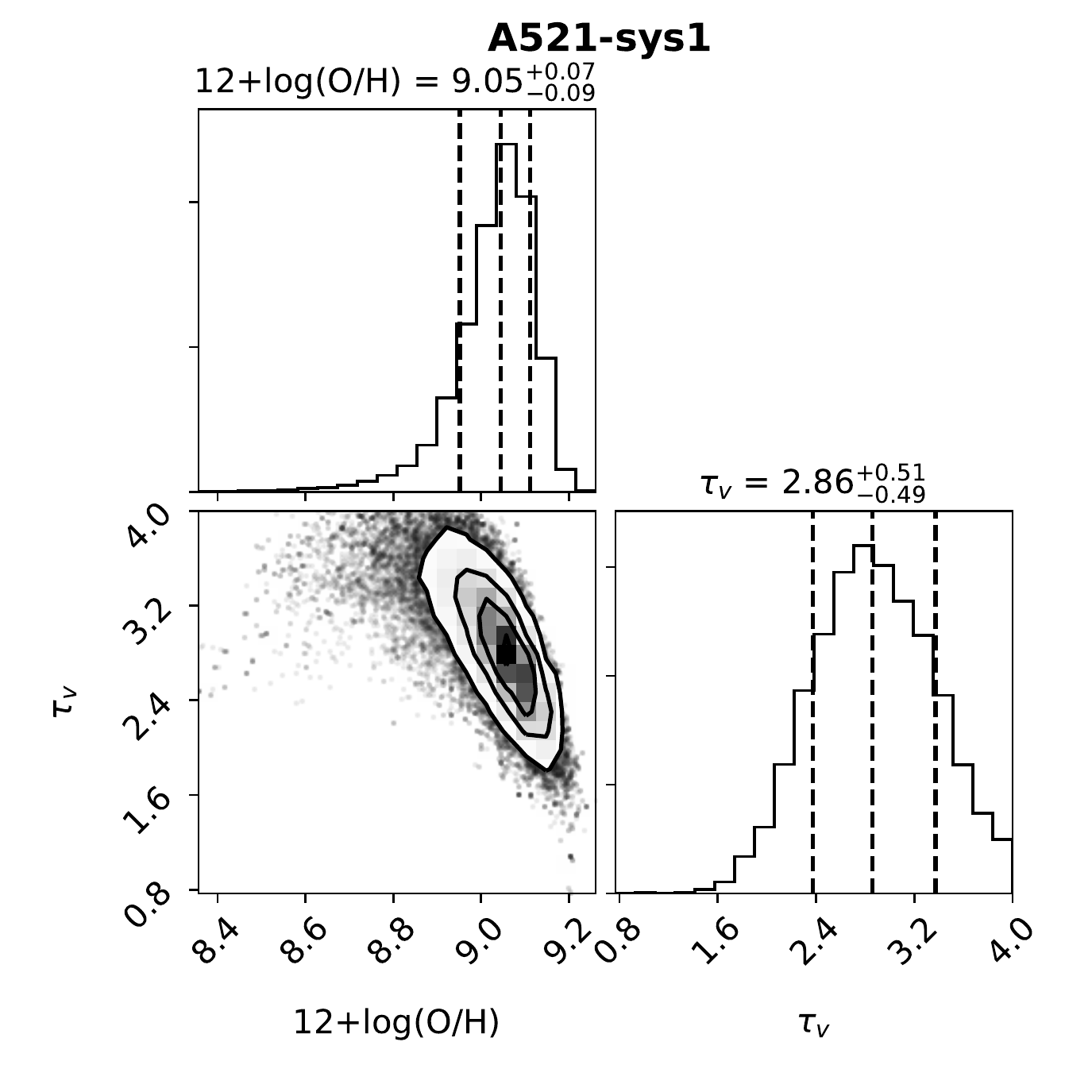}
\caption{Marginalised distributions and joint distribution of metallicity and attenuation for each galaxy obtained from the fit explained in sub-section~\ref{subsec:emission_line_measurements}}
\label{fig:met_cornerplots}
\end{figure}

We present here the marginalised distributions and joint distribution of metallicity and attenuation for each galaxies A370-sys1, A2390-arc, M0416-sys28, A2667-sys1, M1206-sys1 and A521-sys1. The results for AS1063-arc are presented in the main body of this work. M1149-sys1 has only [\oii] available within the MUSE spectral range, reason for which we do not perform this fit in this galaxy.

\section{Individual properties}
\label{sec:properties_each} 

Here we summarize the properties of the individual galaxies derived in this work and compare them with the available studies in the literature for each specific galaxy.

\subsubsection*{AS1063-arc}

This galaxy is a beautiful example of a grand design spiral galaxy whose most striking feature is a gigantic and bright \hii\ region at the edge of one spiral arm. The [\oii] kinematics of the galaxy was briefly studied by  \citealt{Karman2015}, based on the same MUSE observations used in this paper. They found a smooth velocity field with a symmetrical rotation with maximum velocities of up to 200 km\,s$^{-1}$. We confirm this result and also derive the velocity dispersion map that, after being corrected for beam-smearing effects, still displays a higher central velocity dispersion. 
Despite the presence of several star forming clumps visible in the \emph{HST} imaging, none of these coincide with any increases or decreases in the velocity dispersion (Fig.~\ref{fig:2d_kin}). This is true with or without the beam smearing correction. 

\subsubsection*{A370-sys1}

A370-sys1, sometimes called "the dragon", was the first gravitational arc to be spectroscopically confirmed \citep{Lynds1986, Soucail1988}. It extends over 23" in the sky and is composed of several multiple images: the "head" of the dragon is the complete image of the galaxy, and has a somewhat low ($\mu\sim$10) magnification factor, while the "tail" (most elongated structure) is composed by two highly magnified ($\mu\sim$35) multiple images, mirroring each other, and that are only a fraction of the complete galaxy (details in \citealt{Richard2010a}). 
Blue spiral arms are clearly seen in the complete image, with a dozen of strongly star forming regions populating them. As for AS1063-arc, the map shows no particular features at the location of these clumps. The analysis of kinematics also reveals a regular gas rotation, with a twist of the zero-velocity usually associated to the presence of a bar, unknown before this analysis. 

\subsubsection*{A2390-arc}

A2390-arc has been extensively studied in the past both spectroscopically \citep{Pello1991} and photometrically \citep{Cowie2002,Rigby2008,Olmstead2014}. Its kinematics was studied by \citealt{Swinbank2006}, using VIMOS IFU data. The authors derived a maximum velocity (corrected for inclination) of $\sim$187 km$\,$s$^{-1}$, with an inclination of 69 degrees. However, the authors divide this arc in two parts (arcA and arcB, in their figure 3), with arcB corresponding to the $z\sim1$ object seen overlapping with the arc (see Fig.\ref{fig:sample}). The MUSE data also allows to confirm that there is a foreground object at $z\sim1$, but we can still reliably measure the velocity field of A2390-arc at this point. The full arc has a smooth velocity field that is well described by a single velocity model, so we do not divide this arc in two objects as in \citealt{Swinbank2006}. 
The derived kinematic parameters are at odds with the previous analysis by \citealt{Swinbank2006}: we derive a lower inclination, $\sim$57 degrees, and consequently higher maximum velocity of $\sim$220 km$\,$s$^{-1}$. The better quality of the MUSE data, that allows larger radii to be explored as well as the fact that we do not divide this object in arcA and arcB, may explain this difference. 
We derive a mass of $2.39\pm0.5\times10^{10}$ M$_\odot$, with the same order of magnitude as the one derived by \citealt{Olmstead2014} ($0.63\pm0.38\times10^{10}$ M$_\odot$) from rest-frame optical, near-infrared, and mid-infrared photometry. As noted before (Sect.~\ref{subsec:emission_line_measurements}) the lack of visible [\neiii] favours a high metallicity solution ($12+\log$(O/H) = $9.01\pm0.1$). Adopting this high metallicity solution provides a good agreement between the SFR derived from the \citealt{Kennicutt1998} and \citealt{Kewley2004} calibrations is reached ($7.3\pm2.3$ and $7.2\pm5.6$, respectively).

\subsubsection*{M0416-sys28}

This galaxy has the lowest mass on this sample 9.5$\pm$0.25$\times 10^9$ M$_\odot$ and it is also the most compact, making it difficult to fit of the gas kinematics. For example, the transition radius ($r_t$) is not correctly fit for any model, reaching limit values of 12-14 kpc (we impose a limit of 15). We estimate a maximum velocity of 157$\pm$7 \kms and a $V$/$\sigma$ ratio of 6, best fit with an arctangent model. This galaxy is included in the \citealt{Mason2017} sample of lensed galaxies (ID 394) and from H$\alpha$ kinematics the authors derive a maximum velocity of 133$\pm$22 \kms, compatible within errors with our estimate. The authors also derive an intrinsic velocity dispersion of 12$\pm$15 \kms from the integrated spectra, which is in good agreement with our intrinsic dispersion velocity value derived from the 2D map (15$\pm$7 \kms). 

\subsubsection*{A2667-sys1}

The redshift of this galaxy was firstly derived by \citealt{Covone2006} using VIMOS IFU observations. Using \emph{HST} and IRAC data, \citealt{Cao2015} derived its SED and source plane morphology revealing a disc with visible spiral arms, and a negative radial colour gradient, most likely due to a stellar metallicity gradient.

\citealt{Yuan2012} used OSIRIS data to investigate the resolved metallicity of this galaxy through the [\nii]/H$\alpha$ ratio. The central part of the galaxy, that seems to be unaffected by shocks, yields a metallicity of $12+\log$(O/H) = 8.57$\pm$0.03, with a global value of $12+\log$(O/H) = 8.606$\pm$0.051 for the entire galaxy, using the \citealt{PettiniPagel2004} calibrations. The authors also measure a radially increasing [\nii]/H$\alpha$ ratio, but observe that there is evidence of galactic outflows in the outskirts of the galaxy and estimate that shock ionisation (originated by the outflows) contributes significantly to the excitation of [\nii]. \citealt{Yuan2012} conclude that the observed ratio gradient is probably due to the presence of shocks in the outer parts of the galaxy and not by a positive gas-phase metallicity gradient. From the MUSE data, we derive a high-metallicity of $12+\log$(O/H) = 9.08 , in strong disagreement with the $12+\log$(O/H) = 8.6  value previously derived by \citealt{Yuan2012}. Some of this difference can be explained by the different calibrations used. Averaging the [\nii]/H$\alpha$ ratio from \citealt{Yuan2012} over the several spatial regions (Box 1 to 4 in their figure 4) we obtain a value of 0.37, that converted to metallicity using the \citealt{Maiolino2008} calibrations yields a value of $12 +\log$(O/H) = 8.8 , which is still lower than the value derived in this work, but considerably higher ($\sim$0.2 dex) than the one obtained with \citealt{PettiniPagel2004} calibrations. The remaining difference may be explained by the different dust corrections applied in both works. \citealt{Yuan2012} derive a mass of $1.90\pm2.0\times 10^{10}$  M$_\odot$, that agrees within errors with the value of $1.37\pm0.21\times10^{10}$ M$_\odot$ derived in this work. 

On the kinematic side, \citealt{Yuan2012} derive a $V_{\rm max}$ $\sin(i)$/$\sigma$ ratio of $\sim$0.73, which, using the inclination derived in this work ($\sim$21), gives an intrinsic ratio of about 1.05 -- placing it at the limit of rotation-dominated galaxies. In contrast, we obtain a $V_{max}$ /$\sigma$ of 3. As for A2390, this difference can come from the fact that MUSE covers a larger area than OSIRIS (see figure 1 of \citealt{Yuan2012}). Comparing the best 2D kinematic model with the observed one, it seems that the small twist on the zero-velocity curve is not due to lensing effects and might indicate the presence of a small bar. On the velocity dispersion maps corrected for beam smearing ( Fig.\ref{fig:2d_kin}) some residuals are seen (up to $\sim60$ km$\,$s$^{-1}$) but show no significant 2D structure. 

\subsubsection*{M1206-sys1}

MACS1206, nicknamed the \textit{cosmic snake} \citep{Cava2017}, is another example of the "head and tail" structure: it has a long and highly magnified (although not complete) multiple image and a less magnified counter-image, that contains the total spatial information of this galaxy (see \citealt{Eichner2013} for details). From the MUSE and \emph{HST} data, we derive a stellar mass (7.87$\pm0.49\times10^{10}$ M$_\odot$) and, for the first time, the gas-phase metallicity ($12+\log$(O/H) = $8.90\pm0.06$ )  and the star formation rate of (38-43  M$_\odot\,$yr$^{-1}$). This is also a rotation-dominated galaxy, displaying a smooth velocity field. \citealt{Cava2017} performed SED fitting to the \emph{HST} data  and IRAC photometry, deriving a stellar mass of 4.0$\,\times\,10^{10}$ M$_\odot$ and a SFR$_{\rm SED}$ of 30 M$_\odot\,$yr$^{-1}$.

\subsubsection*{A521-sys1}

We present here the first study of the resolved properties in this system. This is a remarkable object, that displays clear spiral arms and numerous star-forming clumps \citep{Richard2010a}. It is also a very good example of a typical disc galaxy at $z\sim1$, with mass and SFR that agree quite well with the main-sequence of star-forming galaxies at these redshifts.

\subsubsection*{M1149-sys1}

Finally, MACS1149, the highest redshift galaxy in the sample, is a spectacular spiral seen in 4 different multiple images. From the MUSE data, only the [\oii] doublet can be seen (besides the \mgii\ and \feii\ emission and absorptions, studied by \citealt{Karman2016}). Also here, the arctangent model provides a better fit to the observed kinematics of the 3 multiple images. \citealt{Yuan2011}, using OSIRIS data, derived the velocity field and metallicity gradient of this galaxy, concluding that it drops from $12+\log$(O/H) = $8.54\pm0.04$ at the centre of the galaxy to $8.05\pm0.05$ in the edges, with a global value of $8.36\pm0.04$. More recently, \citealt{Wang2017} also derived the metallicity of this galaxy using \emph{HST}-grism data (ID 04054), arriving to a value of $12+\log$(O/H) = $8.70\pm0.11$ , that we adopt in this paper.

\bsp	
\label{lastpage}
\end{document}